\documentclass[preprint2]{aastex}\pdfoutput=1

\usepackage{footnote}
\usepackage{graphicx}
\usepackage{subfigure}
\usepackage{natbib}
\usepackage{lscape}
\usepackage{multirow}
\usepackage{color}
\usepackage{longtable}
\usepackage{amsmath}
\makesavenoteenv{tabular}
\shorttitle{Galaxy halo truncation in the Cluster MACSJ1206.2-0847}
\shortauthors{Eichner et al.}
\newcommand{\rt}{r_{\rm t}}
\begin{document}
\title{Galaxy halo truncation and Giant Arc Surface Brightness Reconstruction in the Cluster MACSJ1206.2-0847}

\author{Thomas Eichner\altaffilmark{1,2}, Stella Seitz\altaffilmark{1,2},
  Sherry H. Suyu\altaffilmark{3,4,5}, Aleksi Halkola\altaffilmark{6},
 Keiichi Umetsu\altaffilmark{5}, Adi Zitrin\altaffilmark{7}, Dan
 Coe\altaffilmark{8}, Anna Monna\altaffilmark{2,1}, Piero Rosati\altaffilmark{9}, Claudio Grillo\altaffilmark{10},
 Italo Balestra\altaffilmark{2}, Marc Postman\altaffilmark{8}, Anton
 Koekemoer\altaffilmark{8}, Wei Zheng\altaffilmark{11}, Ole H\o
 st\altaffilmark{10}, Doron Lemze\altaffilmark{11},
 Tom Broadhurst\altaffilmark{13}, Leonidas Moustakas\altaffilmark{14},
Larry Bradley\altaffilmark{8}, Alberto
 Molino\altaffilmark{15}, Mario Nonino\altaffilmark{16}, Amata
 Mercurio\altaffilmark{17}, Marco Scodeggio\altaffilmark{18}, Matthias Bartelmann\altaffilmark{7},
 Narciso Benitez\altaffilmark{15}, Rychard Bouwens\altaffilmark{19},
 Megan Donahue\altaffilmark{20}, Leopoldo Infante\altaffilmark{21},
 Stephanie Jouvel\altaffilmark{12,22}, Daniel Kelson\altaffilmark{23},
 Ofer Lahav\altaffilmark{12}, Elinor Medezinski\altaffilmark{11},
 Peter Melchior\altaffilmark{24}, Julian Merten\altaffilmark{14},
Adam Riess\altaffilmark{8,11}}

\altaffiltext{1}{Universit\" ats-Sternwarte M\" unchen, Scheinerstr. 1, 81679
  M\" unchen, Germany}
\altaffiltext{2}{Max-Planck-Institut f\" ur extraterrestrische Physik,
  Giessenbachstra\ss e, 85748 Garching, Germany}
\altaffiltext{3}{Department of Physics, University of California, Santa Barbara, CA 93106, USA}
\altaffiltext{4}{Kavli Institute for Particle Astrophysics and
  Cosmology, Stanford University, 452 Lomita Mall, Stanford, CA 94035, USA}
\altaffiltext{5}{Institute of Astronomy and Astrophysics, Academia
  Sinica, P. O. Box 23-141, Taipei 10617, Taiwan}
\altaffiltext{6}{Institute of Medical Engineering, University of L\"ubeck, Ratzeburger Allee 160
23562 L\"ubeck, Germany}
\altaffiltext{7}{Institut f\" ur Theoretische Astrophysik, ZAH,
  Albert-Ueberle-Straß e 2, 69120 Heidelberg, Germany}
\altaffiltext{8}{Space Telescope Science Institute, 3700 San Martin
  Drive, Baltimore, MD 21208, USA.}
\altaffiltext{9}{ESO-European Southern Observatory, D-85748 Garching
  bei M\" unchen, Germany}
\altaffiltext{10}{Dark Cosmology Centre, Niels Bohr Institute,
  University of Copenhagen, Juliane Maries Vej 30, 2100 Copenhagen, Denmark}
\altaffiltext{11}{Department of Physics and Astronomy, The Johns
  Hopkins University, 3400 North Charles Street, Baltimore, MD 21218,
  USA}
\altaffiltext{12}{Department of Physics \& Astronomy. University
  College London, Gower Street, London WCIE 6 BT, UK }
\altaffiltext{13}{Department of Theoretical Physics, University of the
  Basque Country, P. O. Box 644, 48080 Bilbao,  Spain}
\altaffiltext{14}{Jet Propulsion Laboratory, California Institute of
  Technology, MS 169-327, Pasadena, CA 91109, USA}
\altaffiltext{15}{Instituto de Astrof\'isica de Andaluc\'ia (CSIC),
  C/Camino Bajo de Hu\'etor 24, Granada 18008, Spain}
\altaffiltext{16}{INAF-Osservatorio Astronomico di Trieste, via
  G.B. Tiepolo 11, 40131 Trieste, Italy}
\altaffiltext{17}{INAF-Osservatorio Astronomico di Capodimonte, via
  Moiariello 16, I-80131 Napoli, Italy}
\altaffiltext{18}{INAF-IASF Milano, Via Bassini 15, I-20133, Milano, Italy}
\altaffiltext{19}{Leiden Observatory, Leiden University, P. O. Box 9513,2300 RA Leiden, The Netherlands}
\altaffiltext{20}{Department of Physics and Astronomy, Michigan State
  University, East Lansing, MI 48824, USA}
\altaffiltext{21}{Departamento de Astrono\'ia y Astrof\'isica, Pontificia
  Universidad Cat\'olica de Chile, V. Mackenna 4860, Santiago 22, Chile}
\altaffiltext{22}{Institut de Cincies de l'Espai (IEEC-CSIC), Bellaterra (Barcelona), Spain}
\altaffiltext{23}{Observatories of the Carnegie Institution of
  Washington, Pasadena, CA 91 101, USA }
\altaffiltext{24}{Center for Cosmology and Astro-Particle Physics, \&
  Department of Physics; The Ohio State University, 191 W. Woodruff
  Ave., Columbus, Ohio 43210, USA}

\begin{abstract}
In this work we analyze the mass distribution of
MACSJ1206.2-0847, especially focusing on the halo
 properties of its cluster members. 
The cluster appears relaxed in its X-ray emission, but has
 significant amounts of intracluster light which is not centrally
 concentrated, suggesting that galaxy-scale interactions are still ongoing
 despite the overall relaxed state. 
The cluster lenses 12 background
galaxies into multiple images and one galaxy at $z=1.033$ into a giant
arc and its counterimage. The multiple image positions and the surface
brightness distribution (SFB) of the arc which is bent around several
cluster members are sensitive to the cluster galaxy halo
properties. We model the cluster mass distribution with a NFW profile
and the galaxy halos with two parameters for the mass normalization and extent of a
reference halo assuming scalings with their observed
NIR--light. We match the multiple image positions at an
  r.m.s. level of $0.85\arcsec$ and can reconstruct the SFB distribution of the arc in several filters to a
remarkable accuracy based on this cluster model. The length scale where the enclosed
galaxy halo mass is best constrained is about 5 effective radii -- a scale
in between those accessible to dynamical and field strong lensing mass
estimates on one hand and galaxy--galaxy weak lensing results on the
other hand. The velocity dispersion and halo size of a galaxy with
$m_{\rm 160W,AB}=19.2$ or $M_{\rm B,Vega}=-20.7$ are $\sigma=150 \rm kms^{-1}$ and
$r\approx 26\pm 6 \rm kpc$, indicating that the halos of the cluster
galaxies are tidally stripped.  We also reconstruct the unlensed source
(which is smaller by a factor of $\sim5.8$ in area), demonstrating the increase
of morphological information due to lensing and conclude that this
galaxy has likely star--forming spiral arms with a red (older) central component.
\end{abstract}

\section{Introduction}
\label{sec:introduction}
For elliptical galaxies the half light radii, central velocity
dispersions and surface brightness within their half light radii form
a fundamental plane \citep{bender1992}. This fundamental plane
relation is very similar for field and cluster galaxies at the same
redshift \citep{andreon1996, saglia2010}. The redshift evolution of the
elliptical galaxies' mass to light ratio is independent of the cluster
velocity dispersion; it is compatible with passive evolution of the
stellar population \citep{bender1998, vanDokkum2007,saglia2010} and
slightly stronger for field galaxies. The effective radii
and velocity dispersions of elliptical galaxies evolve with time, but not
depending significantly on the galaxy environment.\\
Studying elliptical dark matter halos with stellar dynamics,
\cite{thomas05} \& \cite{tho09}
have shown that (1) the stars of elliptical galaxies form at high redshift
(z=3-5), (2) the dark matter halos of (Coma) elliptical galaxies formed
earlier than spiral galaxies of same brightness and environment and
(3) the halos of elliptical galaxies mostly formed at least as early as their
stars (see Fig. 13 of \citealt{wegner2012}).\\
In general, however, galaxy environment plays a major role for the
formation of galaxies and the transforming of galaxy types according
to the morphology-density relation of \cite{dressler80} and their
evolution with redshift \citep{dressler1997}. 
\cite{dressler1997} conclude that
``the formation of elliptical galaxies predates the formation of
rich clusters, and occurs instead in the loose-group phase or even
earlier``. \cite{wilman2012} confirmed this picture in a quantitative way: 
according to their interpretation elliptical galaxies are centrals or they are
satellites which have been centrals in halos before they have been
accreted. Taken together this implies that the central stellar dynamics and the
stellar population content of elliptical galaxies depend on the
present day environment on a minor level.  Elliptical galaxies stay
elliptical galaxies when larger scale halos like groups and clusters
form, but depending on whether they become central or satellite
galaxies their dark matter halos undergo growth or stripping.\\
The stripping of dark matter halos embedded in group and cluster halos
by tidal fields is theoretically expected \citep{merritt83, merritt1984},
 and gets stronger the denser the environment is. Stripping
has also been studied in {\it N}--body dark matter simulations
\citep{ghigna98, limousin2009}. \cite{gao2004} have shown that on average 90 percent
of mass associated with halos accreted at z=1 is removed from the
accreted halos and contribute to the smooth host halo at z=0. Highest
mass accreted halos reach the centers more quickly, due to dynamical
friction, and thus become stripped most quickly. \cite{diemand2007}
have shown that subhalo mass is removed starting from
the outside, in agreement
with the observations that any changes of fundamental plane (FP) relation with environment
can be explained by slight age differences of the stellar populations,
i.e. that the structural parameters of elliptical galaxies do not change
during the build up of groups and clusters. \cite{warnick2008} have shown
that on average surviving subhalos lose about 30 percent of their
mass per orbit in group and cluster halos (this excludes tidally
disrupted halos), where halos with radial orbits may lose 80 per cent
or even more per orbit.  
Their Fig. 4 illustrates the subhalo mass loss sorted as a function of
subhalo distance to the halo center,  for different central
halo masses. Within 10 percent of the virial radius the majority of
subhalos has lost more than 50 percent of its original
mass. \cite{limousin2009} have studied galaxy dark matter halo truncation in
high density environments with hydrodynamical {\it N}--body simulations. They
predict half light radii of galaxies in a Coma and Virgo like cluster
as a function of 3D and 2D projected separation to the
cluster center, finding a measurable effect in both, at a level
stronger than that of \cite{ghigna98}. 
According to their work the total mass of galaxy halos is a few times
larger than its stellar mass in the center and up to about 200 (50)
times larger in the outskirts of the cluster at z=0.7 (z=0).\\
Galaxy halo stripping in clusters has been measured with planetary
nebula kinematics in local galaxies (\citealt{ventimiglia2011} and
references therein). \cite{pu2010} have analyzed the stellar
kinematics of massive local elliptical galaxies and measured halo sizes of
orders of 60 kpc based on the Mgb absorption line strength vs escape velocity relation. These
methods for the analysis of individual galaxy halos do not work for large
samples and larger distances yet.\\
Galaxy halo sizes can also be measured with weak galaxy--galaxy lensing
for field galaxies \citep{schneider1997, hoekstra2004} and
also for cluster galaxies using statistical methods and large
samples. In clusters the effect is stronger per galaxy since the
signal is boosted by the matter of the cluster itself
\citep{geiger1999}, but this imposes also a degeneracy in measuring
the galaxy halos \citep{geiger1999}. Nevertheless halo truncation
has been measured with weak galaxy--galaxy lensing \citep{narayan1998,
geiger1999, natarajan2002a, natarajan2002b, limousin07}, and truncations in half
mass radii by a factor of 4 to field galaxies or more have been reported.\\
\cite{halkola07} have worked out a different idea: Using strong gravitational lensing, they described the
mass distribution in the massive strong lensing cluster Abell 1689
with a smooth dark matter component and a smaller scale component
traced by the cluster galaxies. The combined 'granular' mass
distribution maps multiply imaged galaxies differently than the
best--fitting pure smooth cluster component. Making use of the fundamental
plane and Faber Jackson scaling relations for the cluster galaxies the
properties of a reference halo could be measured. This method finds
the statistically best--fitting reference galaxy halo mass distribution which
reproduces the astrometry of multiply imaged sources best. It relies on
a very precise global mass model (\citealt{broadhurst2005},
\citealt{halkola06}, \citealt{limousin2007}, see also \citealt{diego2005},
\citealt{coe2010}) constrained by a huge number of multiple images (in
this case 32 background galaxies mapped into 107 images) spread over
the Einstein radii corresponding to the various source redshifts.\\
Studying the impact of substructure in the lens  with multiple images
positions does not make use of the full information, since this just
makes use of the differences of deflection angles between multiply
imaged sources and not of higher order or local derivatives of the
deflection angle. This can be done when mapping the full surface
brightness distribution of the images and adjusting the model such
that for every image system of a reproduced source the SFBs
match the observations. \cite{colley1996} were the first to measure the unlensed
surface brightness distribution of the 5 image system in Cl0024 and
thereby helping to constrain the mass distribution of the
cluster. \cite{sei98} analyzed the lensing effect of the cluster
MS1512 using several multiply imaged systems and obtained the surface
brightness distribution of the highly magnified galaxy
cB58 to a unprecedented spatial resolution. In this
  analysis it was important to account for the mass distribution of a
  galaxy perturbing the cB58arc such that it was bent away from the
  cluster center -- although measuring galaxy halos was not the aim of
  this work.\\
Later on \cite{suyuhalkola2010} analyzed the surface brightness
distribution of a source multiply imaged by a galaxy with a satellite
as perturber and could indeed measure the satellite halo size in this way,
showing that the sensitivity of this method can be extended to (still
massive) satellites in favorable lensing systems. 
On cluster lens scale \cite{donnarumma11} used a method similar to
\cite{halkola07} to constrain halo sizes in Abell 611. In this
case one of the sources is mapped into a giant arc system, of which
they used several corresponding surface brightness knots for lens
modeling, thus partially making use of the surface
brightness distribution of the arc in this cluster.\\
In this work we will study galaxy halo truncation in the cluster
MACSJ1206.2-0847, since this is an ideal target for several reasons: MACSJ1206.2 is a massive cluster
at redshift $z=0.439$ (for a summary on properties and lensing, Xray
and SZE results see \citealt{ume12}, \citealt{zitrin2011}). This
cluster shows still signs of its recent assembly, since there is a
'trail' of intra--cluster light along its major axis (in mass and
light), indicating previous tidal stripping down to
the core of galaxies or tidal disruption of galaxies. On the other
hand its central galaxy is almost at rest relative to the center of mass (as
obtained from cluster members' velocities), see Biviano et al. (in
prep.). Further, this
cluster appears relaxed from its Xray contours \citep{ebeling2009,
  ume12}. This means that cluster members orbited each other for
at least a significant fraction of the crossing time, were exposed to
the dense cluster environment and had the necessary (and short) time
to become tidally stripped. Due to its deep multi-band HST photometry
this cluster has many multiple image systems (\cite{zitrin2011}) and
furthermore has a giant arc, which is bent around several cluster
members, making the light deflection of galaxy halos already visible
to the eye. Using the SFB distribution of the arcs and the multiple
images positions, this cluster thus offers the opportunity to provide
very strong constraints on halo sizes.\\ 
This paper is organized as follows: In Section 2 we give an overview
of the data used, in Sect. 3 we present the models for the mass
distribution of the cluster and the halos traced by cluster galaxies,
in Sect. 4 we introduce the scaling relations connecting galaxy
luminosity and dark matter halo properties. In Sect. 5 we obtain a
strong lensing model using only point source constraints from multiple
images and the giant arc. Section 6 then also includes the full
surface brightness distribution of the arc and its counterimage in the
analysis. In Sect. 7 we will discuss our results concerning the scaling
of cluster galaxies' luminosity with their velocity dispersion and
halo sizes and the properties of the unlensed source of the arc's
counterimage. Sect. 8 gives a summary of the work and adds
conclusions. We use WMAP7\footnote{$\rm H_0=71 \rm kms^{-1}Mpc^{-1}$,
  $\Omega_{\rm M}=0.267$, $\Omega_{\Lambda}=0.734$} \citep{wmap7}
cosmology throughout the paper. This
gives a scale of $5.662\, \rm kpc / \arcsec$ at the redshift of the cluster, $z=0.439$.
Einstein radii, convergence and shear values are given in units of
the ratio of the angular diameter distances from the
  lens to the source ($\rm D_{ds}$) and the observer to the source ($\rm D_{s}$), $\rm
D_{ds}D_{s}^{-1}$ if not otherwise stated.
All angles are defined as N over (-E).


\section{Data}
\label{sec:data}
The data used in this work are described in \citet{postman_clash2011},
\citet{zitrin2011} and \citet{ebeling2009}. All raw and reduced HST
imaging data taken by CLASH are public.  We obtain position and shapes
of cluster galaxies with {\sc Sextractor} \citep{sextractor} from the
F606W filter data. The F435W, the F606W and the F814W filter data are
used to extract the surface brightness distribution of the arc and its
counterimage for the lens modeling.  We need a r.m.s.--noise estimate for
each pixel of the giant gravitational arc and its counterimage for the
surface brightness reconstruction.  We obtain the pre-reduced,
publicly available \textit{FLT} images for the F435W, F606W and F814W
filters, respectively.  The pre-reduction, done by \textit{calacs},
includes overscan and bias correction as well as flat-fielding of the
single images. Afterwards, \textit{Multidrizzle} has been used for the
alignment, background subtraction, cosmic--ray rejection and weighted
coaddition of the individual frames and the r.m.s.--noise estimate. The
weighting scheme used is the \textit{ERR}--scheme, where the weighting
is done by the inverse variance of each pixel. From this inverse
variance, we calculate the r.m.s.--noise estimate for each pixel.  For
these frames, we choose a pixel scale of $0.05\arcsec$ resembling the
natural pixel scale of the ACS camera. 
 We verify that
the corresponding star positions in the different filters are accurate
to $\approx 0.5 \rm pix$.

\section{Modeling the cluster and its galaxy component}
\label{sec:modelling theory}
Since we want to measure the parameter values for halo truncation, we
use parametric lens models. The main cluster component is modeled by a
NFW \citep{nfw} halo. Its lensing properties are described in
\citet{wb00} and \citet{golse2002}:
\begin{equation}
\begin{split}
\Sigma(X)=2  r_{\rm s}\delta_{\rm c}\rho_{\rm c}\times& \\ 
\times \left\{ \begin{array}{ll}
  \frac{1}{X^2-1}\left[1-\frac{2}{\sqrt{1-X^2}}\mbox{arctanh} \sqrt{\frac{1-X}{1+X}} \right] & X < 1 \,\\
\frac{1}{3} & X=1 \,\\
\frac{1}{X^2-1}\left[1-\frac{2}{\sqrt{X^2-1}}\mbox{arctan}
  \sqrt{\frac{X-1}{1+X}}
 \right] & X > 1\,.
\end{array} \right. 
\end{split}
\end{equation}
Here $r_{\rm s}$, $\delta_{\rm c}$ and $\rho_{\rm c}$ are the scale
radius and the characteristic overdensity of the halo and the critical
density of the universe for closure at the redshift of the halo. For
the spherical case, $X=\frac{R}{r_{\rm s}}$ denotes the dimensionless
distance in the image plane. 
Following \cite{golse2002, halkola06}, we
  introduce elliptical isopotential contours by
introducing the axis ratio $\rm q=ba^{-1}$ with major
  and minor axes a and b, respectively.  $X=\sqrt{x_1^2/{\rm
    q}+x_2^2{\rm q}}$
then denotes the non--spherical extension of the spherical case above,
with $x_1$ and $x_2$ being the Cartesian coordinates in the major axis
coordinate system. In the following we will only consider the
elliptical case, calling that the NFW profile.
\\
We model the cluster galaxies as \citet{bbs_profile_ref} with their so
called BBS: The density profile is an isothermal sphere with a
``velocity dispersion'' $\sigma$ and a truncation radius $r_t$:
\begin{equation}
\rho(r)=\frac{\sigma^2}{2\pi {\rm G} r^
    2}\frac{\rt^2}{r^2+\rt^2}\,.
\end{equation}
The projected surface mass density is:
\begin{equation}
\Sigma(R)= \frac{\sigma^2}{2 {\rm G} R} \left[ 1 -
  \left(1+\frac{\rt^2}{R^2}\right)^{-0.5} \right] \quad .
\label{theory:BBS-2D-density}
\end{equation}
This gives an enclosed mass within a cylinder of radius $R$ of
\begin{equation}
M(<R)=\frac{\pi \sigma^2}{{\rm G}} \left[R+\rt-\sqrt{R^2+\rt^2}\right] \quad ,
\label{theory:BBS-2D-densityrnclosed mass}
\end{equation}
and a total mass of
\begin{equation}
M_{\rm tot}=\frac{\pi \sigma^2 \rt}{{\rm G}} \quad ,
\label{theory:BBS-2D-density:total mass}
\end{equation}
where G is the gravitational constant and $R$ the 2D-radius. For its
exact lensing properties, see
\citet{bbs_profile_ref}. Following \cite{halkola06},
  ellipticity is again introduced in the potential in the same way as
  in the NFW case.
The truncation radius $\rt$
marks the transition region from a density slope $\rho \sim r^{-2}$ to
a slope of $\rho \sim r^{-4}$. At $\rt$ the projected density is half
the value of the SIS model with the same $\sigma$. For the 3D density
the truncation radius is equal to the half-mass radius of the profile,
see \cite{eliasdottir2007, limousin2009}. For the 2D projected density
the 2D half mass radius is smaller, $r_{1/2,2D}=0.75\rt$.


 \section{Galaxy scaling relations}
\label{sec:scaling relations}

We are not able to precisely constrain galaxy
  halo sizes  for individual cluster members in this cluster.
Therefore we use scaling relations between the different galaxy halos,
based on the luminosity of the individual galaxies
to estimate an average truncation for all halos. As in
\citet{halkola06, halkola07, limousin07} we make use of the
Faber-Jackson \citep{faber76} relation connecting the luminosity
($L$) of early type galaxies with their central stellar velocity
dispersion $\sigma_{\rm star}$ and halo velocity
dispersion\footnote{For this work, we assume these two velocity
  dispersions to be equal.} $\sigma$ with reference values $\sigma^{\star},\,L^{\star}$:
\begin{equation}
\sigma=\sigma^{\star}\left(\frac{L}{L^{\star}}\right)^{\delta}\quad .
\label{theory:Faber Jackson}
\end{equation}
We further assume the truncation radius to scale with luminosity as
\citep{hoekstra2003b, halkola06, halkola07, limousin07} %
\begin{equation}
\rt=\rt^{\star}\left(\frac{L}{L^{\star}}\right)^{\alpha} =\rt^{\star}\left(\frac{\sigma}{\sigma^{\star}}\right)^{\alpha/\delta}
\quad .
\label{truncation_lum_sigma_scaling}
\end{equation}
\begin{table}
\centering
\caption{The scaling parameters for different values of $\delta$,
  $\epsilon$ and $\alpha$.}
\begin{tabular}{cccccccc}
\hline
\multicolumn{4}{c}{Field galaxies} & \multicolumn{4}{c}{Stripped galaxies} \\
 $\delta$ & $\epsilon$ & $\alpha\over\delta$ & $\alpha$ & $\delta$ & $\alpha\over\delta$ & $\alpha$ & $\epsilon_{\rm stripped}$ \\
\hline
0.3   & 0.2 &   2  & 0.6 & 0.30   & 1 & 0.30  &-0.10  \\
0.25 & 0.0 &   2  & 0.5  & 0.25   & 1 & 0.25  &-0.25   \\
0.25 & 0.2 & 2.8  & 0.7  & 0.233 & 1 & 0.233 &-0.30  \\
0.3   & 0.0 & 4/3 & 0.4 &&&&\\
\hline
\end{tabular}
\label{theory:scaling:alpha delta epsilon}
\end{table}
Here, $\sigma^{\star}$ and $\rt^{\star}$ are the 
parameter values for a galaxy halo with reference luminosity
$L^{\star}$. In order to specify the scaling relations,
we need to find appropriate values for $\alpha$ and
$\alpha\over\delta$. The values for the  Faber--Jackson slope $\delta$ quoted in literature
depend on the wavelength range used for the luminosity measurement 
 and on the considered magnitude range
\citep{nigoche2011,focardi2012}. For the B-band
relation we will in the following consider slopes between $\delta=0.3$
(\cite{ziegler1997}) and $\delta=0.25$ 
 \citep{fritz2009, kormendy2013, focardi2012}. 
 Further,
\cite{bernardi2003a} find a value of $\delta=0.25$ for
elliptical galaxies in each of the Sloan Digital Sky Survey $g^*r^*i^*z^*$
bands as well. However, there are indications for
an increase in $\delta$ for fainter elliptical galaxies (see e.g
\citealt{matkovic2005} and references therein). We therefore assume
$\delta$ to be equal to $0.3$ for our analysis. This value has also been
found by \citet{rusin03} from gravitational lensing of field
elliptical galaxies. The exact choice for $\delta$ is not relevant for
our work, since we are not able to distinguish a scaling relation with
a slope of, e.g., $\delta=0.27$ from one with a slope of $0.3$.\\
To limit the reasonable range for the truncation scaling $\alpha$ we
consider the mass to light ratio of galaxies: this total
mass to light ratio is usually described by a power law as well,

\begin{equation}
 {M_{\rm tot} \over L} \propto L^\epsilon \propto \sigma^{\epsilon /
   \delta}\quad .
\end{equation}

Using $M_{\rm tot} \propto \sigma^2 \rt$
(Eq. \ref{theory:BBS-2D-density:total mass}) with
  Eqs. \ref{theory:Faber Jackson} and \ref{truncation_lum_sigma_scaling}, 
we obtain for the same mass to light ratio
\begin{equation}
{M_{\rm tot} \over L} \propto \sigma^{2+\alpha / \delta
  -1/\delta}\quad , 
\end{equation} 
hence, we obtain the following relation of the power law indices
\begin{equation}
\frac{\alpha}{\delta} = \frac{\epsilon}{\delta}-2+\frac{1}{\delta}
\quad .
\label{powerlaw_relations}
\end{equation}
This shows that the scaling relations are fully determined
  by fixing the values for 2 of the parameters $\epsilon$,
  $\alpha$ and $\delta$. Thus, if we fix the $\epsilon$ range for the mass to
  light scaling we also fix the interval for the truncation scaling $\alpha$. 
The ratio for the elliptical galaxies' central dynamical mass and their light
is  $M_{\rm dyn}/L\propto L^{\epsilon_{\rm FP}}$, with a fundamental
plane slope of $\epsilon_{\rm FP}\approx 0.2$ \citep{bender1992}. The exact value depends also on the
filter used to measure the luminosity, see \cite{barbera2011}.  
Strong lensing analyses which measure the central $\rm M_{\rm tot}/L$ also
obtain a scaling of  the central $\rm M_{\rm tot}/L\propto
L^{\epsilon}$ with $\epsilon=0.2$ (see e.g. \citealt{gri09,
  slacs10}). Weak lensing analyses for field galaxies arrive at the
same scaling for the total dark matter to light ratio \citep{fabrice12}.\\
For halos in a dense environment, however, we expect the stripping radius
to be \citep{merritt83}
\begin{equation}
\rt \propto M_{\rm tot} ^{1/3} \quad ,
\end{equation}
and with $M_{\rm tot} \propto \sigma^2 \rt$, we obtain $\alpha/\delta=1$.
 The mass velocity relation then becomes $M_{\rm tot}
\propto \sigma^3$. This gives for the mass to light ratio using
Eq. \ref{theory:Faber Jackson}:

\begin{equation}
\frac{M_{\rm tot, stripped}}{L} \propto L ^{\epsilon_{\rm stripped}} \propto\sigma ^{3-\delta^{-1}} \quad.
\end{equation}

And thus

\begin{equation}
\epsilon_{\rm stripped} = 3\delta-1 = 3\alpha-1 \quad.
\end{equation}

Thus the power law index for
the mass to light ratio for stripped halos as function of light is
negative and of the order of
$\epsilon_{\rm stripped} = -0.3$ to $\epsilon_{\rm stripped}=-0.1$,
depending on the value of $\delta$, see Table
\ref{theory:scaling:alpha delta epsilon}. In summary, we expect the
value of $\epsilon$ to be between $\epsilon=0.2$ and $\epsilon=-0.3$, where the maximum and
minimum values refer to the cases where no halo stripping has yet been
taking place and the case where halo stripping has been completed. 
MACSJ1206.2-0847 shows signs for both relaxation and thus completed halo
stripping and for ongoing build up and thus still ongoing halo stripping.
Therefore,
we choose a value for the  mass to light scaling between that for
isolated field galaxies and the value expected for finalized stripping
in the dense cluster center and we thus take $\epsilon=0$.
 Our choices for $\epsilon$ and $\delta$ lead to the following equation for the truncation scaling:
\begin{equation}
\sigma=\sigma^{\star}\left(\frac{L}{L^{\star}}\right)^{0.3}\quad , \quad
\rt=\rt^{\star}\left(\frac{\sigma}{\sigma^{\star}}\right)^{\frac{4}{3}}
\quad .
\label{theory: galaxy scaling Rusin}
\end{equation}

This scaling relation between the velocity
dispersion and truncation radius is adopted in most parts of
the paper. However, we also investigate whether
the   measurements of the halo sizes changes if we assume
$\delta=0.25,\, \epsilon=0$. We find no significant changes of our
results. Throughout this work, we assume Eq. \ref{theory:
    galaxy scaling Rusin} (or its modification $\delta=0.25,\,
  \epsilon=0$) to hold for all galaxies independent of the distance of the galaxy to the
  cluster center, i.e. a galaxy with velocity dispersion
  $\sigma^{\star}$ (and luminosity $L^{\star}$) always has a size of
  $\rt^{\star}$. In this work, we only investigate the central, dense,
  strong lensing region, meaning that we get an average truncation for
all galaxies in the dense center. We cannot study truncation in less dense
environments by extending the analysis done in this work to
larger distances from the cluster center, since it relies on the strong lensing
effect. Instead the analysis would have to be repeated in the centers of less dense clusters
or groups of galaxies.


\section{Strong lensing model for point-like sources}
\label{sec:point-like strong lensing model}
The first redshift measurement of the
  giant arc as well as the velocity dispersion and redshift of the BCG
was reported by \cite{sand2004}.
The first strong lensing model for cluster MACSJ1206.2-0847 was
published by \citet{ebeling2009}, based on 2 surface brightness peaks
multiply mapped into knots on the giant arc and its counterimage.  The
CLASH data allowed \citet{zitrin2011} to identify 12 multiply imaged
systems lensed into 52 multiple images. Distances for the lensed
galaxies were inferred from spectroscopic redshifts if available or precise photometric
redshifts.  In the following, we use a parametric strong lensing model
for the dark matter and the cluster members close to the
strong lensing area. We describe the model input first, followed by the
results.
\subsection{Model ingredients}
\label{subsec:model ingredients}
For the point-like strong lensing analysis, we need two ingredients:
The point-like multiple image positions and models for the cluster
scale mass distribution and its substructure as traced by the cluster galaxies.
\subsubsection{Multiple image systems}
\label{sec:strong lens modelling image systems}
We start with similar sources as \citet{zitrin2011}, Table 1, but
modify this selection. In Table \ref{modeling:point like image
positions} we present our multiple image identifications, their
positions are shown in Fig. \ref{modeling:point like cluster
image}. The differences to \citep{zitrin2011} are as follows: First,
we keep the systems 2,3,4,5,6,7,8,12,13 unchanged. We split the Arc
system 1 into 3 subsystems at the same redshift using
  corresponding surface brightness peaks, labeled ``1a'',
``1b'' and ``1c'', see
also Fig. \ref{modelling:extended:reconstruction:colors}. Since
systems 2 and 3 are two brightness peaks in the same source, we
replace these systems by numbers 2b and 2c. For the systems 9 and 10, 
\citet{zitrin2011} state an ambiguity of images 9.3, 9.4, 10.3 and 10.4. We implement
these images as 10.3 and 10.4 only: First, the surface brightness
distribution of 10.3 and 10.4 looks more similar to 10.1 and 10.2 than
9.1 and 9.2 and second, also the best-fit model gives a significantly
better fit to this identification of the observations than 9.3 and
9.4. Also, for these systems, we neglect the only probable counterimages
9.5 and 10.5 of \citet{zitrin2011}.
For system 11, we also neglect the candidate images 11.1 and 11.2,
keeping 11.3 to 11.5 as a triple imaged system
only. Our best fit model does indeed not predict the
  multiple images 11.1 and 11.2 and gives model positions 9.5 and 10.5
$6.2\arcsec$ and $9.5\arcsec$ away from the positions given in
  \cite{zitrin2011}, respectively. However there is no certain
  identification possible for these images.
\\
We use the spectroscopic redshift of image systems measured as part of
a VIMOS campaign at the VLT where these are
available. Otherwise, we combine the available photometric redshifts
in \citet{zitrin2011} into an error weighted mean redshift and mean
error for each multiple image system belonging to the same source. The
mean redshift becomes the central value for a Gaussian shaped redshift
prior, and the mean redshift error becomes the $1 \sigma$ width of
this prior.  This gives an approximate, more conservative estimate for the
uncertainties of the redshifts than the r.m.s.--error of the mean. Any
systematic uncertainty in the photometric redshift estimate is equally
present in the estimate of each multiple image, since they have the
same color. Therefore a pure statistical error would underestimate the
true uncertainty of the photometric redshift. These photometric redshifts constraints of the multiple image
 systems are used as priors in the model optimization.
\\
We adopt a
value of $0.5\arcsec$ for the positional uncertainty of the multiple
images. This value is driven by line--of--sight (LOS) structure
and substructure not accounted for in the lens modeling, since the
measurement error of the positions of the multiple images is usually
only a fraction of a pixel.  \cite{jullo2010} estimate the LOS
structure to produce an r.m.s. image position scatter of $\approx 1
\arcsec$ for a cluster like A1689.  \citet{host12} estimates a
relative LOS structure deflection angle depending on the
distance from the cluster center and the redshift of the source to be
$0.5\arcsec$ to $2.5\arcsec$ for typical strong lensing situations.
\begin{table}
\centering
\caption{Multiple image positions}
\begin{tabular}{|c|c|c|c|c|c|}
\hline
Obj & $\Theta_1$\footnotemark[1] & $\Theta_2$\footnotemark[1] & $\rm
z_{input}$ & $\rm z_{model}$ \\
id & ($\arcsec$) & ($\arcsec$) & &  \\
\hline
1a.1 & 12.85 & 19.73 & 1.033\footnotemark[2] & 1.033\footnotemark[2]  \\
1a.2 &  20.76 &     3.46 & 1.033\footnotemark[2] & 1.033\footnotemark[2]  \\
1a.3 &  19.56 &    -6.79 & 1.033\footnotemark[2] & 1.033\footnotemark[2]  \\
1b.1 &  13.72 &    18.91 & 1.033\footnotemark[2] & 1.033\footnotemark[2]  \\
1b.2 &  20.71 &     4.96 & 1.033\footnotemark[2] & 1.033\footnotemark[2]  \\
1b.3 &  19.71 &    -7.54 & 1.033\footnotemark[2] & 1.033\footnotemark[2]  \\
1c.1 &  12.46 &    20.26 & 1.033\footnotemark[2] & 1.033\footnotemark[2]  \\
1c.2 &  19.56 &    -5.84 & 1.033\footnotemark[2] &
1.033\footnotemark[2]  \\
\hline
2a.1 & -35.30 &  -28.95 & 3.03\footnotemark[2] & 3.03\footnotemark[2]  \\
2a.2 & -42.15 &  -14.20 & 3.03\footnotemark[2] & 3.03\footnotemark[2]  \\
2a.3 & -42.65 &   15.40 & 3.03\footnotemark[2] & 3.03\footnotemark[2]  \\
2b.1 & -33.60&    -30.95 & 3.03\footnotemark[2] & 3.03\footnotemark[2]
 \\
2b.2 & -42.15 &   -12.85 & 3.03\footnotemark[2] & 3.03\footnotemark[2]  \\
2b.3 & -42.30 &    14.65 & 3.03\footnotemark[2] & 3.03\footnotemark[2]  \\
2c.1 & -34.00 &   -30.45 & 3.03\footnotemark[2] & 3.03\footnotemark[2]  \\
2c.2 & -42.11 &   -13.15 & 3.03\footnotemark[2] & 3.03\footnotemark[2]  \\
2c.3 & -42.30 &    14.85 & 3.03\footnotemark[2] & 3.03\footnotemark[2]
 \\
\hline
4.1 &   14.37  &   12.57 & 2.54\footnotemark[2] & 2.54\footnotemark[2]  \\
4.2 &  -6.43   &  21.42  & 2.54\footnotemark[2] & 2.54\footnotemark[2]  \\
4.3 & -15.10   &   2.74  & 2.54\footnotemark[2] & 2.54\footnotemark[2]  \\
4.4 &   0.62   &   3.63  & 2.54\footnotemark[2] & 2.54\footnotemark[2]  \\
4.5 &   6.36   & -39.21  & 2.54\footnotemark[2] & 2.54\footnotemark[2]  \\
\hline
5.1 & -21.60 &    17.60 & $1.73\pm0.17$\footnotemark[3] & 1.59  \\ 
5.2 & -22.30 &    -2.80 & $1.73\pm0.17$\footnotemark[3] & 1.59  \\ 
5.3 &  -6.50 &    -30.45 &$1.73\pm0.17$\footnotemark[3] & 1.59  \\
\hline
6.1 &   13.95 &   28.15 & $2.73\pm0.15$\footnotemark[3] & 1.86  \\
6.2 &   22.36 &  -23.50 & $2.73\pm0.15$\footnotemark[3] & 1.86  \\
6.3 &   26.25 &   11.30 & $2.73\pm0.15$\footnotemark[3] & 1.86  \\
\hline
7.1 & -56.30  &  -15.10 & $3.82\pm0.3$\footnotemark[3] & 2.90  \\
7.2 & -55.60  &  -19.30 & $3.82\pm0.3$\footnotemark[3] & 2.90  \\
7.3 & -53.10  &  -24.30 & $3.82\pm0.3$\footnotemark[3] & 2.90  \\
7.4 & -56.29  &  -13.62 & $3.82\pm0.3$\footnotemark[3] & 2.90  \\
7.5 & -56.61  &  -12.68 & $3.82\pm0.3$\footnotemark[3] & 2.90  \\
\hline
8.1 &  -2.67  &   34.72 & $5.46\pm0.29$\footnotemark[3] & 5.03  \\
8.2 &  23.27  &   13.86 & $5.46\pm0.29$\footnotemark[3] & 5.03  \\
8.3 & -16.33  &   -0.46 & $5.46\pm0.29$\footnotemark[3] & 5.03  \\
8.4 &  13.01  &  -40.68 & $5.46\pm0.29$\footnotemark[3] & 5.03  \\
\hline
9.1 &    8.95 & 14.05 & $1.73\pm0.23$\footnotemark[3] & 1.64  \\
9.2 &    2.40 &  16.55& $1.73\pm0.23$\footnotemark[3] & 1.64  \\
\hline
10.1 &   0.35 &     18.95 & $1.34\pm0.26$\footnotemark[3] & 1.69  \\
10.2 &  12.30 &    10.70  & $1.34\pm0.26$\footnotemark[3] & 1.69  \\
10.3 &  -5.55 &     2.00  & $1.34\pm0.26$\footnotemark[3] & 1.69  \\
10.4 &  -2.45 &     2.25  & $1.34\pm0.26$\footnotemark[3] & 1.69  \\
\hline
11.3 &  -10.79 &   19.02 & $1.35\pm0.44$\footnotemark[3] & 1.44  \\
11.4 & -13.87  &   -0.56 & $1.35\pm0.44$\footnotemark[3] & 1.44  \\
11.5 &   2.38  &  -28.57 & $1.35\pm0.44$\footnotemark[3] & 1.44  \\
\hline
12.1 &  -19.04  &  33.42 & $3.84\pm0.52$\footnotemark[3] & 3.28  \\
12.2 & -24.78   &  -7.58 & $3.84\pm0.52$\footnotemark[3] & 3.28  \\
12.3 &  -3.95   & -36.07 & $3.84\pm0.52$\footnotemark[3] & 3.28  \\
\hline
13.1 & -10.99  &  -37.61& $3.18\pm0.99$\footnotemark[3] & 2.34  \\
13.2 & -29.83  &   -1.72& $3.18\pm0.99$\footnotemark[3] & 2.34  \\
13.3 & -28.73  &   17.18& $3.18\pm0.99$\footnotemark[3] & 2.34  \\
\hline
\end{tabular}\\
\footnotemark[1]{relative to the center of the BCG at 12:06:12.134 RA
  (J2000) -08:48:03.35 DEC (J2000)}\\
\footnotemark[2]{spectroscopic redshift, fixed}
\footnotemark[3]{photometric redshift estimate, weighted mean and error}\\
\label{modeling:point like image positions}
\end{table}

\subsubsection{Cluster galaxies tracing dark matter substructure}
\label{sec:point-like modeling:model lenses}
We use the BPZ Photometric redshifts \citep{bpz2000, bpz2004,coe2006}
as described in \citet{postman_clash2011} and spectroscopic
information for this cluster (Rosati et al, 2013, in prep) wherever
available for the cluster member selection.  For simplicity, we
consider as cluster members galaxies with
spectroscopic redshifts between z=0.43 and 0.45; all other galaxies with different spectroscopic
redshifts are excluded.
\\
For galaxies lacking spectroscopic redshifts we use the photometric
redshift estimates and consider all galaxies with a best--fitting photometric
redshift estimate between 0.39 and 0.49 and a $95$ \% confidence
interval width smaller than 0.5 (i.e. $c.l.(95\%)_{\rm max}-c.l.(95\%)_{\rm min}<0.5$) as cluster
members as well. 
From these cluster galaxies, we use only a subsample
which fulfill 2 criteria: First, we only use those within a
$3\arcmin\times3\arcmin$-sized box centered on the BCG
to cover the strong lensing area only. Second, these
galaxies have to trace a sufficiently massive halo to be relevant for
the lens modeling: From the galaxy sample we pick the second
brightest galaxy of this cluster, located at 12:06:15.647 RA
(J2000), -08:48:21.88 DEC (J2000) as the reference galaxy (called hereafter GR), see
Fig. \ref{modeling:point like cluster image}. 
We use the F160W fluxes
of the cluster members in units of GR and use Eq. \ref{theory:Faber
Jackson} to scale the velocity dispersions relative to GR.

\begin{figure}[tbh]
\centering
\includegraphics[width=80mm]{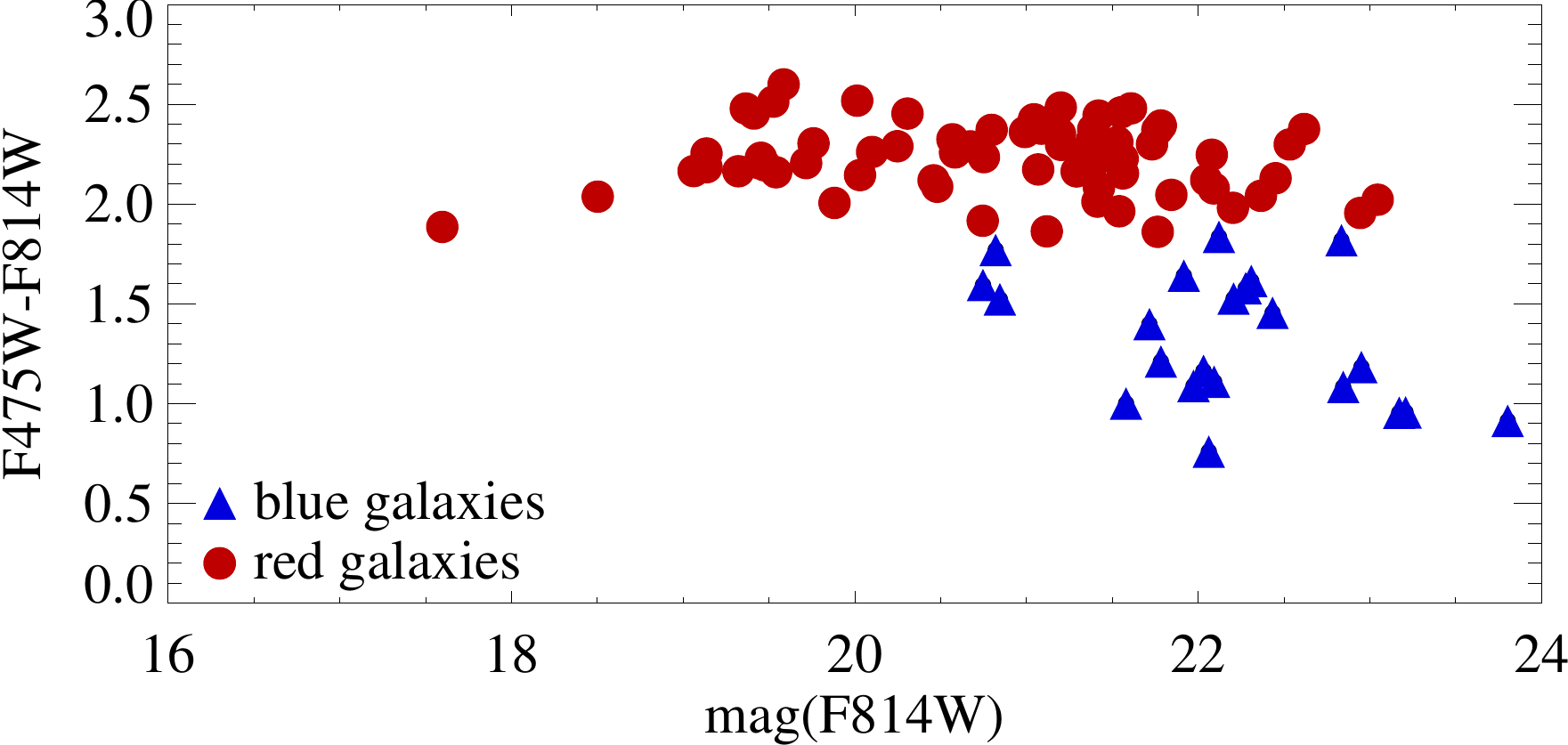}
\caption{The color-magnitude diagram of the selected cluster galaxy
lenses. Plotted is the F475W-F814W color against the F814W magnitude
of the galaxies. We mostly select red galaxies with similar
color. Since we do not select by galaxy color but by photometric and
spectroscopic redshift, we also identify some bluer galaxies as
cluster members, which would not have been possible based on a pure
red sequence cut. The typical error on the magnitude
  and color is smaller than the symbol size. The color indicates the
  SED type of galaxies, separated in red and blue galaxies.}
\label{modeling:point like:galaxy lenses}
\end{figure}

\begin{figure}[tbH!]
\centering
\includegraphics[width=50mm,angle=-90]{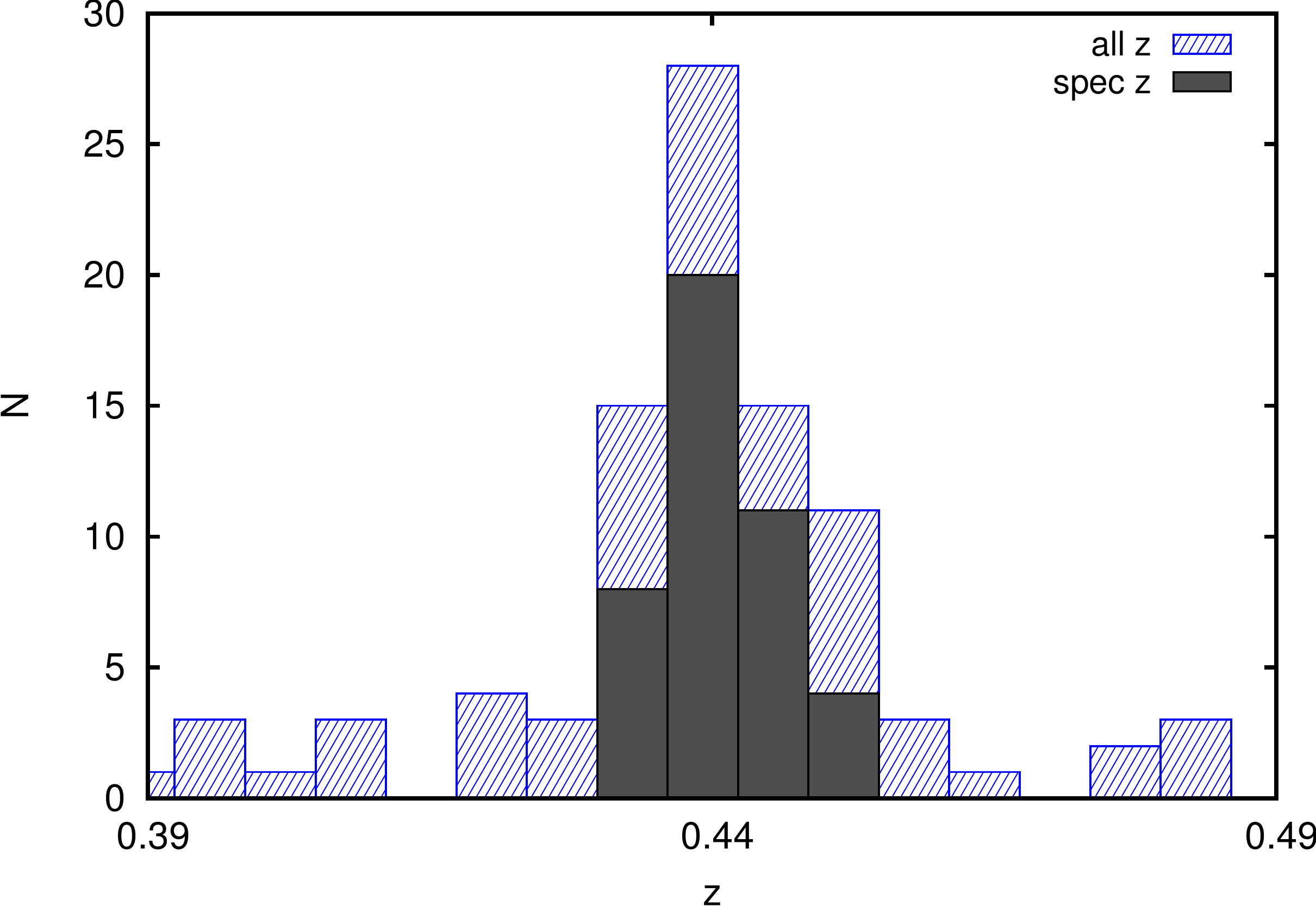}
\caption{The redshift distribution of the modeled cluster members is
  plotted in this figure. The spectroscopically selected members are
  drawn as the gray, solid histogram, all selected
  members are shown as blue, hatched distribution. As can be seen,
  both distributions peak at $\rm z\sim0.44$, giving the redshift of the
  cluster.}
\label{modeling:point like:galaxy redshifts}
\end{figure}

\begin{figure*}
\centering
\includegraphics[width=18cm]{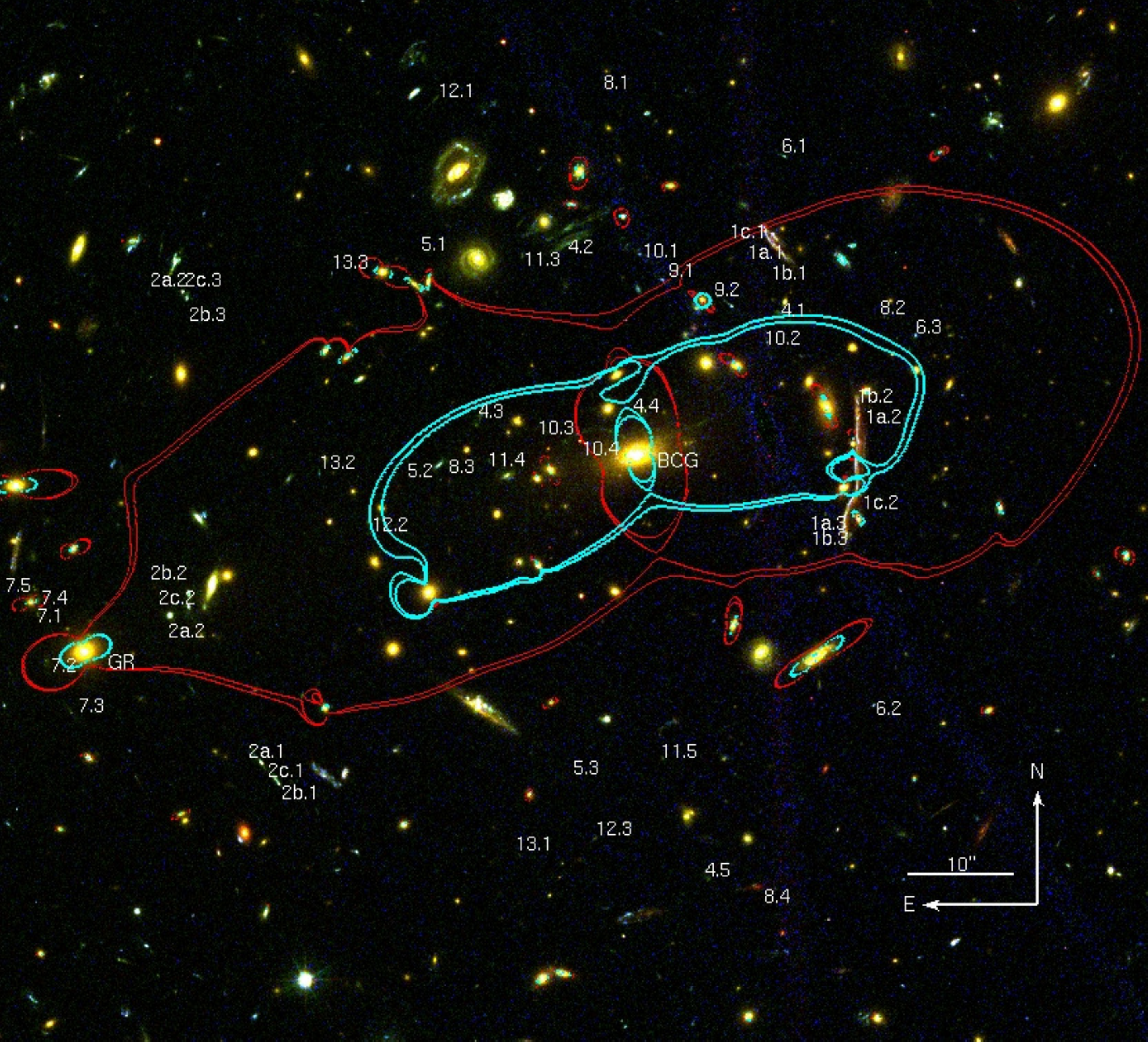}
\caption{A $110\arcsec\times100\arcsec$ cutout of the cluster
center. The multiple image systems are labeled according to Tab
\ref{modeling:point like image positions}. We have added the critical
lines for a source at the redshift of the arc ($z=1.03$) in cyan and for a source
at $z=2.54$ in red. The critical lines are calculated from a pixelated
magnification map, enclosing the high magnification areas of the
image. The BCG and the reference galaxy GR are marked in the
image. North is up and east is left. This color composite image is
made from the F435W, F606W and F814W HST/ACS filter data.}
\label{modeling:point like cluster image}
\end{figure*}

We convert the velocity dispersions in a ``cosmology-free'' Einstein
radius by
\begin{equation}
\Theta_{\rm E}=\frac{4 \pi \sigma^2}{\rm c^2}
\end{equation}
with $\rm c$ being the vacuum speed of light. We explicitly model only those cluster galaxies
which have an Einstein radius larger than 3\% of the Einstein radius of
GR, meaning that we neglect galaxies with an Einstein
radius smaller than $\sim 1 \, \rm pix$.
The redshift distribution of the finally
    selected cluster members, splitted into
    galaxies selected spectroscopically and photometrically, is
    plotted in Fig. \ref{modeling:point like:galaxy redshifts}.
Both in the spectroscopic and the photometric redshifts,
  the cluster is clearly visible as one peak at redshift $\rm
  z=0.44$. 
The cluster members form a red sequence in
  color--magnitude space, see Fig. \ref{modeling:point like:galaxy
    lenses}, with a minor fraction of glaxies being classified as
  blue. The
  distribution of these galaxies in color-magnitude space is shown in
  Fig. \ref{modeling:point like:galaxy lenses}.\\
For the selected cluster members, an Einstein radius of $1\arcsec$ corresponds to a
  velocity dispersion $\sigma=186 \rm kms^{-1}$.
Looking at Eq. \ref{theory: galaxy scaling Rusin} we
  note that we need to measure 2 values to fully determine the halo properties:
  $\sigma^{\star}$ and $\rt^{\star}$. We use 2 different sets of parameters:
$r_{\rm t,1\arcsec}$, for a reference $\sigma=186 \rm kms^{-1}$ which gives the
  value for a galaxy with an Einstein radius of $\Theta_{\rm E}=1\arcsec$, 
and $r_{\rm t,GR}$ which gives the truncation radius for galaxy GR itself.

With this procedure, we obtain 92 galaxies. We take their positions,
orientations and ellipticities from a {\sc Sextractor} \citep{sextractor}
run on the HST/ACS F606W band. A list of all cluster galaxies in our
model is stated in Table \ref{modelling:point
    like:galaxy lenses list}. A comparison with the HST/ACS F814W shows
  consistent values for the orientations and ellipticities of the
  cluster members. 
\\ 
With Eqs. \ref{theory:Faber Jackson} and \ref{theory: galaxy scaling Rusin} we now have
a complete description of all cluster galaxy lenses with only 2 free
parameters, the normalizations of equations \ref{theory:Faber Jackson}
and \ref{theory: galaxy scaling Rusin}. Since we take
$ L^{\star}$ for GR, the only free parameters in our galaxy model
  are $\sigma_{\rm GR}$, thus fully determining Eq. \ref{theory:Faber
    Jackson}, and $r_{\rm t,GR}$ fully determining
  Eq. \ref{theory: galaxy scaling Rusin} for $\sigma_{\rm
    GR}$.\footnote{However, we can equivalently use $r_{\rm
      t,1\arcsec}$ with $\sigma=186 \rm kms^{-1}$ as the full
    determination of Eq. \ref{theory: galaxy scaling Rusin}.} We will
attribute these two parameters to the reference galaxy GR, but we
should however keep in mind that the derived parameters of GR are due
to the combined signal of all the galaxies and that it is irrelevant
which galaxy was chosen as reference. For GR, 
we consistently measure an effective radius $R_{\rm eff}$ of $5 \rm kpc$ to $6\rm
kpc$ from fitting a S\'ersic, \citep{sersic63}, 
a de Vaucouleurs \citep{deVauc48} and a de Vaucouleurs$+$exponential
disc model in the F160W and F814W filters using {\sc Galfit}
\citep{galfit3}. This effective radius agrees well with measurements
(in the HST-F814W and VLT-FORS-I-band filters) of
other elliptical galaxies in various clusters of similar redshift, see
Figure 10 in \cite{saglia2010}.
\\
\subsubsection{Modeling of the cluster component}
\label{sec:point-like modeling:model cluster}
We model the cluster as a NFW \citep{nfw} halo.  We also tried a
non--singular isothermal elliptical (NSIE) profile for the halo, but
doing so
results in worse fits to the positions of the multiple image systems. 
The best fit $\chi^2$ for
the NFW is $\chi^2_{\rm NFW}=227$, while a NSIE cluster scale halo
with the same number of free parameters gives a $\chi^2_{\rm
NSIE}=434$, for the full model using point--like images.  A similar
difference for a NSIE vs NFW model has been reported already for the
stacked weak lensing signal of clusters and groups of galaxies in the
SDSS \citep{mandelbaum06}.
\\
We also add external shear as a free parameter to allow for a
contribution of the large scale environment in the vicinity of the
cluster.\\ This gives in total 6 free parameters for the NFW halo, 2 for the
external shear, 2 for the galaxy lenses, 9 for the source redshifts
and 32 free parameters for the (RA,DEC) source positions of the 16 sources. The
lens model parameters and its priors are posted in Table
\ref{modelling:point like:input parameters priors}. We use uniform priors
with defined minimum and maximum values for each of the parameters.
From the multiple images, we get 104 constraints, leaving this model
with 53 d.o.f.

\begin{table*}
\centering
\caption{The model lens input parameters and priors}
\begin{tabular}{ccccc}
\hline
 parameter & prior & min & max & model result(95\% c.l.) \\
\\
\hline
$\gamma$ & uniform &  0 & 0.4 & $0.20_{-0.03}^{+0.03}$\\
$\Theta_{\gamma}$ & uniform & $-90^{\circ}$ & $90^{\circ}$ &$25.7_{-2.5}^{+3.0 \, \circ}$ \\
$x_{\rm NFW}$\footnotemark[1] & uniform &  $-8\arcsec$ & $ 8\arcsec$ &
$0.19_{-0.47}^{+0.44}\arcsec$\\
$y_{\rm NFW}$\footnotemark[1] & uniform &  $-8\arcsec$ & $ 8\arcsec$ &
$0.78_{-0.23}^{+0.23}\arcsec$\\
$q_{\rm NFW}$ & uniform & 0.35 & 1 & $0.686_{-0.016}^{+0.014}$\\
$\Theta_{\rm NFW}$ & uniform & $-20^{\circ}$ & $44^{\circ}$ & $19.0_{-1.0}^{+1.2
\,\circ}$\\
$\Theta_{\rm E,NFW}$ & uniform &$25\arcsec$ & $200\arcsec$ & $43.8_{-1.4}^{+1.2}\arcsec$\\
$r_{\rm s,NFW}$ & uniform & $50\arcsec$ & $650\arcsec$ & $175_{-20}^{+23}\arcsec$ \\
$r_{\rm t,1\arcsec}$ & uniform &  $11\rm kpc$ & $142 \rm kpc$ &
$31_{-14}^{+36}\rm kpc$\\
$\sigma_{\rm GR}$ & uniform & $59 \rm kms^{-1}$ & $395 \rm kms^{-1}$ &
$236_{-32}^{+29}\rm kms^{-1}$\\
\hline
\end{tabular}\\
\tablecomments{The model lens input parameters and priors are stated. Given are the
  parameter, its prior type, the minimal and maximal allowed value as
  well as
the most likely value and its 95 \% c.l. error.}
\footnotemark[1]{relative to the center of the BCG at 12:06:12.134 RA
  (J2000) -08:48:03.35 DEC (J2000)}\\
\label{modelling:point like:input parameters priors}
\end{table*}
%
%
\subsection{Results of the point-like modeling}
\label{sec:point-like modeling:model output}
Putting all together, we can now reconstruct the lensing signal for
this cluster. We use the strong lensing code {\sc Glee}, a lens
modeling software developed by S. H. Suyu and A. Halkola
\citep{suyuhalkola2010, suyu2012}. This method does not only yield the
best fitting model (using either source plane or image plane
minimization) but in addition includes a Monte Carlo Markov Chain
(MCMC) sampler yielding the most likely parameters with their confidence
limits. 
We obtain the best--fitting cluster model by
  maximizing the posterior probability distribution function. For
  that, the likelihood is multiplied with the priors, see \cite{halkola06,
    halkola2008, suyuhalkola2010}. The likelihood is proportional to
  $\sim\exp(-\chi^2/2)$. The $\chi^2$ is calculated from the
  difference between the observed and the model
  predicted image position:
$$
\chi^2=\sum_{\rm i}\frac{\parallel\boldsymbol{\Theta}_{\rm
    i}-\boldsymbol{\Theta}_{0,\rm i}\parallel^2}{\delta_{\Theta_{\rm i}}^2}\quad ,
$$
where $\boldsymbol{\Theta}_{\rm i}$ and $\boldsymbol{\Theta}_{0,\rm
  i}$ mark the model predicted and observed position of multiple
image i and $\delta_{\Theta_{\rm i}}$ its input uncertainty.
 The MCMC sampling procedure is
  described in \cite{dunkley2005} and \cite{suyuhalkola2010}. We get
  acceptance rates of typically $\sim 0.25$ for the MCMC, the
  covariance matrix between parameters is derived from a previous run
  of the MCMC procedure for the same model parameters. Convergence is
  achieved based on the power spectrum test given in \cite{dunkley2005}.

\subsubsection{Results for the cluster--scale model}
\label{results:point like:cluster}

\begin{figure}[tbh]
\centering
\includegraphics[width=80mm]{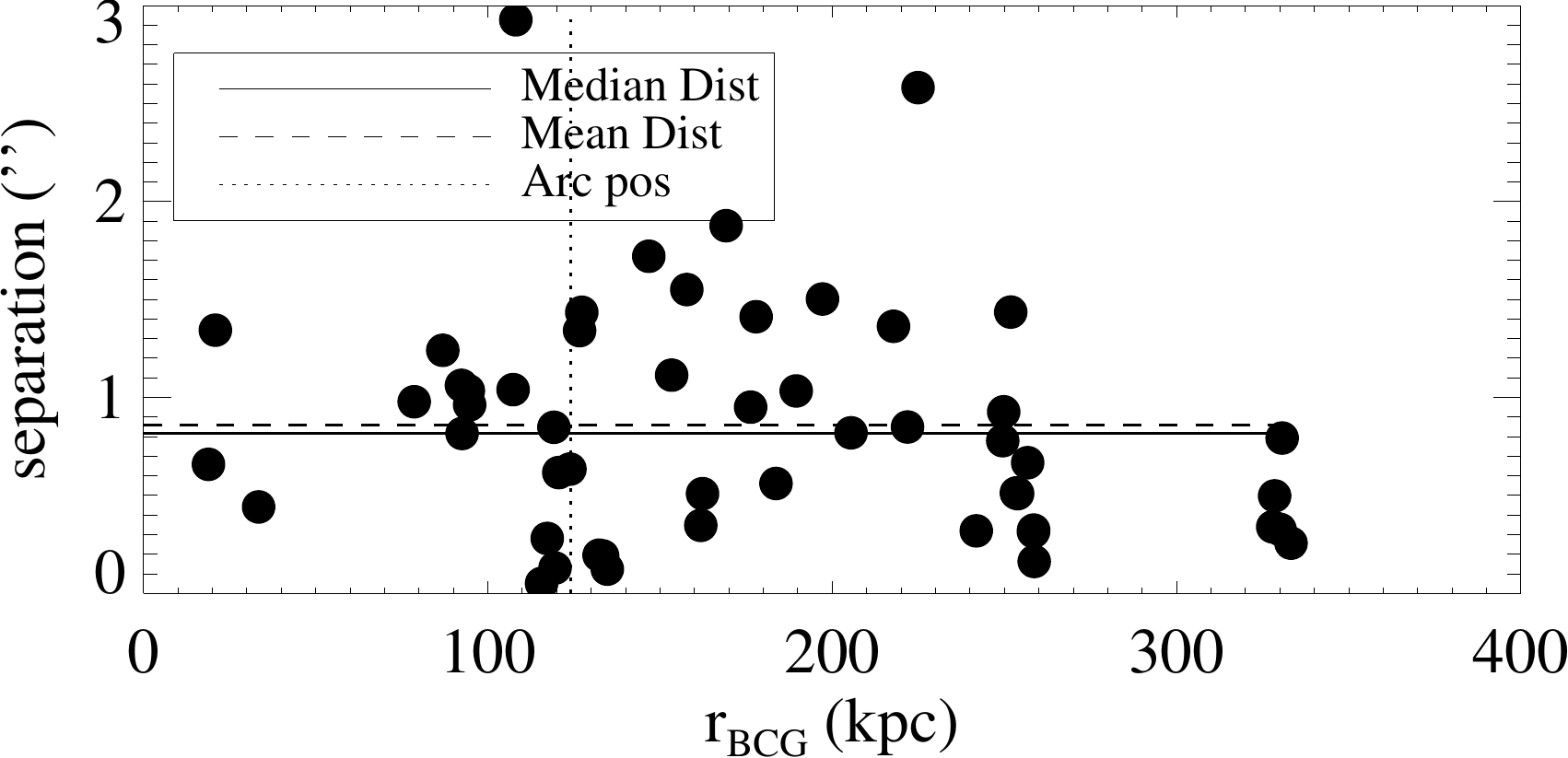}
\caption{The radial error dependence for the best fit model is shown
  in this plot. Plotted is the distance between observed and model
  predicted multiple image position on the y-axis against its
  distance from the center of the BCG. Overplotted are the respective
  median and mean of the images. The vertical dotted
    line marks the mean distance of the giant arc and its counterimage
    to the center of the BCG. There is no radial dependence of the
  error visible in this Model.}
\label{modelling:point-like:output:radial error}
\end{figure}

For the best-fit values\footnote{The error estimates from the MCMC sample will be discussed below},
 we get: $r_{\rm t,1\arcsec}=23.7 \rm kpc$, $\sigma_{\rm GR}=246
\rm kms^{-1}$, $\gamma=0.19$, $\Theta_{\gamma}=26^{\circ}$, $x_{\rm
NFW}=0.15\arcsec$, $y_{\rm NFW}=0.74\arcsec$, $b/a_{\rm NFW}=0.69$,
$\Theta_{\rm NFW}=19^{\circ}$, $\Theta_{\rm E,NFW}=44.1\arcsec$ and
$r_{\rm s,NFW}=174\arcsec$. As explained already the external shear
and the Einstein radius are given in units of $\rm
D_{ds}D_{s}^{-1}$. The redshift estimates of the best-fit model are
given in Table \ref{modeling:point like image positions}. Most of the
redshifts agree with their photometric estimates within the errors,
only system 6 is a clear outlier. The critical lines for the arc
redshift and a redshift of $\rm z=2.54$ are plotted in
Fig. \ref{modeling:point like cluster image}. \\
In Fig. \ref{modelling:point-like:output:radial error}, we show the
differences of the input and model output positions for our best-fit
model. As one can see, the mean and median differences are
$0.86\arcsec$ and $0.82\arcsec$. This justifies the used input
uncertainty of $0.5\arcsec$, since this is a good estimate of the
reconstruction uncertainty.\\
The MCMC sampling provides us with
estimates for the parameter uncertainties.  

\begin{figure*}
\centering
\includegraphics[width=160mm]{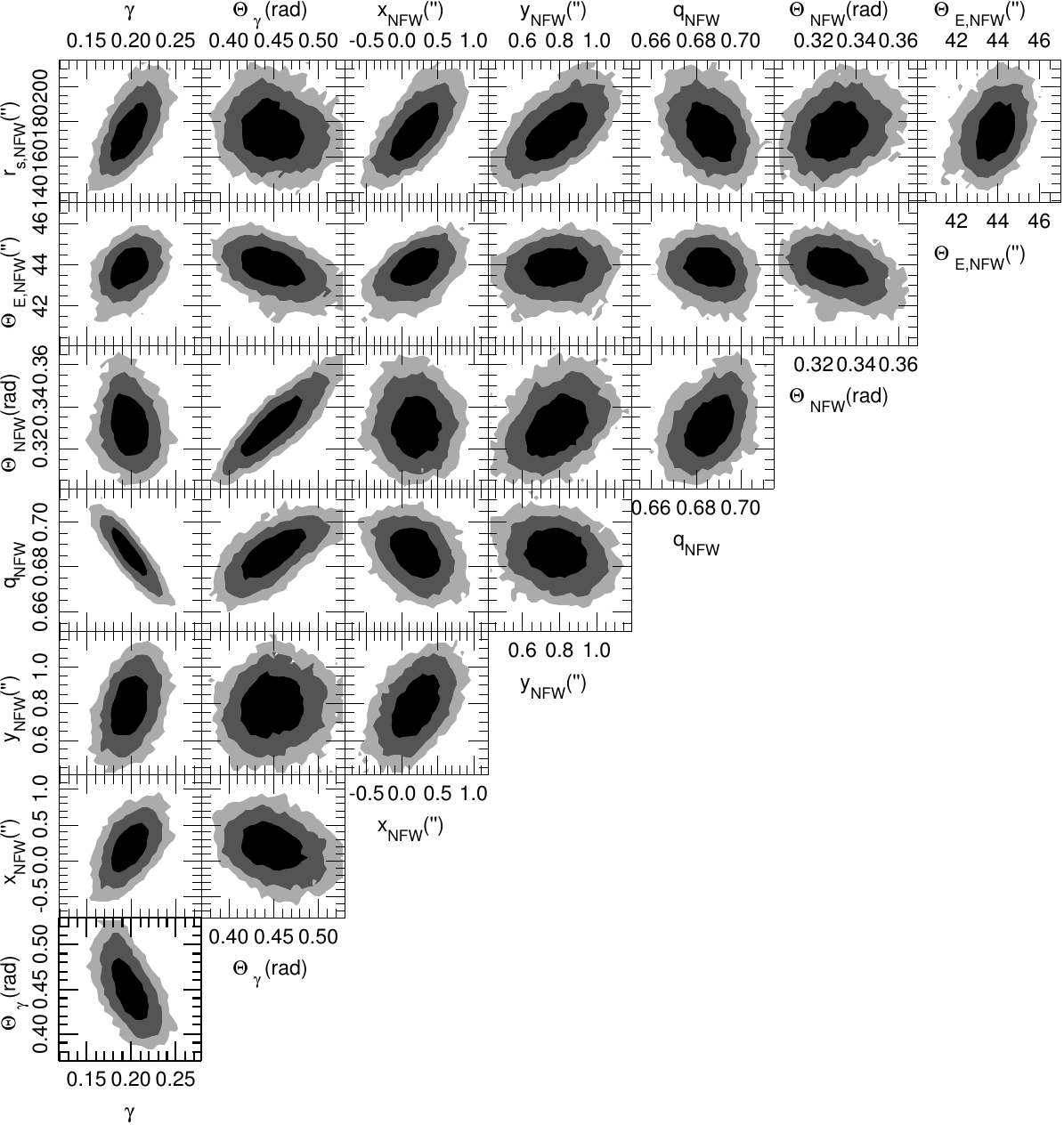}
\caption{Parameter estimates from the MCMC sampling of the
  parameter space. The shaded regions give the 68.3\% 95.4 \% and 99.7
\% uncertainty areas, from dark to light gray, respectively.}
\label{modelling:point-like:output:parameters}
\end{figure*}
\begin{figure}[tbh]
\centering
\includegraphics[width=80mm]{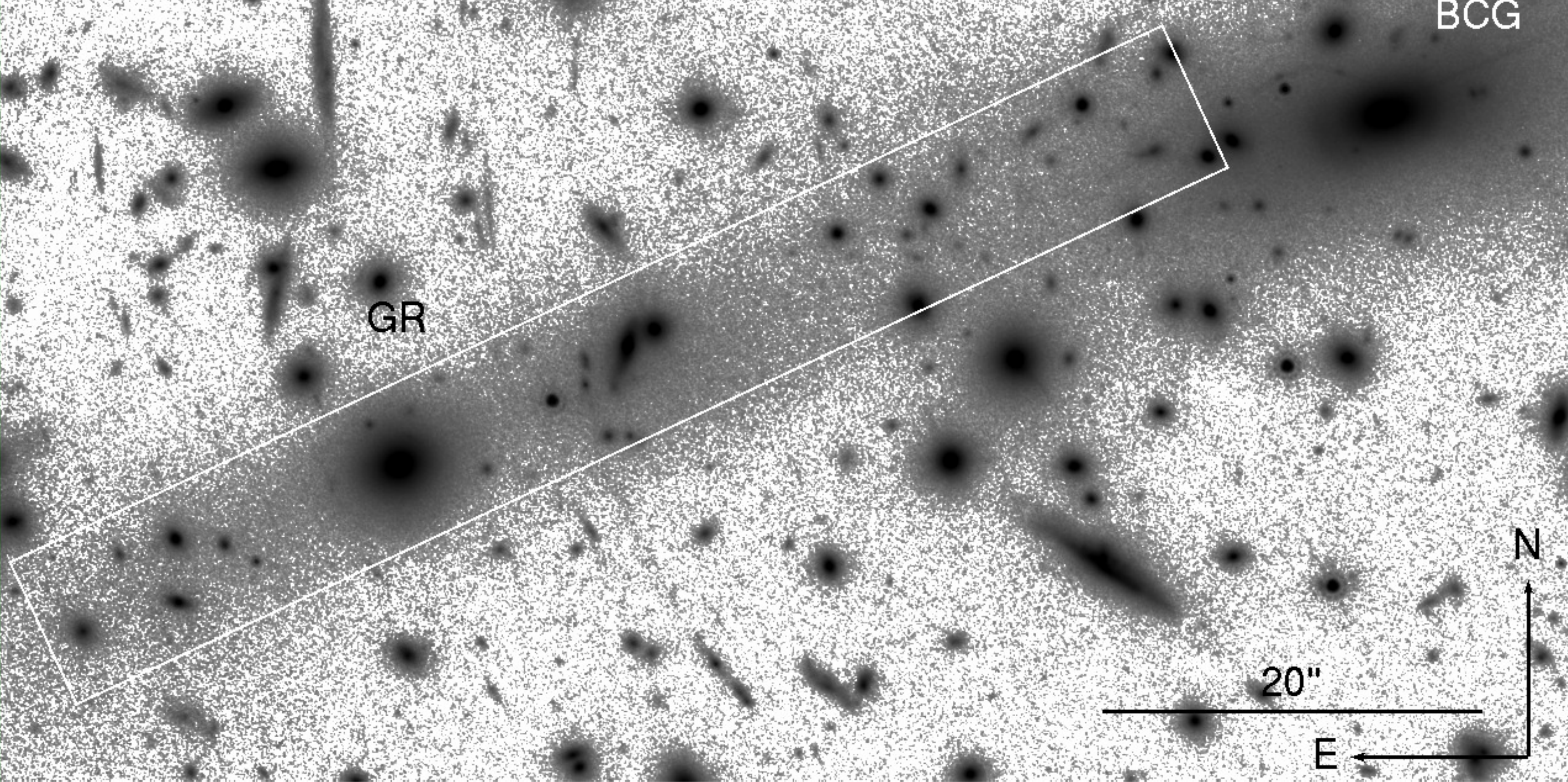}
\caption{The center of the cluster MACSJ1206.2-0847
    as observed with the F160W HST/WFC3 filter. The faint, bar-like
    structure in the intra-cluster light is marked with a white
    box. It extends $\sim1.5\arcmin$ radially outwards from the BCG to the
SE. The mass associated with this intracluster--light acts as further
substructure. We use logarithmic scaling for the fluxes in this image.}
\label{modelling:point-like:output:bar}
\end{figure}

The probability densities for the parameter estimates are shown in
Fig. \ref{modelling:point-like:output:parameters}. We want to discuss
some of the parameters here, quoting the 95 \% confidence intervals:
First, the external shear values are: $\gamma=0.20_{-0.03}^{+0.03}$
and $\Theta_{\gamma}=25.7_{-2.5}^{+3.0 \, \circ}$. This shear can
originate from external structure present in the vicinity of the
cluster or from substructure present in the cluster, but not accounted
for in the model. Indeed, the cluster mass reconstruction map of
\cite{ume12} (see their Fig. 8) shows two additional structures, one
in the southeast, one in the northwest of the cluster center. We take
the 2D mass reconstruction map of \cite{ume12}, and subtract the
surface mass density of their best--fitting cluster NFW-profile,
leaving us with the residual mass map.
We calculate the shear that these additional masses
cause in the cluster center, and obtain values of $\gamma
\lessapprox 0.13$ for $\rm D_{ds}D_{s}^{-1}=1$. This
external structure thus explains a
part of the shear present in the model. 
Additional or external shear can in principle be
  produced by any mass distribution that we do not model
  explicitely. The mass distribution associated to the intra--cluster
  light is such a component: it ranges from BCG towards the galaxy GR
  (in the south-east) and beyond the galaxy GR (see Fig. \ref{modelling:point-like:output:bar}). We tested
  that the presence of this intra--cluster light is not a superposition
  of the light associated with the cluster members: we have subtracted
  a galaxy light model for the galaxies in the
  south-east from the F160W-data; the residual light is not centered on
  any galaxy haloes, hence it cannot be attributed to a galaxy. The
  gravitational shear produced by the mass
  associated with the intra--cluster light is incorrectly
  attributed to the external shear if we do not explicitly model its
  lensing contribution, and thus increases the external shear of the
  lensing model. We employ a test scenario, explicitly modeling a mass
  distribution associated with the intra--cluster light. We used a non--singular,
  highly elongated ($q<0.4$) isothermal ellipsoid with large core radius and small
  truncation radius which roughly resembles a mass bar. The best fit masses
  of this intra-stellar light component are modest (a few times
  $10^{12}M_{\odot}$). The external shear values required in this toy model drop to
  $\gamma=0.13_{-0.04}^{+0.04}$, agreeing with our estimate based on
  \cite{ume12}. We verify that this toy model (approximately
  including the intra--cluster light) results in the same sizes of
  galaxies as our strong lensing model presented in this work.

Second, the cluster--scale NFW halo has the following most likely
parameter estimates: $x_{\rm NFW}=0.19_{-0.47}^{+0.44}\arcsec$, $y_{\rm
NFW}=0.78_{-0.23}^{+0.23}\arcsec$, ${\rm q_{\rm NFW}}=b/a_{\rm
NFW}=0.686_{-0.016}^{+0.014}$, $\Theta_{\rm NFW}=19.0_{-1.0}^{+1.2
\,\circ}$, $\Theta_{\rm E,NFW}=43.8_{-1.4}^{+1.2}\arcsec$, $r_{\rm
s,NFW}=175_{-20}^{+23}\arcsec$.
The results regarding the cluster-scale dark matter halo are within our
expectations:
\begin{itemize}
\item The halo center's position follows the same trend as the X-ray center in
\citet{ebeling2009}, i.e., the center has a slight tendency to move towards
positive values of x and y relative to the BCG center. In total, the
center of mass is shifted by approximately $(0.8\pm0.3)\arcsec$. 
\citet{ebeling2009} report a displacement of the X-ray center from the
BCG center of $(1.7\pm0.4)\arcsec$ in approximately the same
direction implying  that these displacements agree on a $2\sigma$
level. The level of displacement between the BCG and
  the dark matter halo center is comparable to \cite{zitrin2012}.
\item The orientation of the NFW-major axis follows the major axis of
the BCG within $\approx 5^{\circ}$

\item There is some
degeneracy between the orientation of the cluster halo and the
external shear, since both can compensate each other partially. The
same is true for the axis ratio of the halo and the value of the
external shear.
\item For the Einstein and scale radius of the NFW halo, we get:
$\Theta_{\rm E,NFW}=43.8_{-1.4}^{+1.2}\arcsec$, $r_{\rm
s,NFW}=175_{-20}^{+23}\arcsec$. The total mass included within a
cylinder of radius $R$ is presented in
Fig. \ref{modelling:point-like:output:masses}. 
\begin{figure}[tbh]
\centering
\includegraphics[width=80mm]{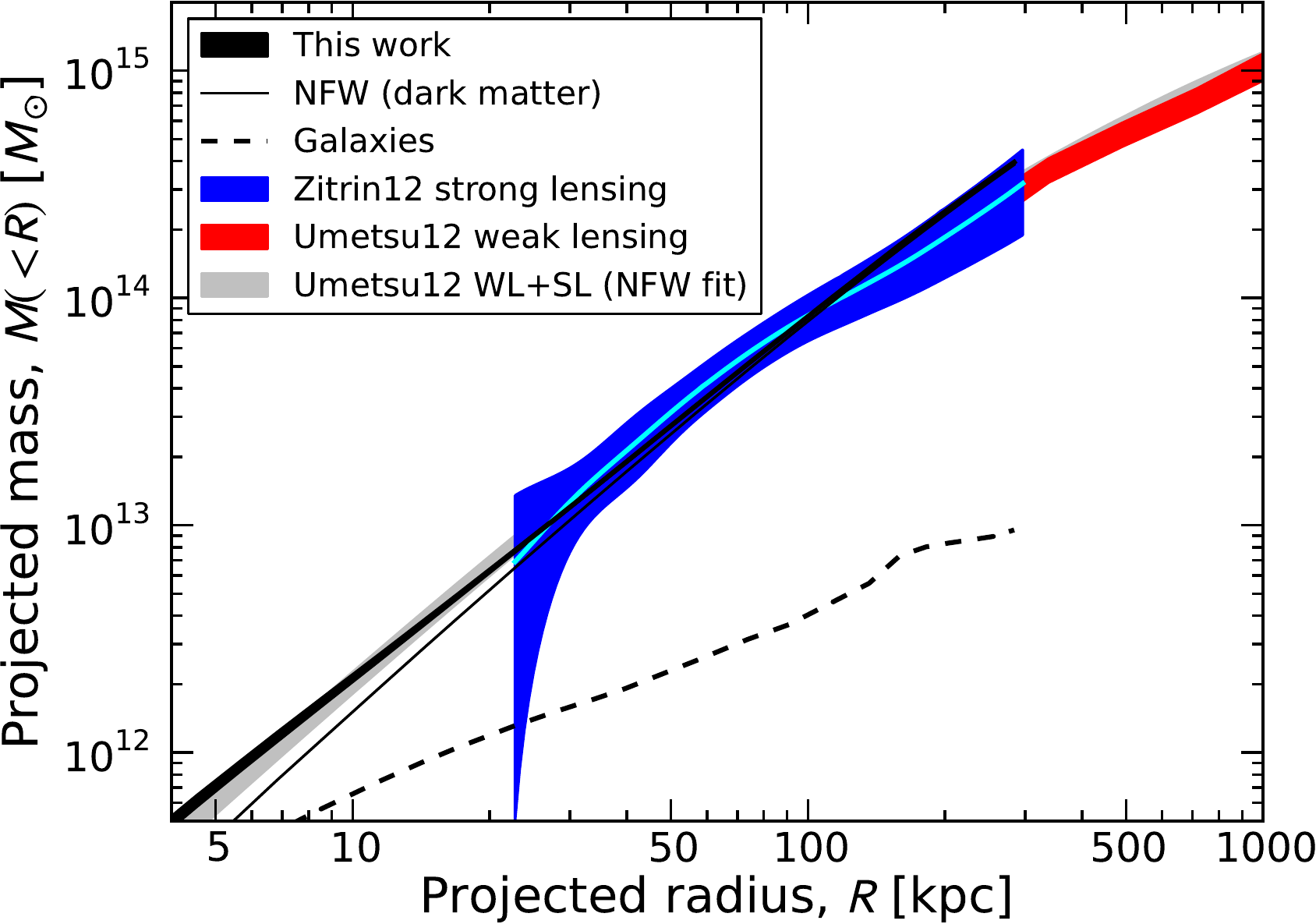}
\caption{The projected mass estimates within circular apertures are
  shown in this figure. The black area shows the 68 \% confidence interval for the
  combined mass, the black solid and dashed lines show the mass contributions for the NFW halo alone
  and the galaxies for the best-fit model, respectively. The
  small uncertainty for the mass estimate comes from the fact that we
  use a parametric model, which needs to reproduce the correct
  Einstein radius, therefore giving too small errors in the
  intermediate radii. We overplot the mass estimates from \cite{ume12}, 
more explicitly their NFW fit to the weak and strong lensing data in
gray, their weak lensing mass estimates alone (red area),
and the \cite{zitrin2011}, \cite{ume12} strong lensing estimate in
blue. In cyan, the best--fitting estimate from
  \cite{ume12} for the same strong lensing model is shown. The mass
estimate in this work agrees in the range of $\sim 4 \rm kpc$ to 
$\sim 150 \rm kpc$ with our previous work.}
\label{modelling:point-like:output:masses}
\end{figure}

Our results for $M(<R)$ agree well with previous results obtained with
various methods and presented in \cite{ume12} (see their Fig. 7). 
The
agreement holds up to $\approx 300 \rm kpc$ which equals the radius probed by
multiple images in this cluster. 
The result from ``Zitrin MCMC''\footnote{See below for the explanation
  of this wording.} agrees within its errors
with all further results shown in \cite{ume12}. Since this is in addition
the only strong lensing result in this work with realistic errors we
only compare to ``Zitrin MCMC'' below.
Our errors on the measured masses are derived
from the mass distribution of 200 random cluster models from the MCMC
points. Since we use a parametric model for the
lens, we only measure the uncertainty within this parametric model, not taking
into account that different parameterizations could give similar good
fits with a slightly different mass profile, hence we are
  underestimating the true error on the radial mass profile. To obtain
  more realistic errors we could take the same approach as it was done
  in \cite{ume12} for the ``Zitrin MCMC'' results, and thus increasing
  our errors by the amount as the difference between \cite{zitrin2011} and
  ``Zitrin MCMC''. Our result
  (black area in Fig. \ref{modelling:point-like:output:masses}) however already now agrees within the
  errors with that of ``Zitrin MCMC'' (blue area, Fig. \ref{modelling:point-like:output:masses}). 
Since the results of the strong lensing analysis of \cite{zitrin2011}
and its improvement in \cite{ume12} have been
presented in detail we here summarize the difference to our method.
In \citet{zitrin2011} both the mass associated with cluster members
and the dark matter of the cluster are modeled starting from the
light distribution of the cluster. The first is obtained by scaling
the galaxy masses with their light and modeling their mass density
profile with a power law (2 parameters). The second is obtained from
smoothing the galaxy light (1 further parameter) and scaling this to
the dark matter with a free amplitude (one further parameter). In
addition there are two free parameters for external shear. By
construction this method does not allow any dark matter not traced by
galaxy light. Also, the radial dark matter profile is closely linked
to the cluster light profile, since any deviation from that can only
be achieved by smoothing. If the concentration of
  the cluster light profile obtained from the smoothed galaxy light is
  different from the concentration of the dark matter this can lead to
  a systematic error of the mass estimate and to a bias in determining the true dark matter
concentration. At least for the number density distribution of cluster
members this seems to be indeed the case: \cite{budzynski2012} find
that the number density profile of cluster members of SDSS clusters
follows an NFW profile but with a factor of 2 lower concentration than in the dark matter
(independent of the mass of the cluster).  In \cite{ume12} the method
of \cite{zitrin2011} has been
generalized by allowing to model the mass associated with the BCG separately. 
In addition, they have altered the covariance matrix such that error estimates 
are increased to account for the too small systematic errors inherent
in a parametric reconstruction. This improved analysis relative to
\cite{zitrin2011} is called ``Zitrin MCMC'' in \cite{ume12}.
\\Our method is different:  
We use a parameterized model for a cluster--scale lens, including it explicitly as an elliptical
  NFW profile (2 main free parameters for the concentration and the
  virial radius, two free parameters for the ellipticity and major
  axis angle, and in principle two free parameters to locate the
  center of mass (the center of mass from the modeling in this cluster however is
  similar to the BCG)). The galaxy scale mass component is
  parameterized with 2 free parameters (halo depth and halo size). 
So formally our method has slightly more free parameters than that of
\cite{zitrin2011} and \cite{ume12}. Both methods are complementary as our method allows to
place halos even if there is no light tracing them (or allows to
off--center halos from their light), where as the \cite{zitrin2011} method
allows for small scale variations in the dark matter, which however
are linked to a smoothed version of the light. As far as the galaxy
matter component is concerned our method describes galaxies as being
isothermal out to large radii (as obtained from strong lensing and
weak lensing analyses of red galaxies, see \citep{slacs4, slacs10})
and allows for a cutoff (smaller than for field galaxies). In contrast
\cite{zitrin2011} can, once tieing the central matter density of galaxies
to their central light, only change the total mass associated with
galaxies by changing their matter density power law slope. This
picture seems to be an inaccurate description 
when tidal stripping of halos is described, since tidal stripping is
not expected to change the central properties,
but to shrink the halos from outside to inside \citep{gao2004b}.
The accuracy that can be obtained with our method is larger
(The image plane reproduction error is
$1.76\arcsec$ in \cite{ume12} whereas it is $0.85\arcsec$ in our
work). This is likely not the case because
of the increased number of free parameters, but because the galaxy component is modeled in a better
way. Our approach for modeling the galaxy component is also followed 
by \cite{zitrin2013} in their strong lensing model for the
mass distribution of MACS J0416.1-2403. In this work \cite{zitrin2013} also
compare the performance for a cluster component obtained with a mass
follows light approach with an elliptical NFW component (leaving the
galaxy component the same) finding the later to provide the better
fit. 

\item We fit a circular
NFW\footnote{We give the values for an overdensity of
  $\Delta=200$. The conversion to \citet{ume12}, who use $\Delta=132$,
is $\rm c_{132}\sim1.2\rm c_{200}$.}
halo to the total azimuthally averaged mass in
Fig. \ref{modelling:point-like:output:masses} to estimate the
concentration $c_{\rm 200}$ and $r_{\rm s,NFW}$ from the total included
mass with a least square fit. We get a concentration of $c_{\rm
  200}=3.7\pm0.2$ and a scale radius of $r_{\rm s,NFW}=677\pm48 \rm
kpc$. When we exclude the central $70 \rm kpc$ from the fit, we get $c_{\rm
  200}\approx3.2$ and $r_{\rm s,NFW}=827 \rm kpc$.
 Our radially averaged mass distribution agrees with the results
of \citet{ume12} in the center. Our scale
radius value is an extrapolation beyond the scales of strong lensing datapoints. Since
\citet{ume12} do a combined strong and weak lensing analysis
constraining the profile on a much larger scale than our work can do,
confidence intervals for these two parameters are smaller 
than ours and their conclusions are much more firm. Regarding results
of MACSJ1206.2-0847's mass--concentration relation we therefore refer the reader to
the work of \cite{ume12}.
\end{itemize}

\subsubsection{Results for galaxy halos tracing the cluster--substructure}

\begin{figure}[tbh]
\centering
\includegraphics[width=80mm]{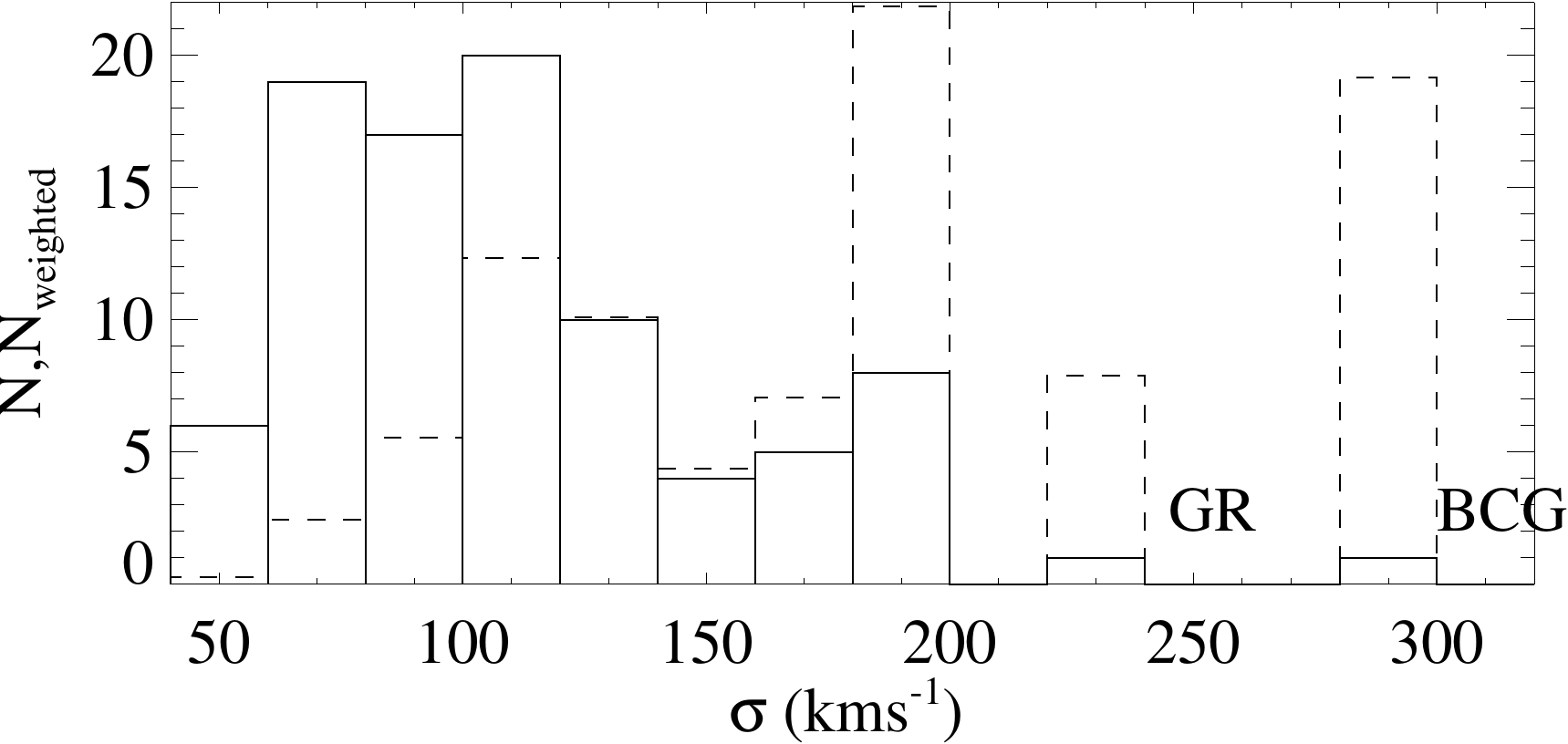}
\caption{The velocity dispersion distribution for the galaxy cluster
  MACSJ1206.2-0847 is shown here for the best fit. Marked are the
  brightest cluster galaxy (BCG) and the second brightest galaxy
  (GR) which is used as a reference for the Einstein radius scaling
  in this work. For the same galaxies, the dashed
    histogram gives the weighted velocity dispersion distribution. As a weight, the mean
    deflection angle of a galaxy on all multiple images is used. As
    can be seen, the galaxies with lower velocity dispersions get
    down-weighted, meaning that they contribute on a minor level to the
    summed galaxies' lensing signal. The BCG has a
      velocity dispersion of $\sim290\rm kms^{-1}$ from the best fit
      scaling law. This agrees with \cite{sand2004} who measure a
      stellar velocity dispersion of $\sigma\sim250\pm50\rm kms^{-1}$
      in the central $\sim 1.5\arcsec$ of the BCG.}
\label{modelling:point-like:output:Einsteinradii}
\end{figure}

\begin{figure}[tbh]
\centering
\includegraphics[width=80mm]{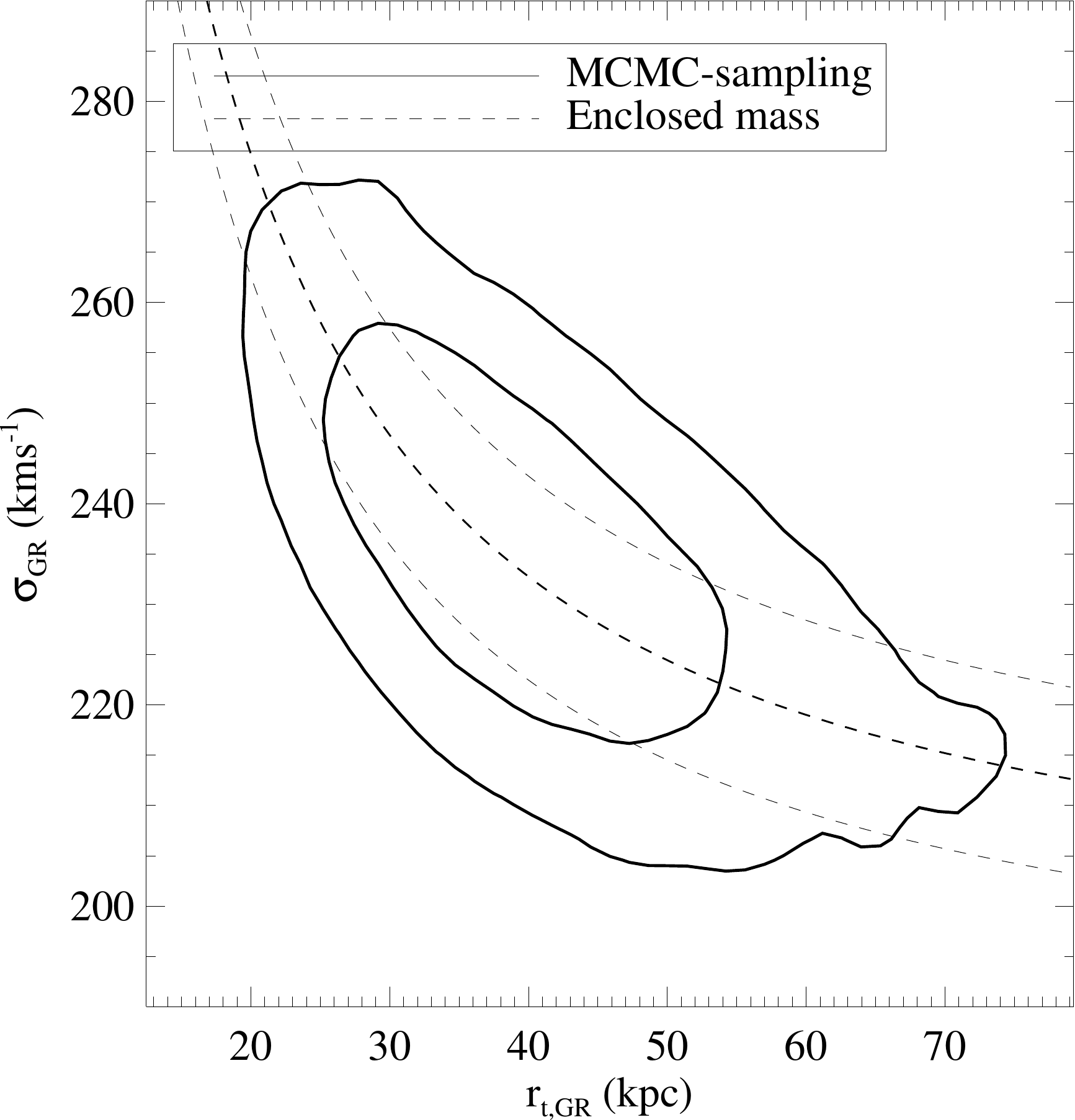}
\caption{Here we show the probability contours for the 2 parameters governing the
  profile of the GR for the point source modeling: The truncation radius
  $r_{\rm t,GR}=r_{\rm t,1\arcsec}(\sigma_{\rm 
    GR}(186\rm kms^{-1})^{-1})^{1.333}$ and the velocity dispersion of the GR $\sigma_{\rm 
    GR}$. We also show the best fit for the enclosed mass within an
  effective radius as dashed lines, which gives a
  radius of $R_{\rm mass,p}=26.6\rm kpc$ and a enclosed mass of $M(<R_{\rm mass,p})=7.3\pm0.6\times10^{11}M_{\odot}$ for the GR.}
\label{modelling:point-like:output:Tevsrt*}
\end{figure}

Using the F160W flux of the
galaxies and scaling relations, the mass distribution of the galaxies
is described as a function of the two (free) parameters, the velocity
dispersion of GR $\sigma_{\rm GR}$, and the normalization of the
truncation radius scaling $r_{\rm t,1\arcsec}$. 
This truncation scale $r_{\rm t,1\arcsec}$ is not to be confused with $r_{\rm t,GR}$, which 
gives the truncation radius for galaxy GR and is shown in Fig. \ref{modelling:point-like:output:Tevsrt*}.
For these 2 values, we
get the most likely values of:
$r_{\rm t,GR}=41_{-18}^{+34}\rm kpc$ and $\sigma_{\rm
  GR}=236_{-32}^{+29}\rm kms^{-1}$. 
\\
We apply the Faber-Jackson relation and show 
the velocity dispersions for all cluster members
galaxies as a histogram in
Fig. \ref{modelling:point-like:output:Einsteinradii}.
Since the lenses' impacts scale like $\propto \sigma^2$, 
most of the low velocity dispersion galaxies
have a minor influence on the lensing signal. There is however a
secondary effect, i.e. that the
deflection angle that a galaxy can impose on the LOS to a
multiple image position depends also on the transverse distance to
it. We therefore now weight each cluster galaxy by
the mean deflection
angle it imposes on all multiple images and obtain the effective velocity
dispersion histogram for the cluster members, also shown in Fig. \ref{modelling:point-like:output:Einsteinradii}. It shows that the major
impact is caused by galaxies with velocity dispersion between $100\rm kms^{-1}$
and $200\rm kms^{-1}$ ($55\%$ of cluster galaxies light deflection for multiple
images) or $250\rm kms^{-1}$ ($60\%$).
\\
For the galaxies, we get the following scaling law on a 95\% CL basis:
\begin{equation}
r_{\rm t}= 31_{-14}^{+36}\rm kpc\left(\frac{\sigma}{186\rm
    kms^{-1}}\right)^{4\over 3}\quad .
\label{results:point like: galaxy scaling law}
\end{equation}
 We translate the output of the MCMC sampling for the truncation radius of
a galaxy with $1\arcsec$ cosmology free Einstein radius into ($1 \sigma$
and $2 \sigma$) confidence contours for $\sigma_{\rm GR}$ and $r_{t, \rm
GR}$ and show them in Fig. \ref{modelling:point-like:output:Tevsrt*}. 
If we would be able to
constrain only the mass $M(<R_{\rm mass,p})$ within one scale $R_{\rm mass,p}$
(as it is the case for strong lensing analysis of galaxies with one
multiple image or one Einstein radius only) then the contours would
extend to infinite truncation radius and also smaller minimum value,
given by Eq. \ref{theory:BBS-2D-densityrnclosed mass} as

\begin{equation}
\sigma^2=\frac{G M(<R_{\rm mass,p})}{\pi} \left[R_{\rm mass,p}+\rt-\sqrt{R_{\rm mass,p}^2+\rt^2}\right]^{-1}
\quad .
\end{equation} 
Hence the contours in Fig. \ref{modelling:point-like:output:Tevsrt*}
demonstrate that the degeneracy
between the two free parameters is broken (albeit not yet
completely). This implies that not only the enclosed mass at some
radius but also the gradient of the mass profile at this radius must
be constrained by the observables, i.e. there must exist a scale
$R_{\rm mass,p}$, where the profile is best determined, i.e where the
enclosed mass is most equal for all $\sigma_{\rm GR}$ and $r_{t, \rm
GR}$ pairs of the Chain output. We use
Eq. \ref{theory:BBS-2D-densityrnclosed mass} for all MCMC sample output
pairs and find this scale to be $R_{\rm
mass,p}=4.7\arcsec\widehat{=}26.6\, \rm kpc$. The enclosed mass at
this scale becomes ${\rm M}(<R_{\rm
mass,p})=7.3\times10^{11}M_{\odot}$ for the most likely $\sigma_{\rm
GR}$ and $r_{\rm t, \rm GR}$ pair. The curve of this constant enclosed
mass is added as thick dashed line in
Fig. \ref{modelling:point-like:output:Tevsrt*}. As expected it traces
the degeneracy in the $\sigma_{\rm GR}$ and $r_{\rm t, \rm GR}$ parameter
space.
\\
We then use Eq. \ref{theory:BBS-2D-densityrnclosed mass} at this fixed
enclosed mass radius and calculate the mass within $R_{\rm
mass,p}=4.7\arcsec$ for each pair in the MCMC sample. From this
distribution of enclosed masses, we take the central 68 \% as the
error interval and get an enclosed mass of $ M(<R_{\rm
mass,p})=7.3\pm0.6\times10^{11}M_{\odot}$ at the fixed enclosed mass
radius of $R_{\rm mass,p}=4.7\arcsec$. These 68\% upper and
lower confidence values are plotted as dashed lines in Fig.
\ref{modelling:point-like:output:Tevsrt*}.
\\
Thus we conclude that our lens model is indeed not only sensitive to the
total mass associated with galaxies but also to the size of the
galaxy dark matter halos. There remains a degeneracy between halo
velocity dispersion and truncation radius at a level of a factor of 2 for
the truncation radius. For the reference halo GR within radius
$R_{\rm mass,p}=4.7\arcsec\widehat{=}26.6\, \rm kpc$ the
enclosed mass is ${\rm M}(<R_{\rm mass,p})=7.3\pm0.6\times10^{11}M_{\odot}$
\\ 
For galaxies with different luminosity and thus velocity dispersion
and truncation radius the radius where the mass is best known and the
mass within this radius scales like $R_{\rm mass,p} \propto
r_{\rm t} /r_{\rm t, \rm GR}$ and ${\rm M}(<R_{\rm mass,p})\propto \sigma^2
r_{\rm t}/ (\sigma_{\rm GR}^2 r_{\rm t,\rm GR})$.
\\
To constrain the truncation scaling even further, we need to trace the
lensing signal at various galaxy distances more densely. This is
achieved with the pixel by pixel image reconstruction of the giant arc
since every pixel has a different distance to the centers of
the surrounding galaxies.
%


\section{Strong lensing modeling of the full surface brightness of
  the giant arc and its counterimage}
\label{sec:extended strong lensing}
We aim to further constrain the scaling relation for
the truncation radius in this section. For that, we take a different
approach, reproducing the full surface brightness of the giant arc and
its counterimage. The full surface brightness not only contains
information about the deflection angle, but also about its derivative,
making it a good tool to explore galactic halo truncation in this system.

\subsection{Setup of the Model}
\begin{figure}[tbh]
\centering
\subfigure[]{\includegraphics[height=80mm]{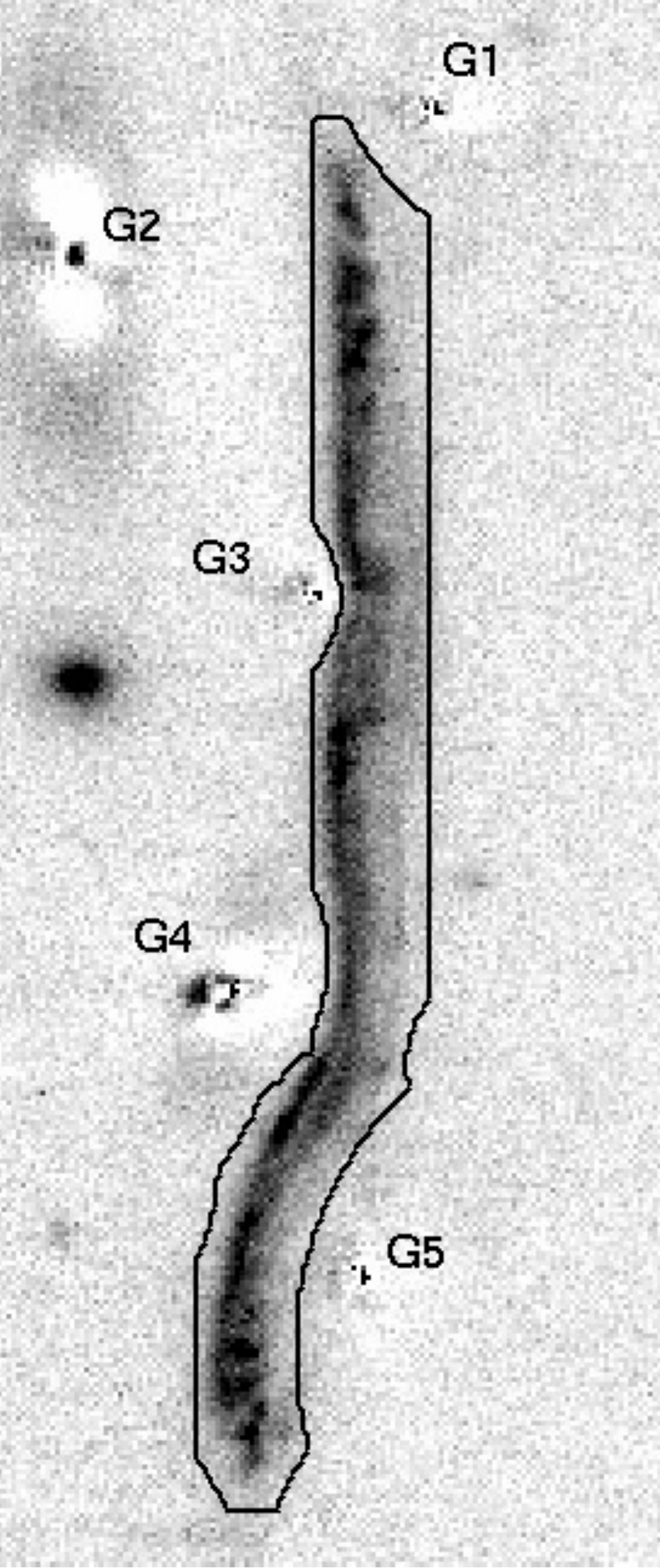}\label{modelling:extended:F814Wmask
arc}}
\subfigure[]{\includegraphics[height=80mm]{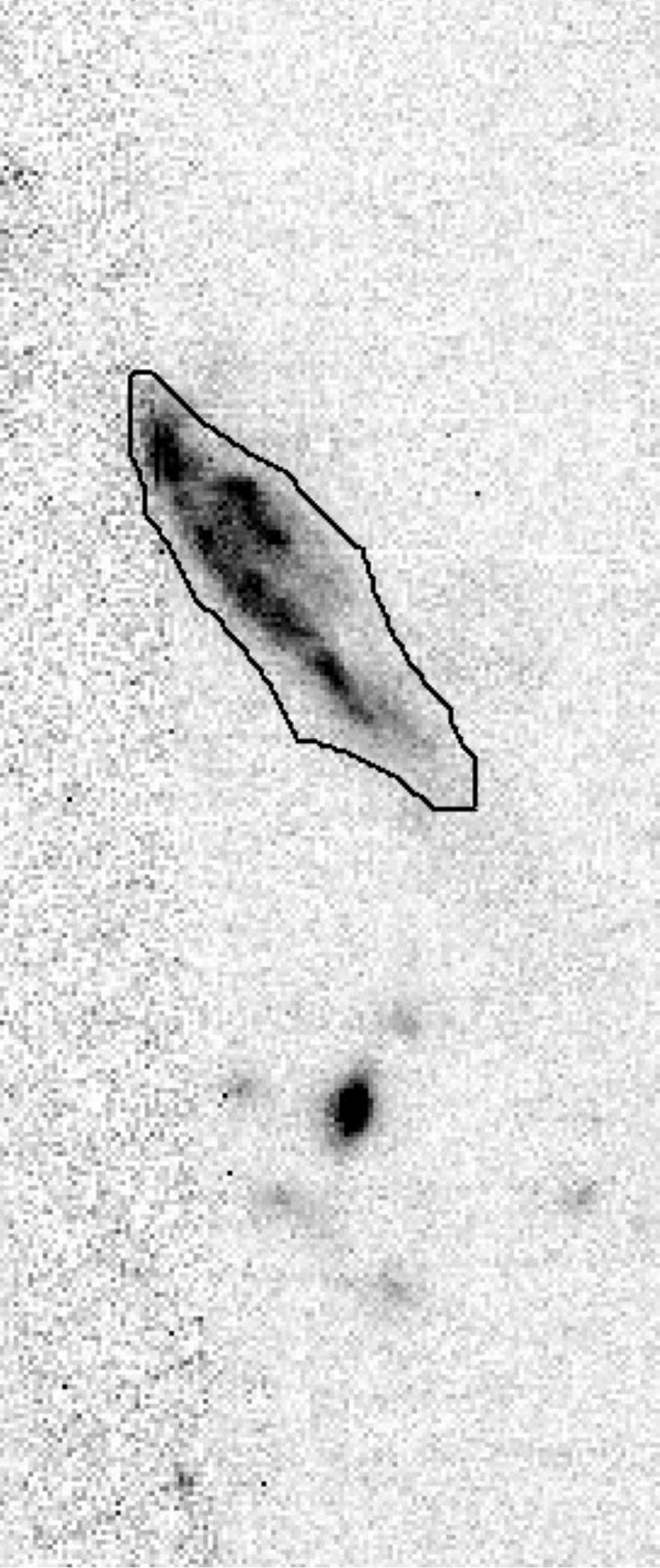}\label{modelling:extended:F814Wmask
counter}}
\caption{This frame shows the used region around the giant
  arc \ref{modelling:extended:F814Wmask arc} and its counterimage
  \ref{modelling:extended:F814Wmask counter} in this
  galaxy cluster. The mask is outlined in black. The underlying image
  is the F814W observed image for this cluster. The galaxies
  marked with G1 to G5 in Fig. \ref{modelling:extended:F814Wmask arc}
  have been subtracted to minimize possible contamination
of the arc light from the galaxies. One pixel corresponds to $0.05\arcsec$. North is up and
east is left.}
\label{modelling:extended:F814Wmask}
\end{figure}

We use data from the F435W, F606W and F814W bands for the extended
image reconstruction. We take different filters to
  minimize effects of light pollution of the surrounding galaxies. The
  cluster galaxies are significantly dimmer in the F435W filter,
  therefore minimizing the possibility of galaxy light disturbing the arc
  light. Since the arc is already faint in this filter (The average
  signal--to--noise ratio in the used mask area is $\sim0.5$), we
  do not consider even bluer bands. We
  also include a redder filter (F606W) in which the arc but also the
  surrounding galaxies become brighter. We add the F814W filter with an even
  brighter arc. In this filter the systematic uncertainty from the
  subtraction of the surrounding galaxies' light gets comparable to
  the noise in the arc region, hence we refrain from investigating even
  redder bands. We apply {\sc Galfit}  to
subtract the light of the surrounding galaxies G1 to G5, see
Fig. \ref{modelling:extended:F814Wmask}. For the F435W and
F606W-filter data, we fit a de Vaucouleurs profile \citep{deVauc48} as
a light model to the data and subtract it. For these 2 filters, the
subtracted fluxes at the position of the arc are small compared to the
intrinsic noise of the images for these pixels, so the
impact of the exact details of the subtracted galaxy's light model are
small. This is not the case for the F814W filter, therefore we create
a best-fit de Vaucouleurs, a best-fit S\'ersic \citep{sersic63} and a
best-fit King profile for galaxies G1 to G5. From these 3 light
models, we create a mean model and subtract that from the observed
image. To account for the systematic error introduced by the light
subtraction in the F814W filter, we add the difference of the maximum and
minimum value in each pixel for the 3 models to the error image
derived before. We limit the analysis to a small region around the arc
and its counterimage for computational reasons. This masked region is
shown in Fig. \ref{modelling:extended:F814Wmask}. The region is chosen
by eye based on the arc visible in the F814W filter and used in all 3
bands.

As a systematic test, we choose the region to be reconstructed also by a
$\rm S/N > 2$ cut on the F814W
  frame. Before the modeled area is selected, the signal--to--noise
  map is block--smoothed with a length of 7 
pixels. This leads to a slightly different
selection of the modeled region. However, the changes introduced on
the truncation law by changing the mask are small, as described below. 
\\
For the source reconstruction, we use a $9\times9$ pixel grid with a
free pixel scale and source plane position, therefore the physical
size of the reconstructed source is unrestricted by the number of
source pixels. We compare different numbers of source pixels later
on. For details of the extended surface brightness reconstruction, see
\citet{suyu2006, suyuhalkola2010}. It uses a linear
  inversion method \citep{war03} in a Bayesian framework
\citep{suyu2006}.  We search for the
most probable solution of the nonlinear lens mass parameters by maximizing the
 posterior in reconstructing the source (see Eq. 11 of \citealt{suyuhalkola2010}). 
 The lens parameter space is sampled by MCMC methods. 
We tried both the curvature and gradient forms
 of regularization, and find that the resulting lens parameters are insensitive to the choice of
 regularization.
\\

\subsection{Results for the full surface brightness
    reconstruction}
\label{Sec:extended_reconstruction:results}
We now concentrate on modeling the galaxies G1 to G5 around the arc which are
already subtracted in Fig. \ref{modelling:extended:F814Wmask}. 
We fix all parameters (shear, cluster halo, source redshifts, galaxy
parameters) to its best-fit values from Sec. \ref{sec:point-like
  modeling:model output}, and now only model galaxies G1 to G5. For the
galaxies G1, G2, G4, and G5, we allow each galaxy its own orientation
and Einstein radius, keeping a joint truncation scaling law following
Eq. \ref{theory: galaxy scaling Rusin} for these galaxies. The values
derived in Secs. \ref{sec:point-like modeling:model lenses} and
\ref{sec:point-like modeling:model output}, used as
  starting values, are stated in Table
\ref{modelling:extended:input parameters}.

\begin{table*}
\centering
\caption{Galaxies G1 to G5; results from the point-like model in
  Sec. \ref{sec:point-like modeling:model output}}
\begin{tabular}{ccccccccc}
\hline
 & z & $\Theta_1$\footnotemark[1] & $\Theta_2$\footnotemark[1] & q & 
$\Theta_{\rm pt}$ & $\sigma_{\rm pt}$ & $r_{\rm t,pt}$ & $M_B$ \\
 & & $(\arcsec)$ & $(\arcsec)$ & & $(^{\circ})$ & $(\rm kms^{-1})$ &
$(\rm kpc)$ &  \\
\hline
G1 & 0.4449\footnotemark[2] & 21.592 & 5.996 & 0.79 & 18.5 &
$121_{-15}^{+16}$ & $13_{-6}^{+15}$ & -19.46\\
G2 & $0.46\pm0.06$\footnotemark[3] & 17.846 & 4.499 & 0.68 & -47.3 &
$190_{-25}^{+26}$ & $24_{-11}^{+28}$ & -21.06\\
G3 & $0.53\pm0.04$\footnotemark[3] & 20.365 & 1.021 & 0.91 & -68.9 & $143_{-20}^{+19}$ &
$16_{-8}^{+19}$ & - \\
G4 & 0.4380\footnotemark[2] & 19.473 & -3.083 & 0.80 & 25.5 & $139_{-19}^{+19}$ &
$16_{-7}^{+18}$ & -19.94\\
G5 & 0.4446\footnotemark[2] & 20.862 & -6.007 & 0.71 & -74.9 & $104_{-14}^{+14}$ &
$11_{-5}^{+12}$ & -18.94\\
\hline
\end{tabular}\\
\tablecomments{The errors give 95\% confidence, derived from the respective
  errors in Sec. \ref{sec:point-like modeling:model output}. $M_B$ is
  calculated independently from the HST photometry, assuming a galaxy
  redshift of z=0.44.}
\footnotemark[1]{relative to the center of the BCG at 12:06:12.134 RA
  (J2000) -08:48:03.35 DEC (J2000)}\\
\footnotemark[2]{spectroscopic redshift}
\footnotemark[3]{photometric redshift estimate, 95\% confidence}\\
\label{modelling:extended:input parameters}
\end{table*}

\begin{figure}[H!tb]
\centering
\subfigure{\includegraphics[bb=3 570 80 665, clip=true,
    scale=.8]{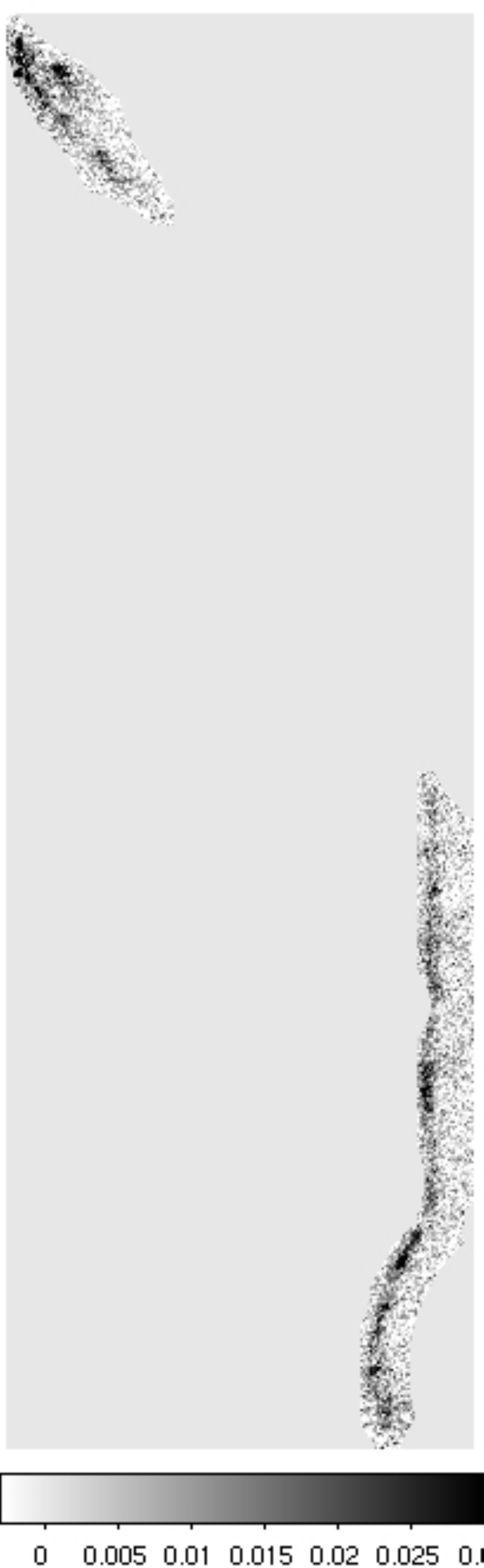}\label{modelling:extended:best-fit:F390Wdata
c}}
\subfigure{\includegraphics[bb=3 570 80 665, clip=true,
    scale=.8]{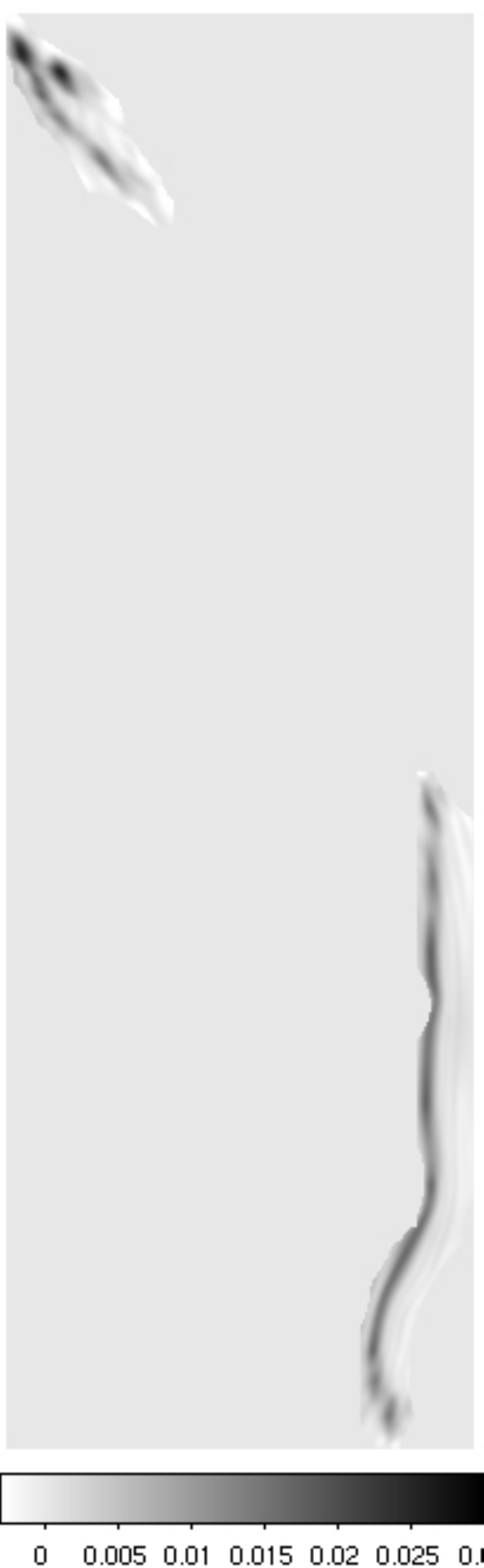}\label{modelling:extended:best-fit:F390Wmodel
c}}
\subfigure{\includegraphics[bb=3 570 80 665, clip=true,
    scale=.8]{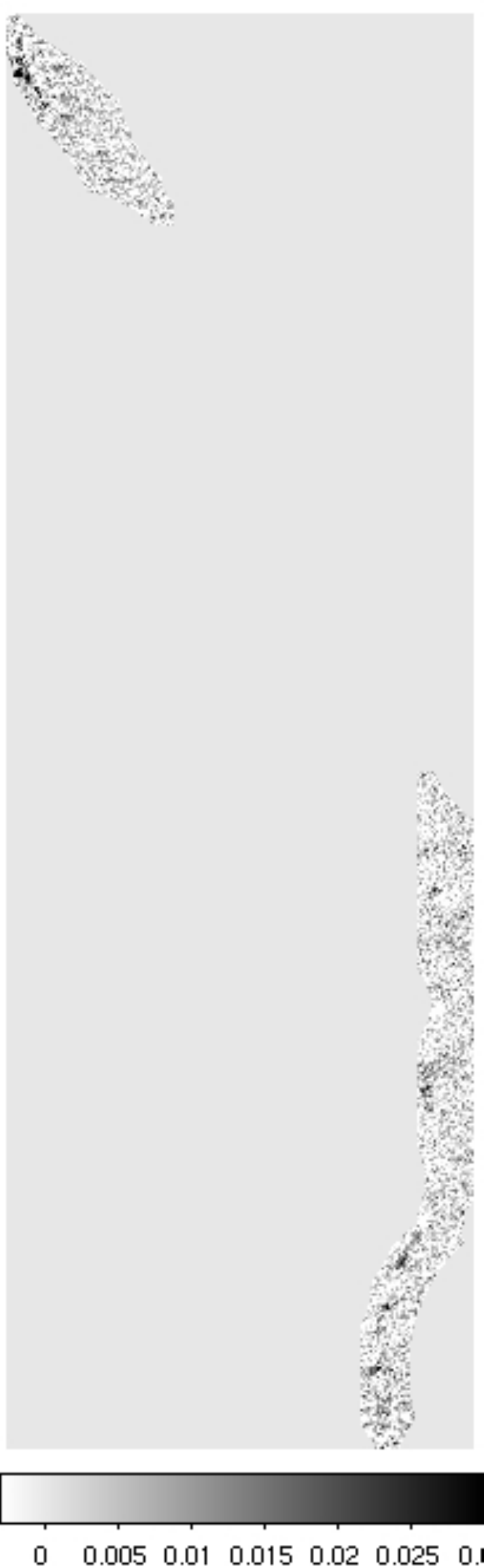}\label{modelling:extended:best-fit:F390Wresidual
c}}
\setcounter{subfigure}{0}
\subfigure[]{\includegraphics[bb=145 45 200 350, clip=true, scale=.8]{{MACSJ1206.2-0847_high_z_errorr_all_scaling.0.8b_free_Te_T_test_es001_im_1}.eps}\label{modelling:extended:best-fit:F390Wdata}}
\subfigure[]{\includegraphics[bb=145 45 200 350, clip=true, scale=.8]{{MACSJ1206.2-0847_high_z_errorr_all_scaling.0.8b_free_Te_T_test_es001_im_2}.eps}\label{modelling:extended:best-fit:F390Wmodel}}
\subfigure[]{\includegraphics[bb=145 45 200 350, clip=true, scale=.8]{{MACSJ1206.2-0847_high_z_errorr_all_scaling.0.8b_free_Te_T_test_es001_im_3}.eps}\label{modelling:extended:best-fit:F390Wresidual}}
\caption{The arc and its counterimage reconstruction in the F435W
  filter are shown in this plot. From left to right the data, the model and the
  residuals are given. The top row shows the counterimage, the bottom row shows
  the giant arc. The levels of gray are the same in each image. For
  this figure, a source size of $20\times20$ pixels is used.}
\label{modeling:extended:best-fit:F390W}
\end{figure}

\begin{figure}[H!tb]
\centering
\subfigure{\includegraphics[bb=3 570 80 665, clip=true,
    scale=.8]{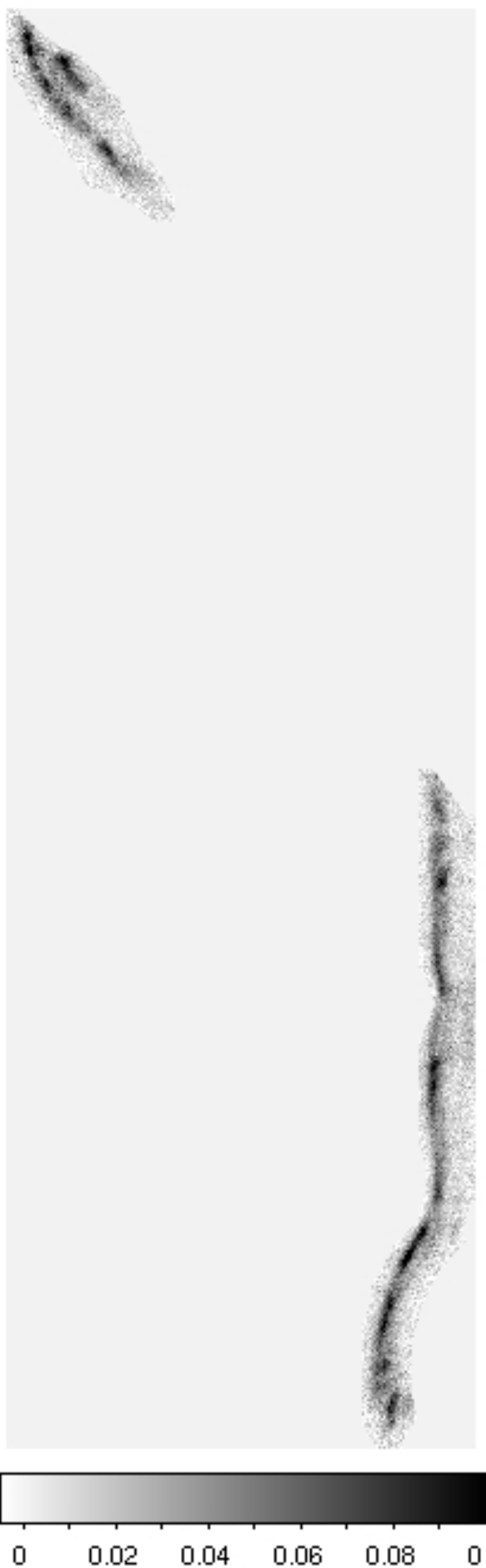}\label{modelling:extended:best-fit:F606Wdata
c}}
\subfigure{\includegraphics[bb=3 570 80 665, clip=true,
    scale=.8]{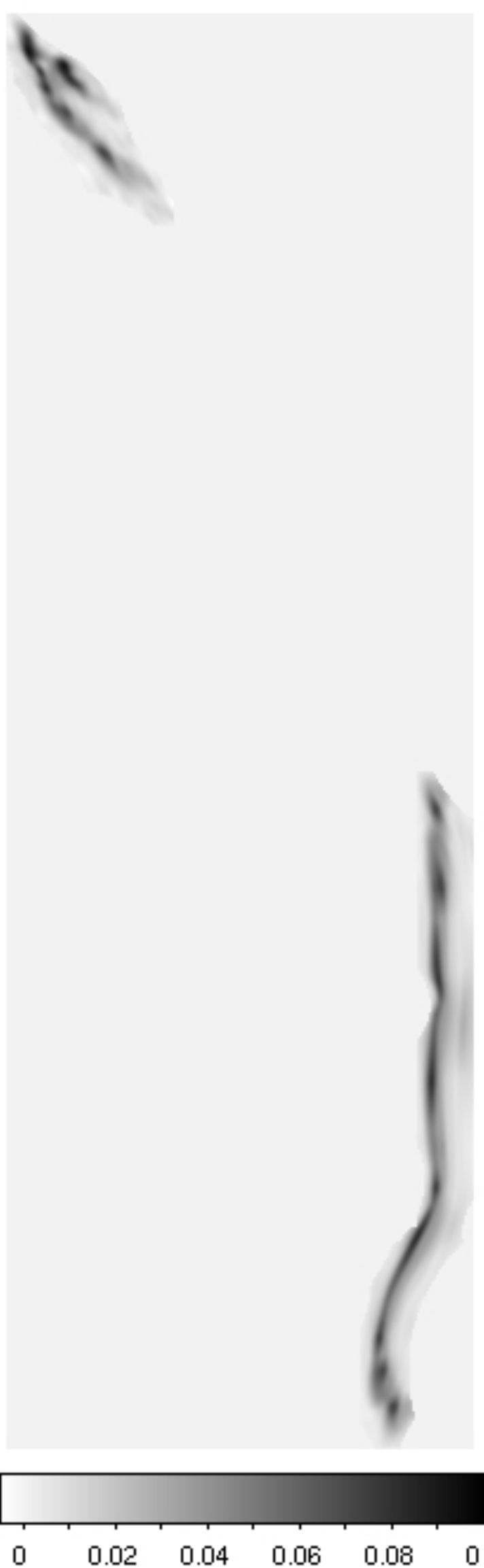}\label{modelling:extended:best-fit:F606Wmodel
c}}
\subfigure{\includegraphics[bb=3 570 80 665, clip=true,
    scale=.8]{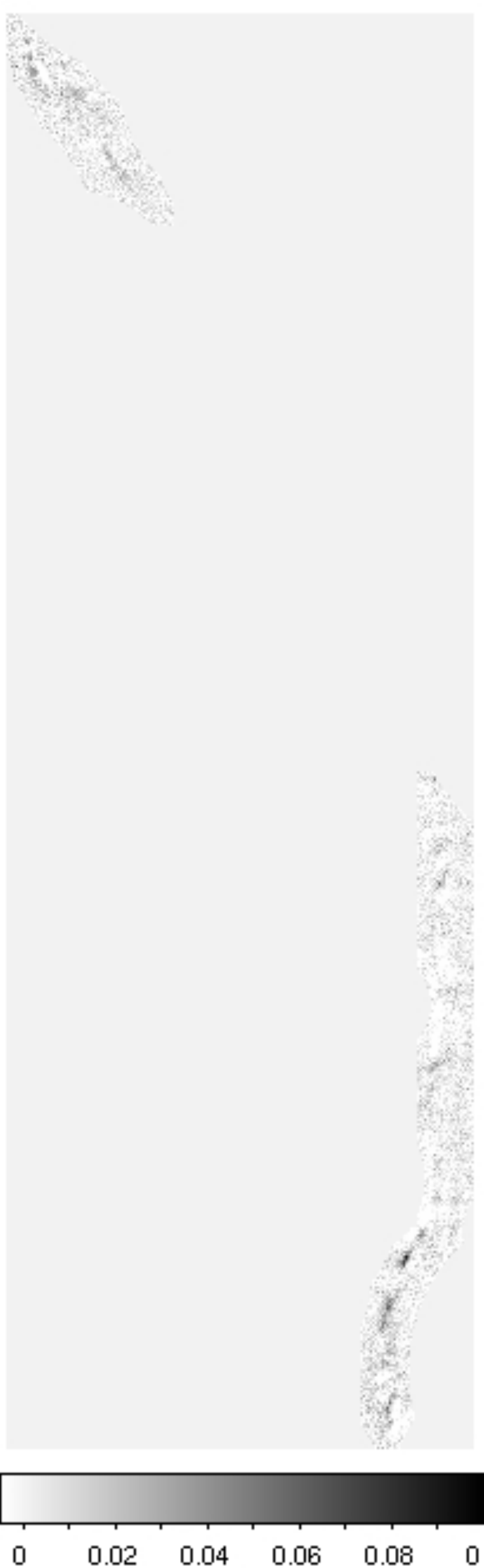}\label{modelling:extended:best-fit:F606Wresidual
c}}
\setcounter{subfigure}{0}
\subfigure[]{\includegraphics[bb=145 45 200 350, clip=true, scale=.8]{{MACSJ1206.2-0847_high_z_errorr_all_scaling.0.8b_free_Te_T_test_es002_im_1}.eps}\label{modelling:extended:best-fit:F606Wdata}}
\subfigure[]{\includegraphics[bb=145 45 200 350, clip=true, scale=.8]{{MACSJ1206.2-0847_high_z_errorr_all_scaling.0.8b_free_Te_T_test_es002_im_2}.eps}\label{modelling:extended:best-fit:F606Wmodel}}
\subfigure[]{\includegraphics[bb=145 45 200 350, clip=true, scale=.8]{{MACSJ1206.2-0847_high_z_errorr_all_scaling.0.8b_free_Te_T_test_es002_im_3}.eps}\label{modelling:extended:best-fit:F606Wresidual}}
\caption{same as Figure \ref{modeling:extended:best-fit:F390W}, this
  time for the F606W filter}
\label{modeling:extended:best-fit:F606W}
\end{figure}

\begin{figure}[H!tb]
\centering
\subfigure{\includegraphics[bb=3 570 80 665, clip=true,
    scale=.8]{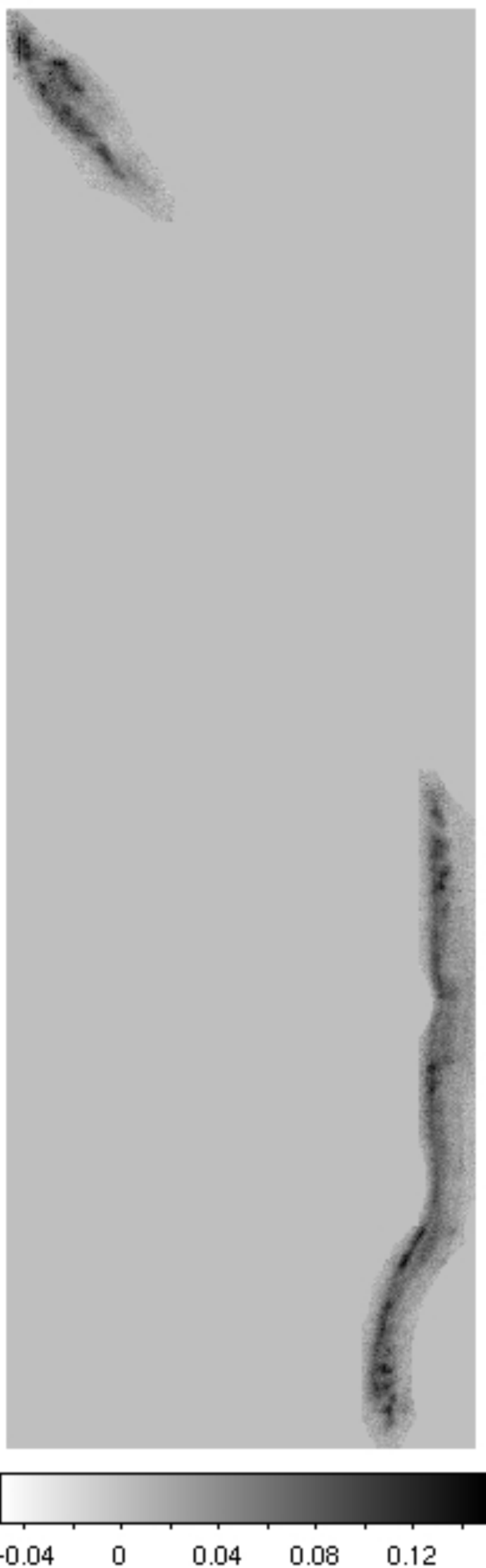}\label{modelling:extended:best-fit:F814Wdata
c}}
\subfigure{\includegraphics[bb=3 570 80 665, clip=true,
    scale=.8]{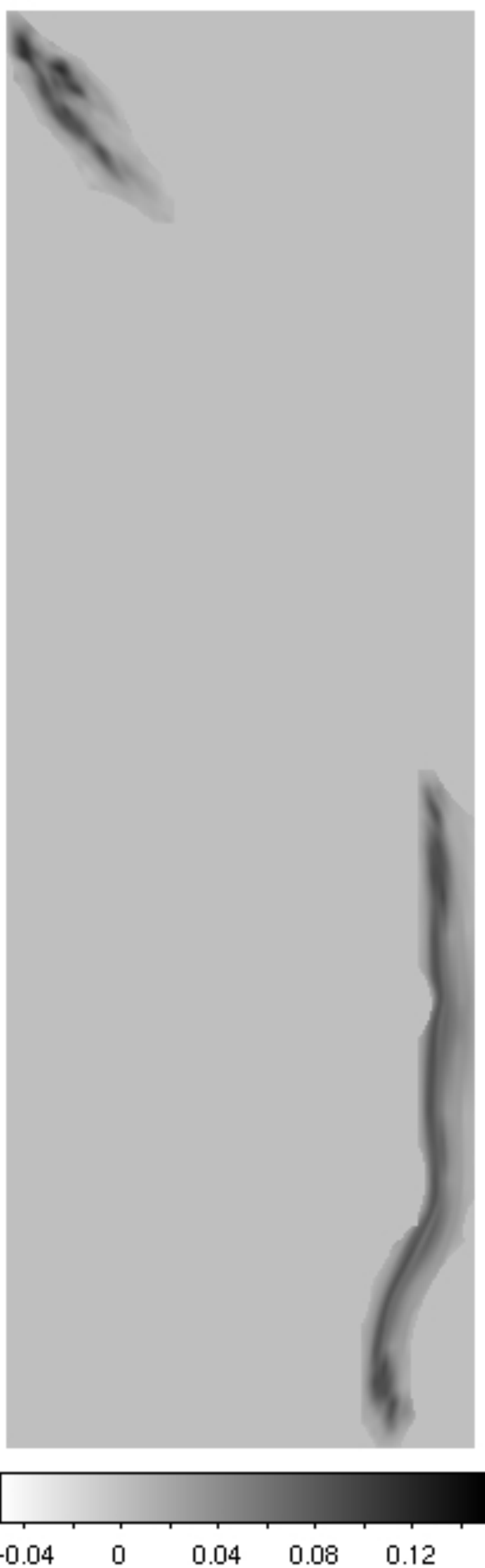}\label{modelling:extended:best-fit:F814Wmodel
c}}
\subfigure{\includegraphics[bb=3 570 80 665, clip=true,
    scale=.8]{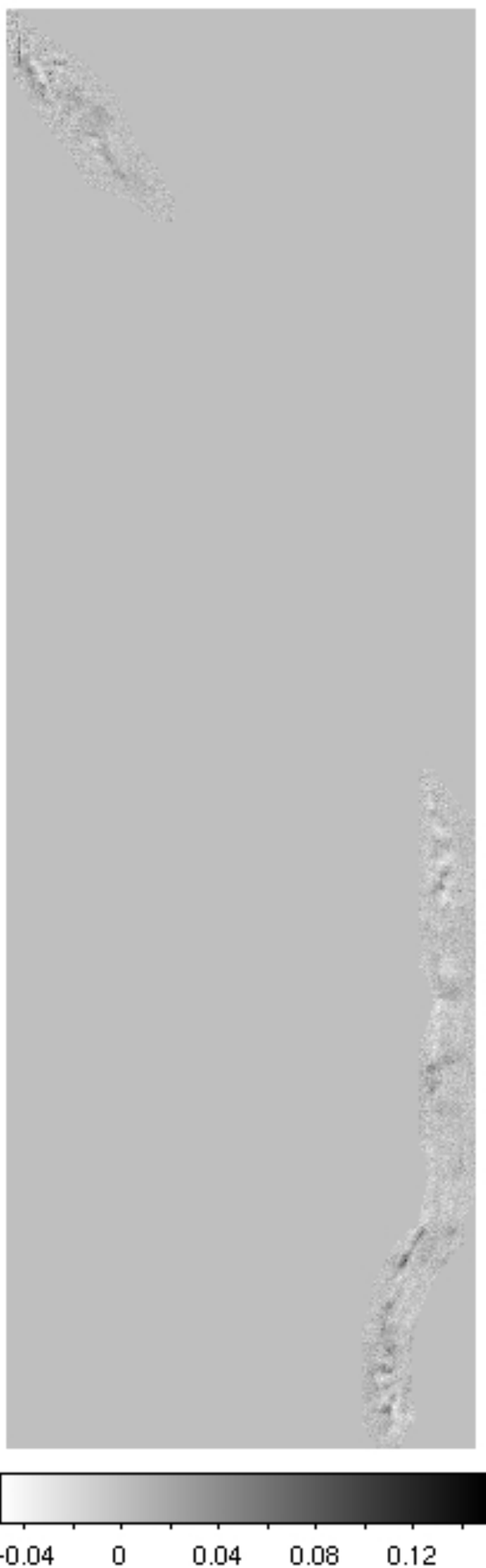}\label{modelling:extended:best-fit:F814Wresidual
c}}
\setcounter{subfigure}{0}
\subfigure[]{\includegraphics[bb=145 45 200 350, clip=true, scale=.8]{{MACSJ1206.2-0847_high_z_errorr_all_scaling.0.8b_free_Te_T_test_es003_im_1}.eps}\label{modelling:extended:best-fit:F814Wdata}}
\subfigure[]{\includegraphics[bb=145 45 200 350, clip=true, scale=.8]{{MACSJ1206.2-0847_high_z_errorr_all_scaling.0.8b_free_Te_T_test_es003_im_2}.eps}\label{modelling:extended:best-fit:F814Wmodel}}
\subfigure[]{\includegraphics[bb=145 45 200 350, clip=true, scale=.8]{{MACSJ1206.2-0847_high_z_errorr_all_scaling.0.8b_free_Te_T_test_es003_im_3}.eps}\label{modelling:extended:best-fit:F814Wresidual}}
\caption{same as Figure \ref{modeling:extended:best-fit:F390W}, this
  time for the F814W filter}
\label{modeling:extended:best-fit:F814W}
\end{figure}

We do not enforce the scaling law on G3, since it is doubtable whether
it is a cluster member or not (it has a different photometric redshift
and is formally not in our cluster member catalog).  Therefore G3 is
modeled with 3 free parameters: its orientation, Einstein radius and
truncation radius. We obtain a best fit model using this 12 free
parameters, optimizing the F435W, F606W and F814W filter data
simultaneously.
\\
The best-fit data, model and residuals for each of the 3 filters are shown in
 Figs. \ref{modeling:extended:best-fit:F390W},
 \ref{modeling:extended:best-fit:F606W} and
 \ref{modeling:extended:best-fit:F814W}. 

The statistical error is
estimated again using a MCMC sampling of the parameter space. The most
likely values and the errors for $r_{\rm t,1\arcsec}$ and the truncation
radius for  each of the galaxies can be seen in Table
\ref{modelling:extended:most likely values}. 

\begin{table}[H!tb]
\tablewidth{75mm}
\centering
\caption{Most likely values and errors for the full surface brightness
  model of the arc and its counterimage}
\begin{tabular}{ccccc}
\hline
 & $r_{\rm t,1\arcsec}$ & $\Theta$ & $\sigma$ & $r_{t}$\footnotemark[1] \\
& ($\rm kpc$) &$(^{\circ})$ & ($\rm kms^{-1}$) & ($\rm kpc$) \\
\hline
G1 & \multirow{4}{*}{$34.2_{-1.2}^{+1.2}$} &$-1.5_{-3.7}^{+3.3}$ &  $130_{-11}^{+10}$ & $21_{-4}^{+4}$\\
G2 &  &$-49.9_{-0.8}^{+0.8}$ &  $165_{-2}^{+2}$ & $29_{-2}^{+2}$\\
G4 &  &$-1.4_{-2.3}^{+2.3}$ & $143.1_{-1.2}^{+1.2}$ & $24.1_{-1.5}^{+1.5}$\\
G5 &  &$-41.2_{-2.7}^{+2.5}$ & $114.9_{-1.5}^{+1.5}$ & $17.9_{-1.5}^{+1.5}$\\
\hline
\end{tabular}\\
\tablecomments{Given are the 95\% c.l. errors. The best fit
    cluster model from Sec. \ref{results:point like:cluster} is used
    as the cluster model.}
\footnotemark[1]{calculated for the galaxies from the scaling law}
\label{modelling:extended:most likely values}
\end{table}

\begin{figure*}
\centering
\subfigure[]{\includegraphics[height=170mm]{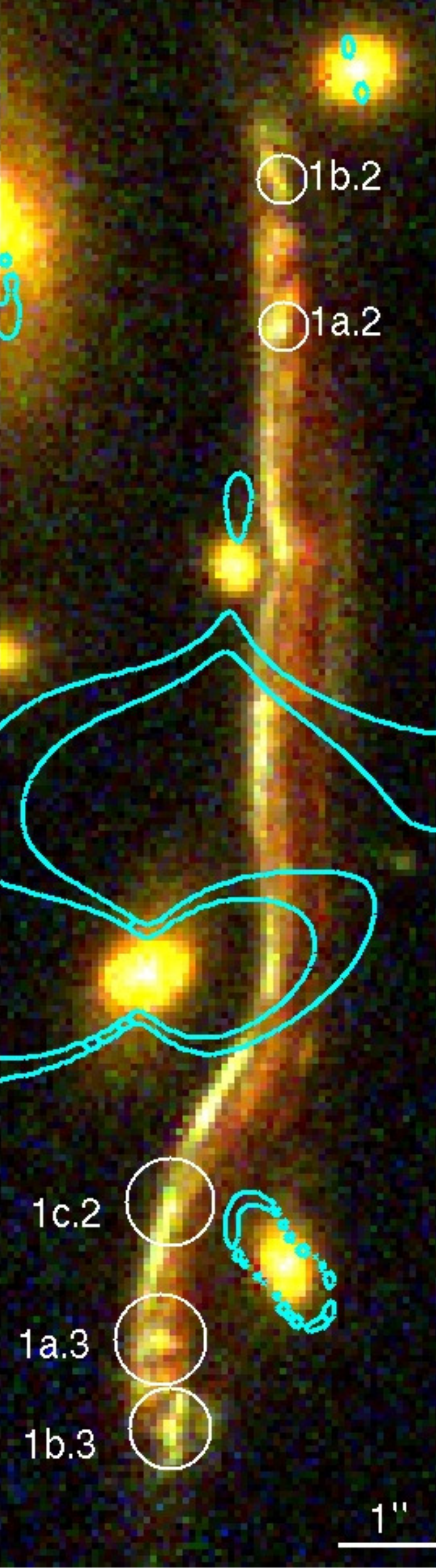}\label{modelling:extended:crit line arc}}
\begin{minipage}[b][170mm][c]{55mm}
\subfigure[]{\includegraphics[width=50mm]{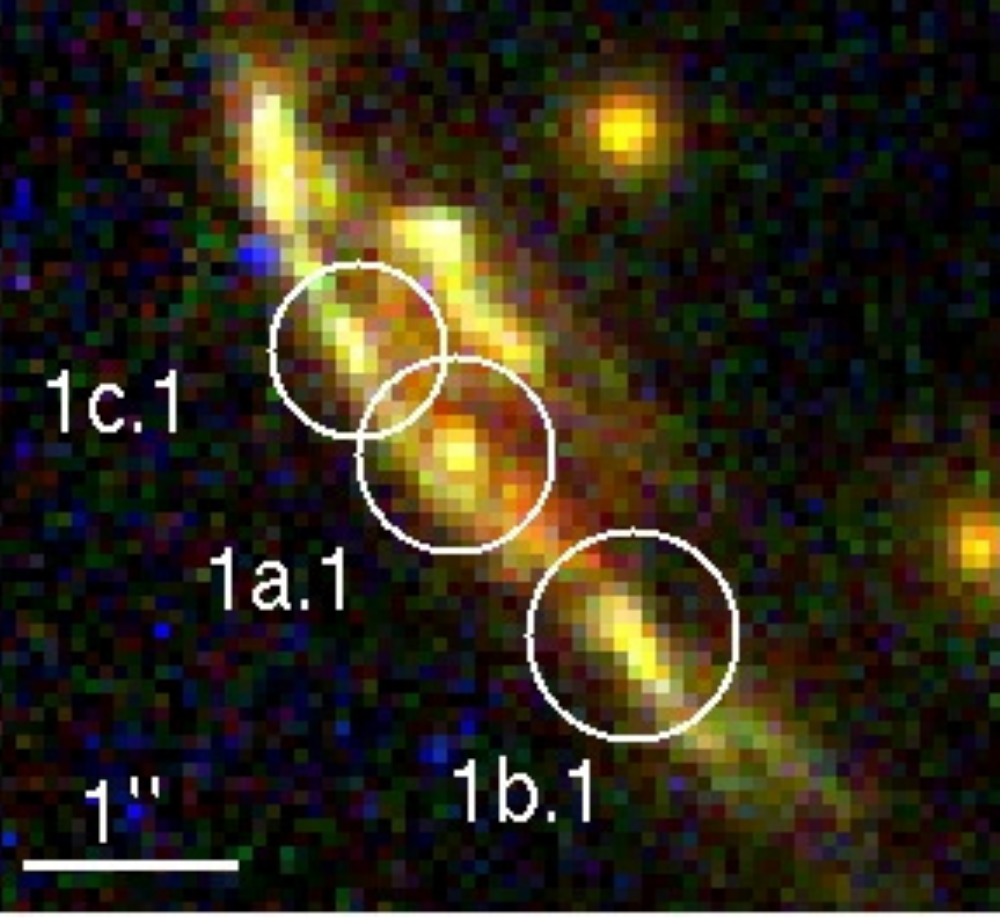}\label{modelling:extended:crit line count}}
\begin{minipage}[b]{55mm}
\subfigure[]{\includegraphics[width=25mm]{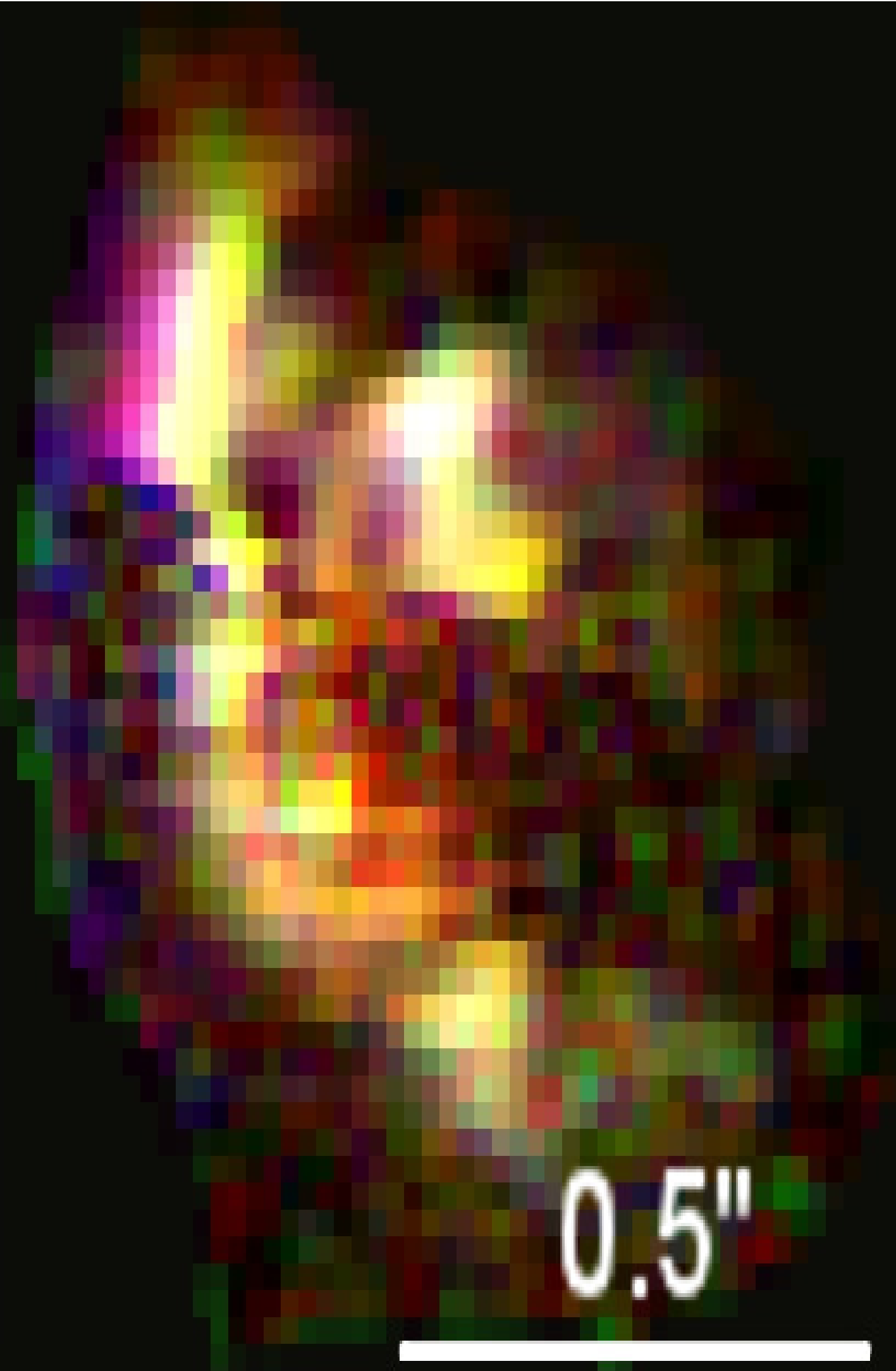}\label{modelling:extended:source
50 pix}}
\subfigure[]{\includegraphics[width=25mm]{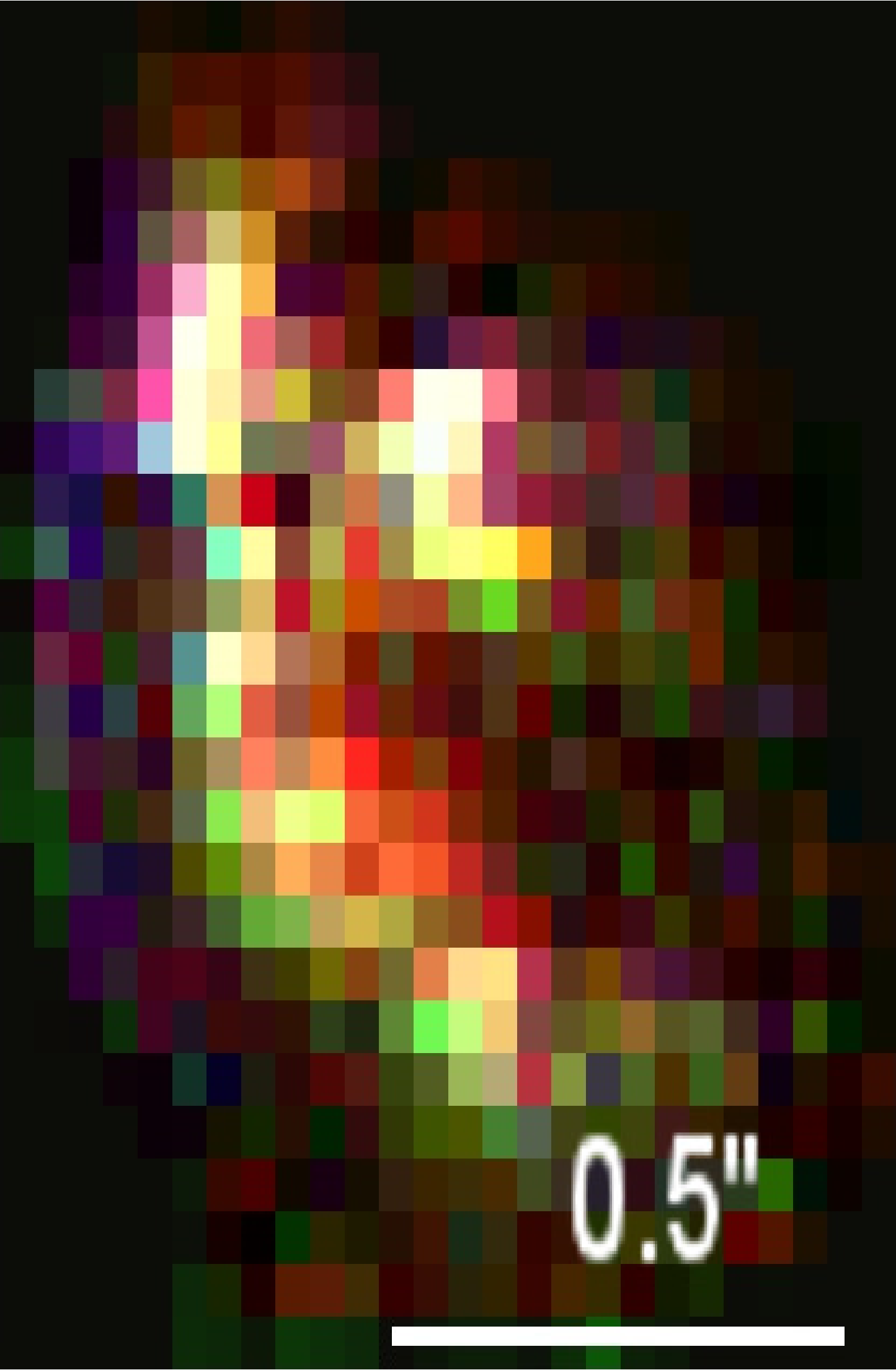}\label{modelling:extended:source
25 pix}}
\end{minipage}
\subfigure[]{\includegraphics[width=50mm]{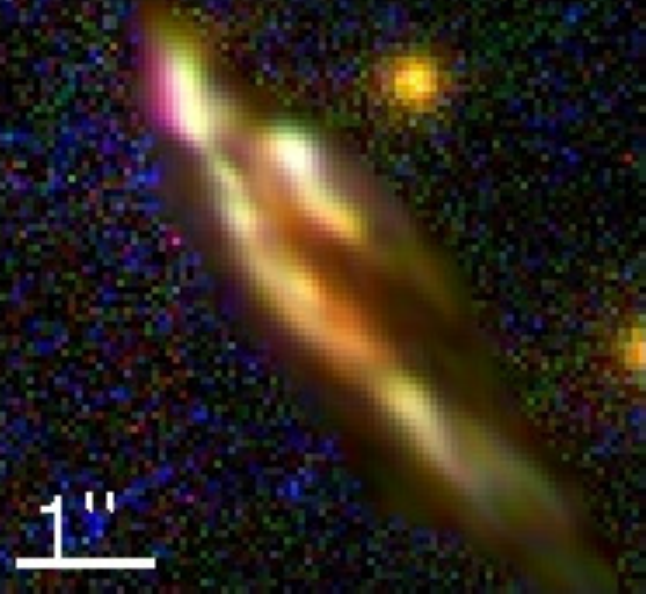}\label{modelling:extended:counterrec color}}
\end{minipage}
\subfigure[]{\includegraphics[height=170mm]{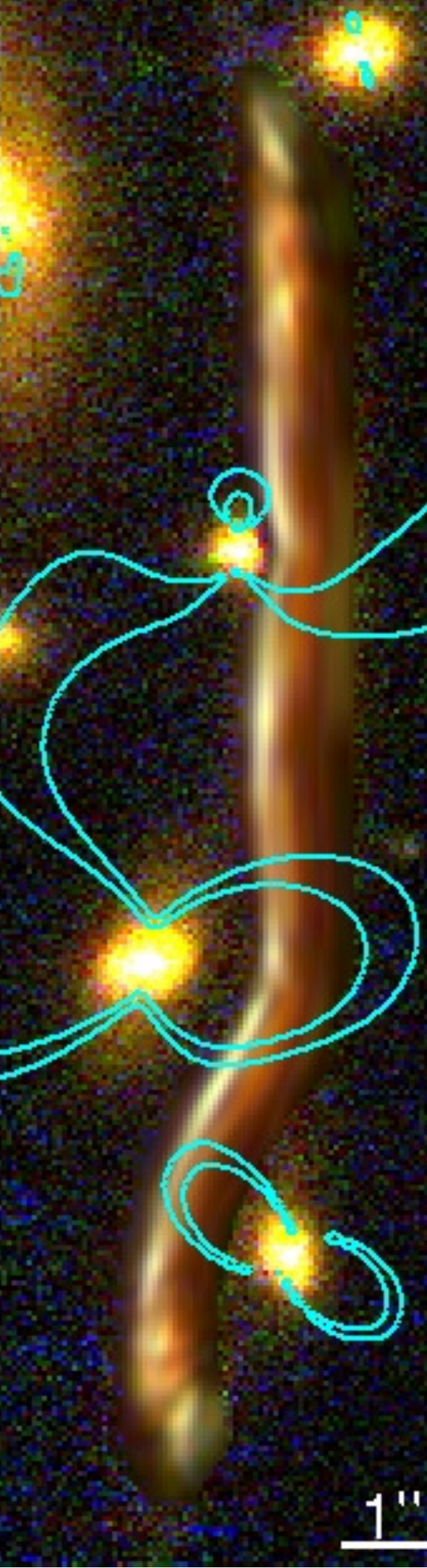}\label{modelling:extended:arcrec color}}
\caption{This figure shows, from left to right: The
    observed arc on the left. In the middle, from top to bottom: The
    observed counterimage, the reconstructed source with different
    resolutions and the model
    counterimage. On the right: The model arc. All images are
    combinations of the F435W, F606W and F814W bands, respectively. On
    the left, the numbers mark the multiple image input positions on
    the arc for the point-like model. We overplot the critical line
    structure in cyan on the left for the point-like model, on the
    right for full surface brightness
    reconstruction, respectively. The critical lines are calculated
    from a pixelated
  magnification map, the lines define regions above a absolute
  magnification value of 100, not taking parity into account. For the sources, the left source (\ref{modelling:extended:source
50 pix}) shows the
  source galaxy at a 50 pixel grid, giving a better than HST
  resolution, the right hand source (\ref{modelling:extended:source
25 pix}) shows the same source at approximate
HST resolution.}
\label{modelling:extended:reconstruction:colors}
\end{figure*}

\begin{figure}[tbh]
\centering
\includegraphics[height=80mm]{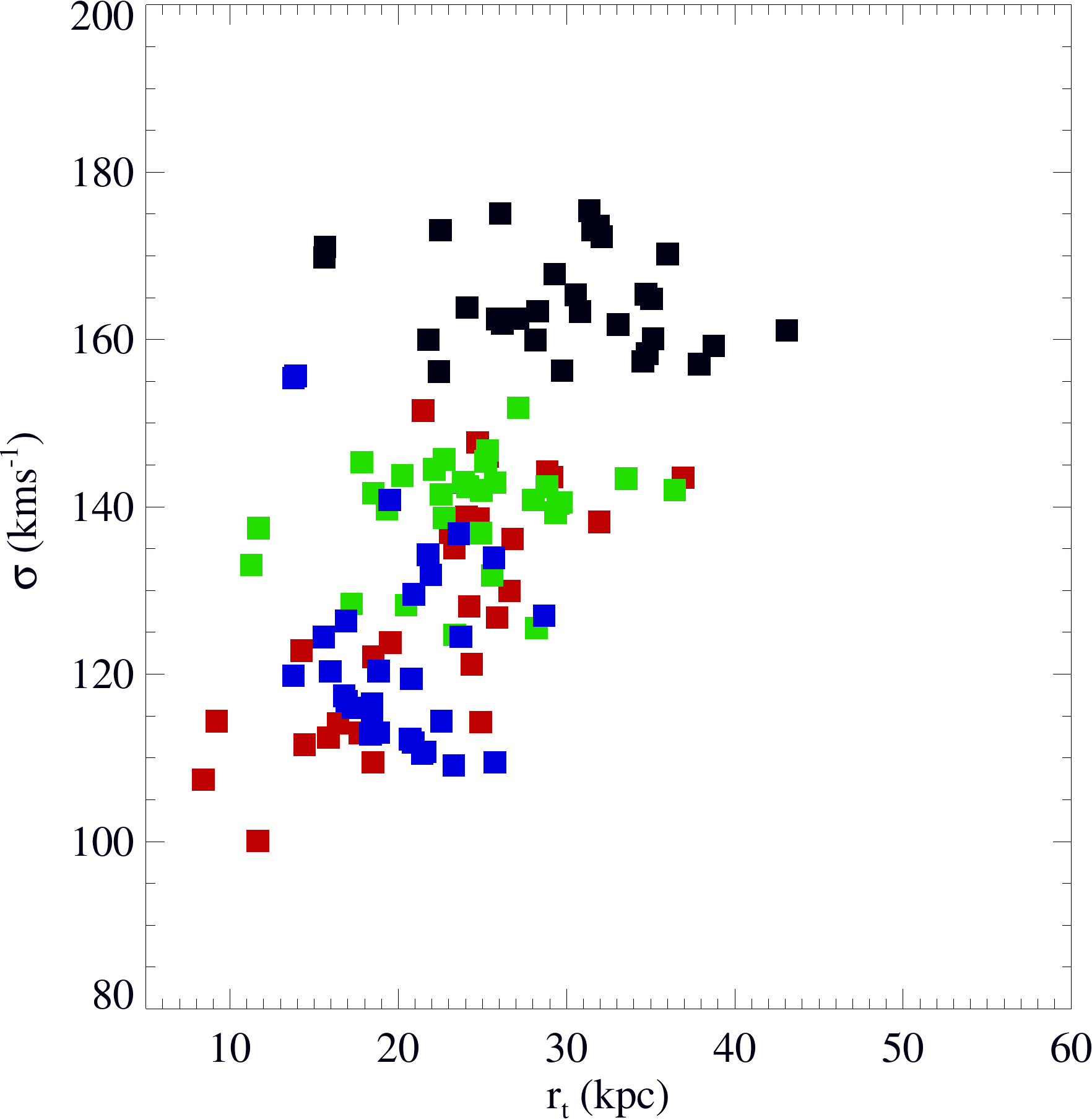}
\caption{The velocity dispersions and calculated truncation radii for
  the galaxies G1, G2, G4 and G5 for the different cluster
  realizations. Each cluster representation has one entry for each
  galaxy. The color coding is the following: red: G1, black: G2, green:
  G4, blue: G5.}
\label{modelling:extended:systematics:diff clusters}
\end{figure}

The truncation for the
individual galaxies is still following
Eq. \ref{theory: galaxy scaling Rusin} with $\sigma^{\star}=186\rm
kms^{-1}$.  For every galaxy we give its most likely values and the
95\% c.l. errors. The truncation uncertainties for each of the
galaxies are derived from the uncertainties on the Einstein radii and
the truncation scaling law. Especially by comparing Tables
\ref{modelling:extended:input parameters} and
\ref{modelling:extended:most likely values}, 
we note that the truncation scaling amplitude and the
Einstein radii for the galaxies agree with each other within the errors, but giving
tighter constraints from the extended image reconstruction. The
orientations of the galaxies in Tables \ref{modelling:extended:input
parameters} and \ref{modelling:extended:most likely values} change by
$\approx 20$ to $30^{\circ}$, meaning that there is a misalignment
between light and total mass for these galaxies. This misalignment
value is slightly higher than the $\approx 18^{\circ}$ found by
\citet{slacs7} on isolated early type strong lensing galaxies.
\\
\cite{suyuhalkola2010} quote a misalignment of their satellite light
and dark matter major axis of about
$50^{\circ}$. \cite{knebe2008} show from {\it
    N}--body simulations that satellite
  halos as a whole prefer to be radially aligned with respect to the
  centers of their host halos, but not the satellites' inner
  parts (which predominantly trace the light distribution). This leads
  to a misalignment between light and dark matter of satellite galaxies. Our misalignment is
not as high, but nevertheless it would
be worth to study  how tidal effects can alter the major axis of dark
matter halos.
\\
In Fig. \ref{modelling:extended:reconstruction:colors}
the observed arc (Fig. \ref{modelling:extended:crit line arc}) and its
counterimage (Fig. \ref{modelling:extended:crit line count}) are shown
in the left column and the top row of the middle column; alongside with this,
the same is shown for a replacement of the arc and its counterimages
with its full surface brightness reconstruction from its best-fit
models in the left column (Fig. \ref{modelling:extended:arcrec color})
and the bottom row of the middle column
(Fig. \ref{modelling:extended:counterrec color}). The angular scales are
given in the figures. The reconstructed source
can also be seen in this Figure as the two panels in the middle column
(Figs. \ref{modelling:extended:source 50 pix} and \ref{modelling:extended:source
25 pix}).
It is fully lensed into the
counterimage and only partly lensed into the arc itself. There are 2
versions of the source, one with $50\times50$ pixels, giving a
resolution superior to HST/ACS and a $25\times25$ pixels source,
giving the same source as it would be observed at approximate HST/ACS
resolution. Both sources show the same
field of view of $0.94\arcsec$ in x and $1.42\arcsec$ in y direction,
respectively. To estimate the magnification of the counterimage, we
map the masked area in Fig. \ref{modelling:extended:F814Wmask
counter} ($A_{\rm CI}=6.3\rm arcsec^2$) back into the source plane and get
an area of $A_{\rm sr}=1.1 \rm arcsec^2$. Therefore, the  magnification of the
counterimage is $\mu_{\rm counterimage}=5.8$. We repeat this with the
signal-to-noise based mask mentioned above ($A_{\rm CI}=5.2\rm
arcsec^2$,$A_{\rm sr}=0.9 \rm arcsec^2$) and get the same value for
the magnification. Also, a direct calculation of the Jacobian matrix at the
position of the counterimage gives a similar value.
\\
While the above statements are made for the best fit cluster model we
now marginalize over the variety of cluster distributions compatible
with the observations. To estimate the uncertainty related with the
cluster model, we repeat the extended model analysis for 30 random
cluster representations. These representations are taken from the MCMC
sampling calculated in Sec. \ref{sec:point-like modeling:model output}
to estimate the error. The results are presented in Table
\ref{modelling:extended:most likely values clusters}. 
\begin{table}
\tablewidth{75mm}
\centering
\caption{Most likely values and errors for the full surface brightness
  model of the arc and its counterimage, taking different cluster
  models into account}
\begin{tabular}{ccccc}
\hline
 & $r_{\rm t,1\arcsec}$ & $\Theta$ & $\sigma$ & $r_{t}$ \\
& ($\rm kpc$) &$(^{\circ})$ & ($\rm kms^{-1}$) & ($\rm kpc$) \\
\hline
G1 & \multirow{4}{*}{$35\pm8$} &$1.2\pm20.6$ & $128 \pm 18$ & $22\pm7$
\\
G2 & &$-47.0\pm6.1$ & $165\pm6$ &$30\pm6$ \\
G4 & &$9.3\pm17.6$ & $140\pm6$ & $24\pm6$ \\
G5 & &$-45.3\pm19.1$ & $124\pm13$ & $20\pm4$ \\
\hline
\end{tabular}\\
\tablecomments{From the
    MCMChain used to calculate the errors in Sec. \ref{results:point
      like:cluster}, 30 random cluster representations are taken. The
    analysis outlined for the best-fit cluster model is repeated for
    each of the random cluster models. The errors give the r.m.s
    errors on the galaxies' parameters, and are therefore marginalized over these different cluster models.}
\label{modelling:extended:most likely values clusters}
\end{table}
We see that the
errors on the parameter estimates are increased
compared to Table \ref{modelling:extended:most likely
values} by taking the uncertainties from the cluster model into
account. For the truncation, we get slightly tighter constraints than the
point-like model described in Eq. \ref{results:point like: galaxy
  scaling law} in  Sec. \ref{sec:point-like modeling:model
output}. 
We get: 
$$\rt=35\pm 8\rm kpc\times \left(\frac{\sigma}{186\rm
    kms^{-1}}\right)^{4\over3}\quad .$$
The velocity dispersions and truncation radii for galaxies
G1, G2, G4 and G5 for the different clusters are plotted in
Fig. \ref{modelling:extended:systematics:diff clusters}.

\subsection{Tests for systematic errors}
The statistical error for the truncation scaling
in this galaxy cluster is on
the order of 25\%, making this method in principle a good tool
to study truncation of galaxies.
\\
We now investigate the robustness of the truncation and Einstein radii
results derived in 
Sec. \ref{Sec:extended_reconstruction:results} against possible sources of
systematic errors. Possible systematic effects might stem from the
treatment of the data of the filters or the frames itself, the
analyzed arc region,
the number of source pixels or the forced scaling law.
First, we repeat the analysis in each of the filters individually. The
results for the different filters are summarized in Table \ref{modelling:extended:galaxies parameter table}:
All values agree with each other within the 95\%
c.l. intervals, implying that the surface brightness distribution in
different filters gives consistent results regarding the halo
truncation. Since the F435W band data have lower signal to noise for
the arc than the data in the two redder filters considered in this work, the
best fit parameters for the the model using all 3 filter data
simultaneously are driven by the two redder bands.
\\
Next, we change the investigated region around the arc based on a
$2\sigma$ cut of a smoothed signal to noise map in the F814W filter.
We again use the data of all three filters at the same time. For the
mask based on the signal to noise level we get slightly different but
consistent values for the truncation scale and the individual Einstein
radii, see Table \ref{modelling:extended:galaxies parameter
table}(``mask2''). \\ 
Next we use different numbers of source
pixels. For the analysis, we use only the F814W filter and the
standard mask. Starting from a $8\times8$ pixel grid and going up to a
$13\times13$ grid, we calculate the best fit for each model. The
results are again given in Table \ref{modelling:extended:galaxies
parameter table}(``sr pix''). We get a systematic uncertainty from the
source pixel size comparable to the statistic uncertainties for the
best fit cluster model when we fix the cluster
  potential. We verify that this is also true for much different numbers
of source pixels. Using a $25\times25$ and $30\times30$ pixel grid
we get values consistent with the ones stated in Table
\ref{modelling:extended:most likely values clusters}.\\
Recent spectroscopic results indicate that G3 could
  be a member of the galaxy cluster. Hence we repeat the above
outlined analysis including G3 as a cluster member allowing for a
free central velocity dispersion and orientation, but
  forcing it to follow the same scaling law for the truncation as
  G1, G2, G4 and G5. Doing this, there is no change in the
truncation scaling or a decrease of the errorbars.
\\
Finally we investigate how the truncation results depend on the assumed
Faber-Jackson index $\delta$. We use $\delta=0.25$ instead of
$\delta=0.3$, still keeping $\epsilon=0$. We restart the modeling for the point-like images,
fixing the global parameters and then turn again to the extended image
modeling. The corresponding truncation radii are shown in the last
column of Table 
\ref{modelling:extended:galaxies parameter table}(``FJ,$\delta=0.25$").
Here, the truncation law gets:
$$\rt=41.8\rm kpc\times \left(\frac{\sigma}{186\rm
    kms^{-1}}\right)^{2}\quad .$$
The individual velocity dispersions and derived truncation radii, however
agree with the ones derived before within the errors, see Tables
\ref{modelling:extended:most likely values clusters} and \ref{modelling:extended:galaxies parameter table}.
 This means, there is no indication for the preferred exponent of the scaling
 law in this work since both scaling laws give
   similarly good fits.
\\
Our tests show that the systematic errors are smaller than the ones
from the uncertainty of the cluster potential, making our estimates
robust with respect to systematic effects. In summary, we conclude
that if we vary the weighting of the extended image input data (SFB in
different filters), the masking regions or modeling details as the
assumed Faber Jackson index then these changes the estimated halo
sizes less than our ``statistical errors'' due to different global
halo models from the MCMC sample.
\begin{table*}
\centering
\caption{Parameter results for the systematic tests}
\begin{tabular}{cccccccc}
\hline
&&F435W\footnotemark[1] & F606W\footnotemark[1] & F814W\footnotemark[1]
& mask2 & sr pix\footnotemark[2] & FJ,$\delta=0.25$ \\
\hline
$r_{\rm t,1\arcsec}$& ($\rm kpc$) & $35.1_{-4.7}^{+6.3}$ & $36.0_{-2.0}^{+2.1}$ &
$36.9_{-2.0}^{+2.2}$ & 35.6 & $34.9\pm0.9$  & 41.8 \\
$\frac{\alpha}{\delta}$ & & $\frac{4}{3}$ & $\frac{4}{3}$ &
$\frac{4}{3}$ & $\frac{4}{3}$ & $\frac{4}{3}$  & 2.00 \\
$\sigma_{\rm G1}$ & ($\rm kms^{-1}$) & $115_{-6}^{+5}$ & 
$126.4_{-1.4}^{+1.4}$ & $129.1_{-1.4}^{+1.3}$ & 124 &
$124\pm8$  & 133 \\ 
$r_{t,G1}$ & ($\rm kpc$) & 18.2 & 22.6 & 21.6 & 20.1 &  $20.2\pm1.6$
&  21.5 \\
$\sigma_{\rm G2}$ & ($\rm kms^{-1}$) & $161_{-7}^{+6}$ &
$161_{-2}^{+2}$ & $166_{-2}^{+2}$ & 164 & $162\pm3$
& 165 \\
$r_{t,G2}$ & ($\rm kpc$) & 28.4 & 31.1 & 30.1 & 29.3 & $29.0\pm0.8$  & 32.5 \\
$\sigma_{\rm G4}$ & ($\rm kms^{-1}$) & $132_{-5}^{+4}$ &
$140.7_{-2.5}^{+2.4}$ & $141.9_{-1.2}^{+1.2}$ & 139 &
$140.7\pm1.8$  & 143 \\
$r_{t,G4}$ & ($\rm kpc$) & 21.9 & 25.9 & 24.6 & 23.7 & $24.1\pm0.6$ &
24.7 \\
$\sigma_{\rm G5}$ & ($\rm kms^{-1}$) & $117.9_{-1.5}^{+1.5}$ &
$117.9_{-1.5}^{+1.5}$ & $113.4_{-1.5}^{+1.5}$ & 119 &
$117.9\pm3.5$&116 \\
$r_{t,G5}$ & ($\rm kpc$) & 19.3 & 20.8 & 18.0 & 19.3 & $19.0\pm0.8$ &
16.2 \\
\hline
\end{tabular}\\
\footnotemark[1]{The errors given are the 95 \% c.l. on the input parameters}\\
\footnotemark[2]{given are the r.m.s. errors}
\tablecomments{We omit errors for the truncation radii of the
  individual galaxies since these can be derived from the truncation
  law for the individual filters. We omit all errors for the mask2 and
FJ,$\delta=0.25$ models since these are similar to the ones stated in
Table \ref{modelling:extended:most likely values}.}\\
\label{modelling:extended:galaxies parameter table}
\end{table*}
%
%
\section{Discussion}
\subsection{Lens modeling and cluster mass distribution}
Using positions of multiply imaged galaxies we measured the mass distribution in
the center of MACSJ1206.2-0847 based on a parameterized model, where
the smooth dark matter was described with an elliptical NFW-profile
and the matter traced by cluster galaxies was described with
singular truncated isothermal
ellipsoids. Using scaling relations between luminosity 
and velocity dispersion and between luminosity and truncation radius, the
essential halo parameters (velocity dispersions and truncation radii)
of all galaxies' dark matter halos are modeled with just 2 free parameters. 
The best fit model reproduces
the observed multiple image positions with a mean accuracy of $0.85\arcsec$. 
The level of the positional mismatch is in agreement with expectations
from unaccounted substructure or LOS contamination. 
For the same cluster \citet{zitrin2011} get a slightly higher value of
$\approx 1.3\arcsec$ for the average image--plane
  reproduction uncertainty per image. \\
In general the match of multiple image position seems to depend on the number of multiple
images that have been identified
\citep{zitrin2011A,richard2010Locuss,lim07b,halkola06}. Given the
number of multiple image systems a mean image plane distance below
$1\arcsec$ is a rather good value. 
\\
Finally we find that the model
would become better and require a more reasonable value for the external
shear if we account for the intra-cluster light which has an almost
rectangular shape and a major axis in the direction of the major
cluster axis, indicating stripped stars. This offers prospects
to constrain the properties (e.g.. mass) of the intra--cluster light
component, which is however beyond the scope of this work.
\\
Our total mass profile agrees with that from the previous work of
\cite{zitrin2011} and \cite{ume12}. Regarding values for
concentration and scale radius for the total cluster mass distribution
we  refer the reader to the work of \cite{ume12} since in this
work the mass profile has been constrained on much larger scale (using
strong- and weak-lensing shear and magnification information). 
\\
In addition to previous work we pay special attention to match the
extended surface brightness distribution of the giant arc and its
counterimage as observed in the F435W-, the F606W- and the F814W-filters. 
This helps us to constrain the velocity dispersion and truncation
parameters of cluster galaxy halos considerably beyond the result obtained from our point
source modeling alone. We ensured that the results are robust
regarding modeling details and regarding the exact 
information used from the extended light distribution of the arc.
\subsection{Halo velocity dispersion versus Faber-Jackson relation}

\begin{figure}[h!]
\centering
\includegraphics[width=80mm]{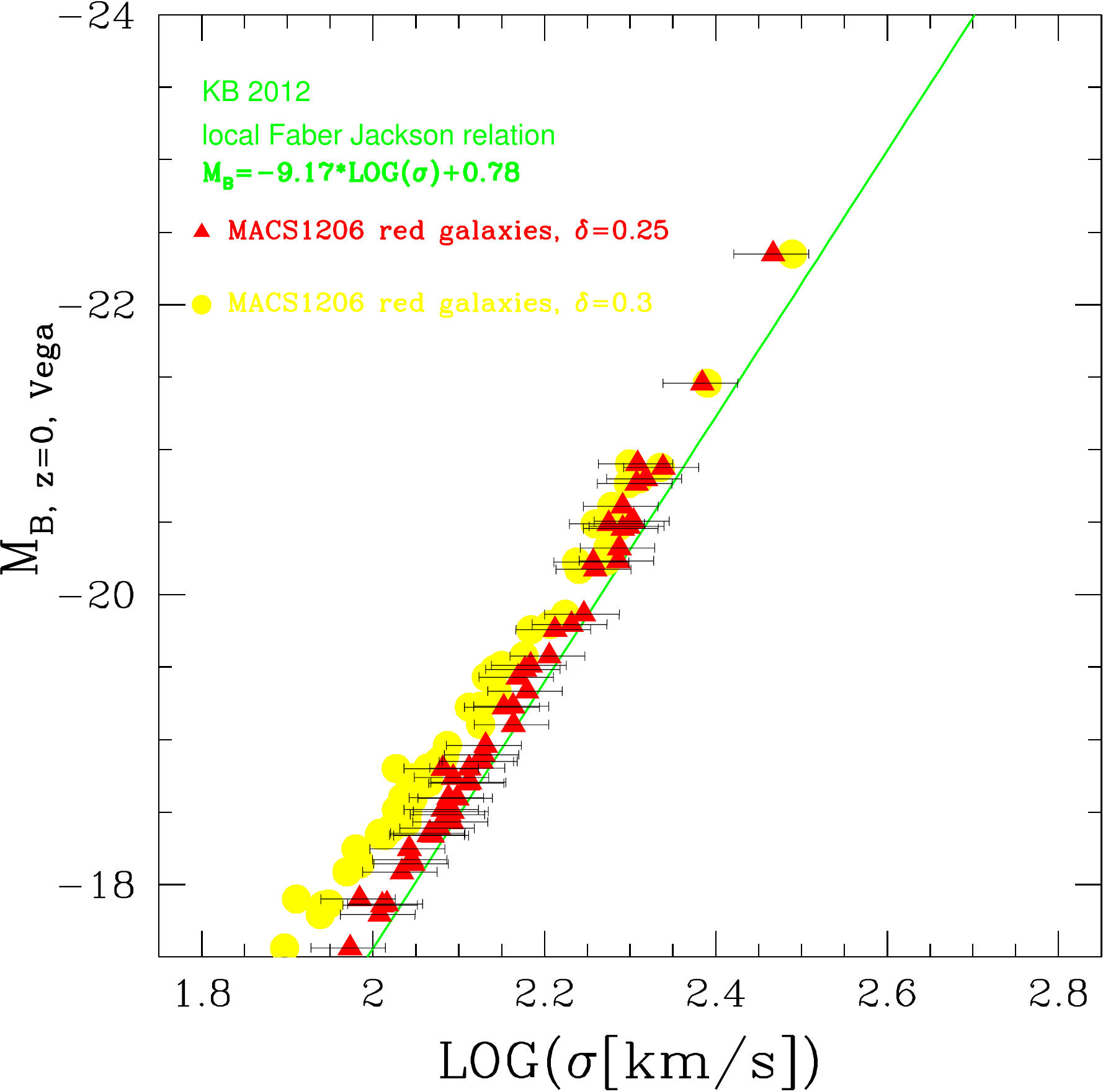}
\caption{This Figure shows with a green line the local Faber-Jackson
  relation in absolute Vega B-Magnitudes vs the central stellar
  velocity dispersion from \cite{kormendy2013}. The red triangles show the
absolute B-magnitude of MACSJ1206.2 red cluster members corrected for the
luminosity evolution to redshift zero by a factor of 1.8 versus the
halo velocity dispersion obtained from the lens modeling. Note that
we do not model each galaxy separately but only the amplitude of the
relation for the assumed scaling law (in this case $\delta =
0.25$). The filled yellow circles show the same galaxies for the assumed scaling
law of $\delta = 0.30$.
The scatter around the $\delta=0.25$-slope is due to the fact
that the luminosity-$\sigma$ scaling was applied using the NIR
F160W-data and not the restframe B-magnitude obtained from the
SED-fitting. The small scatter demonstrates that the SEDs of the red
galaxies are fairly uniform.} 
\label{modeling:point-like:FJ_vs_lensing}
\end{figure}

The amplitudes for the luminosity vs velocity dispersion scaling law
(and the luminosity vs truncation radius scaling law) were constrained
without any reference to optical galaxy properties. 
We obtain for the relation between the apparent AB-magnitude in the
$F160W$-filter and the halo velocity dispersion
\begin{equation}
m_{160, AB}=-8.333 \log (\sigma \rm [kms^{-1}]) + 37.39 
\label{Hband-halo-velocity-dispersion}
\end{equation}
In the above relation the value for the slope was assumed and the
zeropoint determined. The lensing
  derived velocity dispersion in this work agrees with the measured
  stellar velocity dispersion for the BCG. Recent measurements also indicate
  an agreement of the lensing derived and measured velocity dispersion
for GR.\\
It is known from field
elliptical strong lenses that multiple image systems can be well
reproduced assuming an isothermal total mass profile with an amplitude
given by the central stellar velocity dispersion. This isothermality
is measured out to two Einstein radii (\cite{slacs3,sdss1538, sdss1430}).
However, since Einstein radii of elliptical galaxies are
typically of the order of the effective radius, the mass
distribution is only measured out to one effective radius with strong
lensing of field elliptical galaxies. This is the scale where the
stellar mass is still dominating or at most the dark matter and
luminous matter are of the same order. 
We want to compare the lensing derived
  Faber-Jackson relation from this work with a local estimate from
  \cite{kormendy2013}. For that, we need the absolute B-band
  magnitudes $M_B$ for the cluster members and evolve these to
  $z=0$.
For all galaxies in our cluster
member catalog we fit the spectral energy distribution (SED) using
their full 16-filter photometry (see Fig. 10 of \citealt{postman_clash2011})
and assuming that they are at $z=0.44$. We in this way obtain for each
cluster member the SED-type and  an estimate for the restframe 
absolute magnitude in the Bessel B-band, $M_B$  (in the Vega system). 
We then use redshift evolution
of the elliptical galaxies fundamental planes mass to light
ratio, which we assume to be due to aging of the
stellar population (luminosity evolution). \cite{saglia2010} measured this in the
EDISC sample with cluster (and field) elliptical galaxies
and obtained an evolution of the mass to light ratio of cluster
elliptical galaxies of $\Delta \log M/L_B =-1.6*(1+z)$ which gives a flux dimming by a factor of 1.8
from $z=0.44$ to $z=0$. We plot the luminosity evolved absolute B-band
magnitudes of red cluster members versus their halo velocity dispersion
in Fig. \ref{modeling:point-like:FJ_vs_lensing}. The velocity
dispersion results for the $\delta=0.3$ case are shown in yellow, and
those for the $\delta=0.25$ case in red. We do not change the halo velocity dispersion when
evolving the cluster elliptical galaxies to redshift zero, since
at fixed stellar mass there is hardly any evolution of the stellar
velocity dispersion from redshift $0.44$ to zero according to Fig. 22
of \cite{saglia2010}. We assume the
same to hold also for the halo velocity dispersion.
We also draw errors of 10 \% for the velocity dispersion to guide
the eye, since this is the accuracy at which we can determine the
amplitude of the luminosity versus velocity dispersion scaling.
In the same Figure we added the local Faber-Jackson relation from
\cite{kormendy2013} as a green line. Its slope (in our
notation) is $\delta_{FJ}=0.273$ and thus in between our assumed
$\delta = 0.25$ (red triangles) and $\delta = 0.3$ (yellow circles) cases.
Both results agree within their errors with the Faber-Jackson relation,
although the $\delta=0.3$ case is shifted to lower velocity
dispersions at the faint end.\\ 
Up to now, we assumed the stellar and halo velocity
  dispersions to be equal. In the following, we
  want to address the possible difference between stellar
  velocity dispersion and dark matter halo velocity dispersion.
We have shown in Sec. \ref{sec:point-like modeling:model output} that we
constrain the mass profile of our cluster galaxies most strongly at a
scale of $\sim5$ effective radii. 
This is where dark matter dominates and thus we now
can compare the halo velocity dispersion derived from lensing with the
stellar velocity dispersion amplitude. An estimate for the stellar
velocity dispersion amplitude can be obtained from the Faber-Jackson
relation \citep{faber76} or the Fundamental Plane \citep{bender1992}.
Stars in elliptical galaxies are
dynamically colder than their dark matter halo (see
\cite{ortwin_giant_ellitpcals}) and their velocity dispersion is
linked to the maximum
circular halo velocity as $\sigma_{\rm stars}= 0.66 v_{\rm circ}^{\rm
  max}$ (at least for the sample of elliptical galaxies investigated in
\cite{ortwin_giant_ellitpcals}, see their Eq. (2)). Therefore we would expect the
halo velocity dispersion to be $\sigma_{\rm halo}=1.07 \sigma_{\rm
  stars}$. Our best fit halo velocity dispersions in
  Fig. \ref{modeling:point-like:FJ_vs_lensing} are slightly smaller
than those of the stars according to the FJ relation
derived in \cite{kormendy2013}, but considering the uncertainty on the measured
  halo velocity dispersion, this is not significant.
We would need a more
precise global cluster model (to decrease the error on the halo velocity
dispersions) and spectroscopic stellar velocity dispersions for the
red cluster members to measure the relation between halo and stellar
velocity dispersion more precisely.

\subsection{Halo truncation and stripped mass fraction}
\begin{figure}[tbh]
\includegraphics[height=80mm]{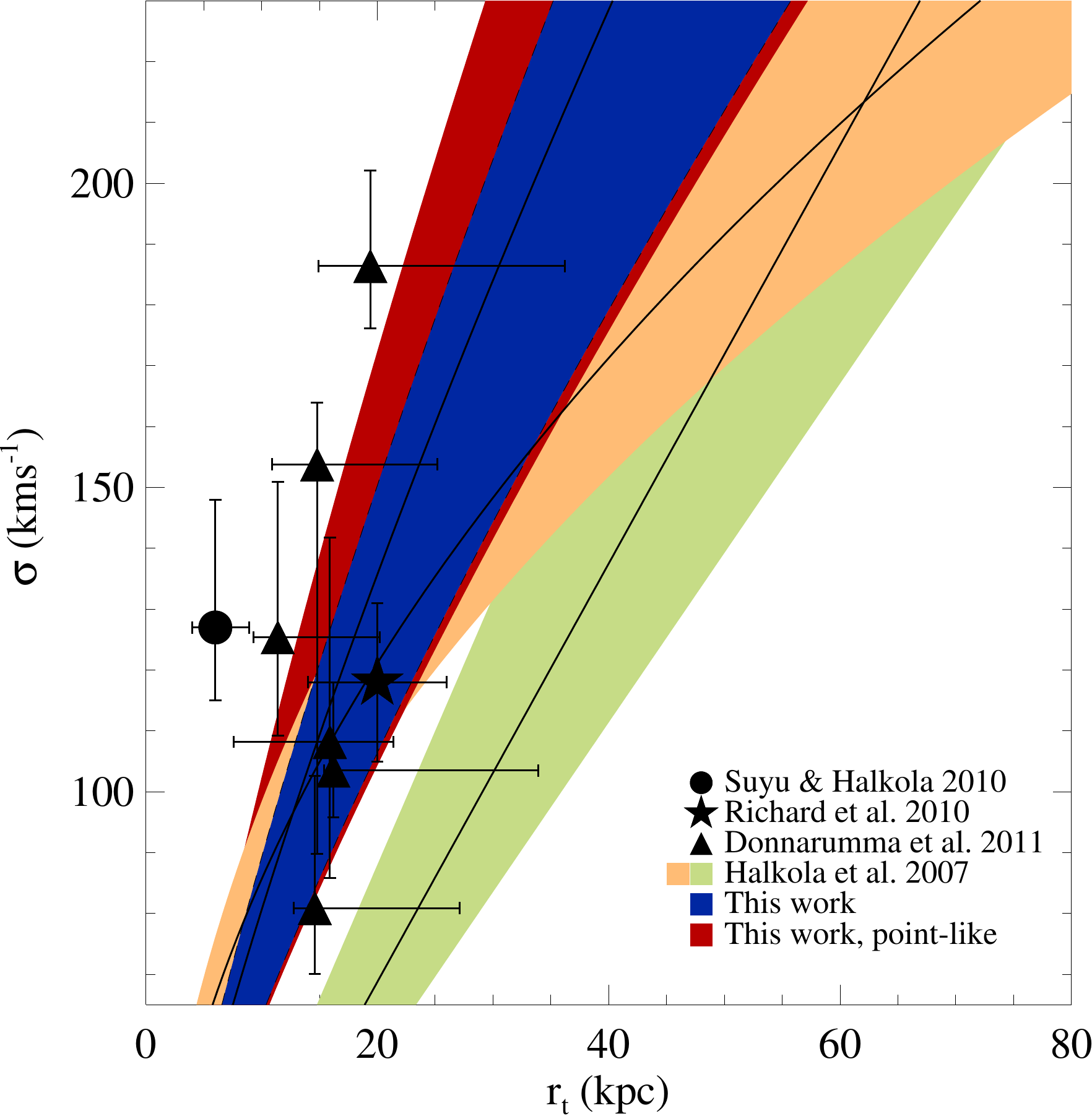}
\caption{This figure shows our results for the halo truncation radius
  vs velocity dispersion for the point source modeling (red region
  marks the $68\%$ confidence interval) and the SFB-modeling of the arc (best fit is
  the black line, and the 1 sigma confidence region is shown in blue). 
The triangles mark constraints (and their 1 sigma errors) for individual galaxies obtained by
  \cite{donnarumma11}, the star marks the result for one galaxy from
  \cite{richard2010}, the point is taken from
 \cite{suyuhalkola2010}. The light green and light orange marks the 
$1 \sigma$ confidence intervals obtained from \cite{halkola07} for two different
scaling relations, $\rt\sim\sigma$ and $\rt\sim\sigma^2$, as analyzed in their work.}
\label{discussion:literature truncation}
\end{figure}

The truncation radius vs velocity dispersion relation for the halo of
cluster members is
\begin{equation}
\rt=(35\pm8 \rm kpc)\left(\frac{\sigma}{186\rm
    kms^{-1}}\right)^{\frac{4}{3}}
\quad ,
\label{rtvssigma}
\end{equation}
from the full surface brightness reconstruction of the extended arc
and its counterimage, based on 4 nearby cluster galaxies. We get a very similar relation for the
point-like modeling, which includes all cluster members statistically.
We have shown in Fig. \ref{modelling:point-like:output:Einsteinradii} that 
the galaxies contributing most strongly to our point-like
halo truncation measurement have velocity dispersions between
$100\rm kms^{-1}$ and $200\rm kms^{-1}$. 
In Eq. \ref{rtvssigma}
the exponent $\frac{4}{3}$ is assumed to be known and the
amplitude is determined.  
As can be seen in Fig. \ref{discussion:literature
  truncation} the errors on this relation in the range of $100\rm kms^{-1}$ and $200\rm kms^{-1}$ 
are quite large, hence different exponents for the truncation vs
velocity dispersion law fit the multiple image positions equally well, 
as long similar values for the actual truncation radii of the most
relevant individual galaxies are predicted.
If the exponent was
changed to 2 the results are still very similar for the majority of
galaxies and we get a similar fit quality.
Our velocity dispersion vs truncation
radius relation is shown in Fig. \ref{discussion:literature
  truncation}  where the error intervals obtained
from the point like modeling are in red and the
 errors for the extended SFB modeling are in blue.
\\
Since the halo velocity dispersion is not a direct observable a more
practical relation than Eq. \ref{rtvssigma} is to rephrase the upper equation as a
function of apparent $m_{AB, 160}$ magnitude,
\begin{equation}
\log r_{\rm t}[\rm kpc]=\log(35\pm8)-0.16 m_{160, AB} +2.96
\label{discussion: halo truncation:logrt vs M160}
\end{equation} 
such that it gives a recipe to model the galaxy halos
also for other clusters at the same redshift. To obtain a redshift
independent relation we transform Eq. \ref{discussion: halo
  truncation:logrt vs M160} to relate the truncation radius of
each galaxy directly to its absolute B-band magnitude (in Vega). We obtain:
\begin{equation}
\log \rt [\rm kpc]=\log(35\pm 8)-0.16 M_{B, Vega}-3.372
\label{discussion:halo truncation:log rt vs MB}
\end{equation} 
\\
This equation holds for the red galaxies in
  Fig. \ref{modeling:point like:galaxy lenses}.
We now compare our results with previous work on the 
truncation of galaxies halos in clusters of galaxies:
\cite{halkola07} do a statistical analysis of all galaxies in the
strong lensing regime of the cluster A1689. 
Although they include galaxies in the modeling with (Fundamental
plane and Faber-Jackson) velocity
dispersion estimates from about $300\rm kms^{-1}$ down to about $20\rm
kms^{-1}$ (see Fig. 5 \citealt{halkola06}) in their sample it seems that their
sensitivity for halo truncation is mostly due to massive galaxies with
a velocity dispersion of $220\rm kms^{-1}$. This can be seen in
Fig. \ref{discussion:literature truncation} which shows
that the halo truncation size for the two parameterizations
($s\propto \sigma$ and and $s\propto \sigma^2$)
agrees for $\sigma=220$km/s galaxies where the halo size then is equal to
about $65$kpc with a one sigma error of about $15\rm kpc-20\rm kpc$. 
Besides this their Fig. 1 shows that their $\chi^2$ starts to
rise steeply only for halo sizes smaller than $30$kpc. This implies
that their result is in agreement with ours.
\\
The work of \cite{richard2010} and \cite{donnarumma11} allows a more
direct comparison to our results since they analyze a situation more
similar to ours. Their cluster galaxies have mostly low velocity
dispersion (triangle and stars in Fig. \ref{discussion:literature truncation}) and they typically have a
projected distance to the cluster center of the order of $\approx
10\arcsec$. 
\\
 Our median ``lensing-weighted''
cluster galaxy distance to the cluster center is $\sim26\arcsec$ (the 4
cluster members close to the arc have a distance of $\sim20\arcsec$
which is 6\% of the virial radius of this cluster, \citealt{ume12}).
This means that our galaxy
sample and that of \cite{richard2010} and \cite{donnarumma11} is
likely to have undergone a similar amount of stripping (assuming that
the central cluster density and the collapse state of their clusters
is similar to ours). The results of \cite{richard2010} and
\cite{donnarumma11} are inserted into Fig. \ref{discussion:literature
truncation} and are in agreement with ours.
\\
\cite{suyuhalkola2010} measure the individual truncation of a
satellite halo embedded in a group (for which we estimate a velocity
dispersion of about $400\rm kms^{-1}$ to $500\rm kms^{-1}$ based on their lensing model)
where the satellite is only $\sim26 \rm kpc$ away in projection from the group center.
They estimate the velocity dispersion of the satellite galaxy to be
around $120$km/s and have a truncation radius of only $4-9$kpc at
$95\%$ confidence. Their result shows that indeed halo truncation can be
severe close to centers of groups (and thus even more for clusters).
\\
With a different method, \cite{limousin07} measure the
truncation of cluster galaxies with weak lensing for 5 different
clusters and get similar results within the errors. 
\cite{pu2010} investigate 3 nearby group members using dynamical
modeling. They use a common cutoff-radius for all three galaxies with
velocity dispersions between $\sigma \approx 200 \rm kms^{-1}$ and
$\sigma \approx 300 \rm kms^{-1}$, somewhat higher than our
sample. Their best-fit value is $\rm R_c =60 kpc$ which would agree
with our measurement if we extrapolate to higher velocity
dispersions. 
\\
We compare our value for the
truncation radius with the half mass radius derived in
\citet{limousin2009} from simulations of halo stripping in 2
numerically simulated clusters, one with a
similar virial mass as MACSJ1206.2-0847. Our galaxy G4 in Table
\ref{modelling:extended:most likely values clusters} has a
truncation radius of $24\pm6 \rm kpc$ and a R-band rest-frame
luminosity of $\rm L_{R,rf}\approx 3*10^{11}L_{R,\odot}$. At this
luminosity, \citet{limousin2009} get a half mass radius of $\rm
r_{1/2}\approx20 kpc$ for a galaxy close to cluster center in
projection, which agrees well with our result.
\\
We can infer the amount of stripped dark matter for cluster galaxies if
we compare their truncation radii with the truncation radii of the
corresponding galaxies in the field. \cite{fabrice12} measure a truncation radius of
$s=245_{-52}^{+64}h_{100}^{-1}\rm kpc$
for a  reference galaxy with $\sigma=144\rm kms^{-1}$, with red SED
and in underdense environments.  For the same velocity dispersion our
cluster galaxies have a truncation radius of $\rt=25\pm6\rm \, kpc$. Consequently the
ratio for the total halo mass in the field and in the cluster for
this kind of galaxy are
$\rm M_{tot,field}/M_{tot,cluster}=13.9_{-4.4}^{+4.9}$.
In the last step we have assumed that ``the velocity dispersion''
(i.e. kinematics of stars and central dark matter particles) of a
halo does not change when it is stripped during cluster infall.
Models of massive galaxies \citep{pu2010} indeed suggest that a change in
the halo truncation radius (as long as it happens beyond $\sim5 R_{\rm
  eff}$)
has no detectable influence on the stellar kinematics inside $\sim 5
R_{\rm eff}$.
(J. Thomas, private communication).
The truncation radius for GR is $\sim5$ times higher
  than the effective radius of this galaxy. \cite{romanishin1986} give
  a relation for the absolute B-band magnitude $M_{\rm B}\sim-2.06
  \log R_{\rm eff}$. This means that $R_{\rm eff}$ drops faster with fainter
  $M_B$ than $\rt$ in Eq. \ref{discussion:halo truncation:log rt vs
    MB}, implying that the $\rt/R_{\rm eff}$ rises for smaller
  fluxes and hence stripping of the galaxies does also not affect the
  kinematics of the lower luminosity galaxies.

The large mass loss of the cluster galaxies (close in projection to the
cluster center) agrees with results from numerical modeling of the
stripping (see also introduction), which shows that mass losses up to
$90\%$ are common for cluster galaxies close to the cluster center \citep{warnick2008}.
\\
If we assume that all cluster galaxies considered in our model have
halo masses of only $10\%$ of their infall mass then the total
stripped mass amounts to
 $\rm M_{\rm stripped}=5.1_{-1.5}^{+1.8}\times10^{13}\rm M_{\odot}$
out to a projected radius of $\approx100\rm kpc$. The total mass estimate at the same radius
is $7.11_{-0.03}^{+0.04}\times10^{13}\rm M_{\odot}$. Within a
projected radius of $\approx 400 \rm kpc$, the ratio of stripped to total cluster mass
gives values of 25 to $50\%$.  This will be an upper value, since
the fractional stripped galaxy halo masses will be smaller in the
outskirts. Nevertheless it implies that a significant fraction
of the smooth dark matter component in the cluster core
originates from cluster members stripped
during the formation and relaxation of the cluster.

\subsection{The SFB-distribution of the source of the giant arc}

\begin{figure}[h]
\centering
\subfigure[]{\includegraphics[width=45mm]{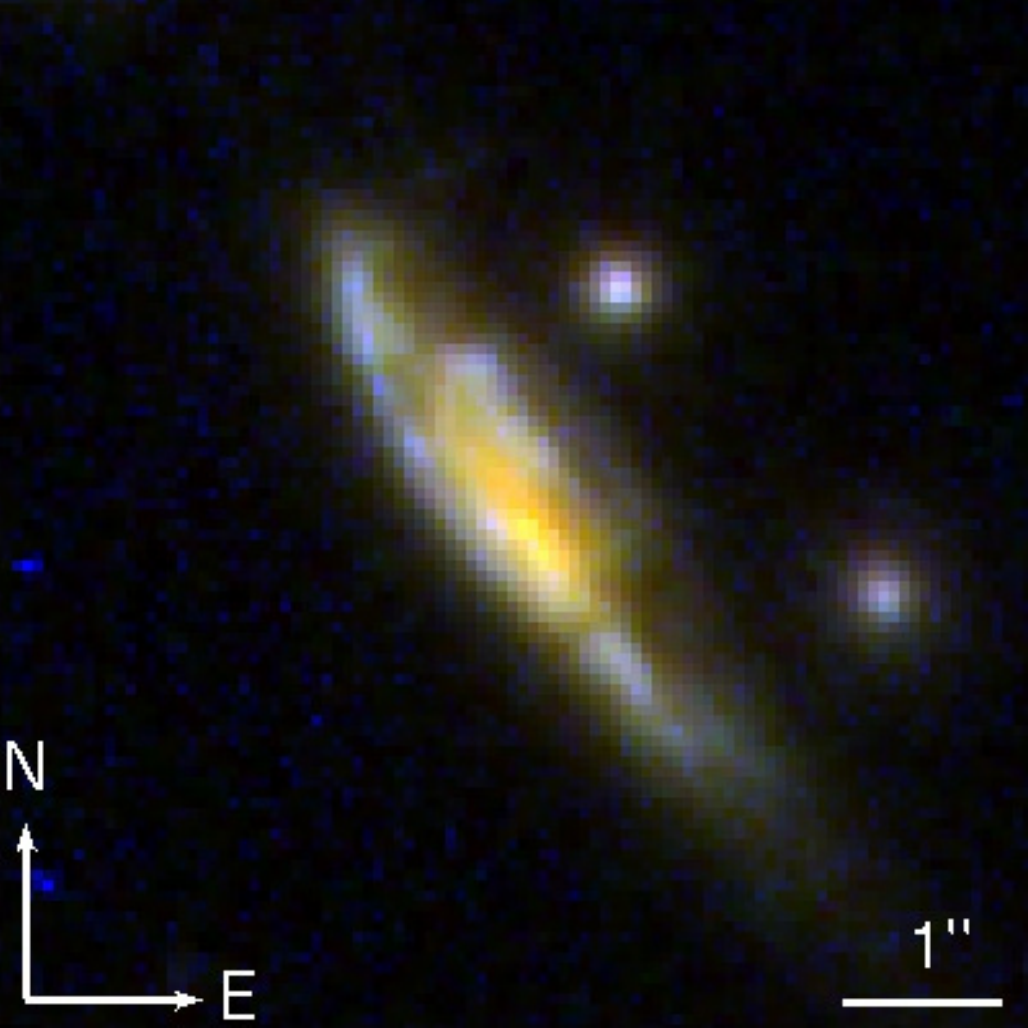}}
\subfigure[]{\includegraphics[width=18.7mm]{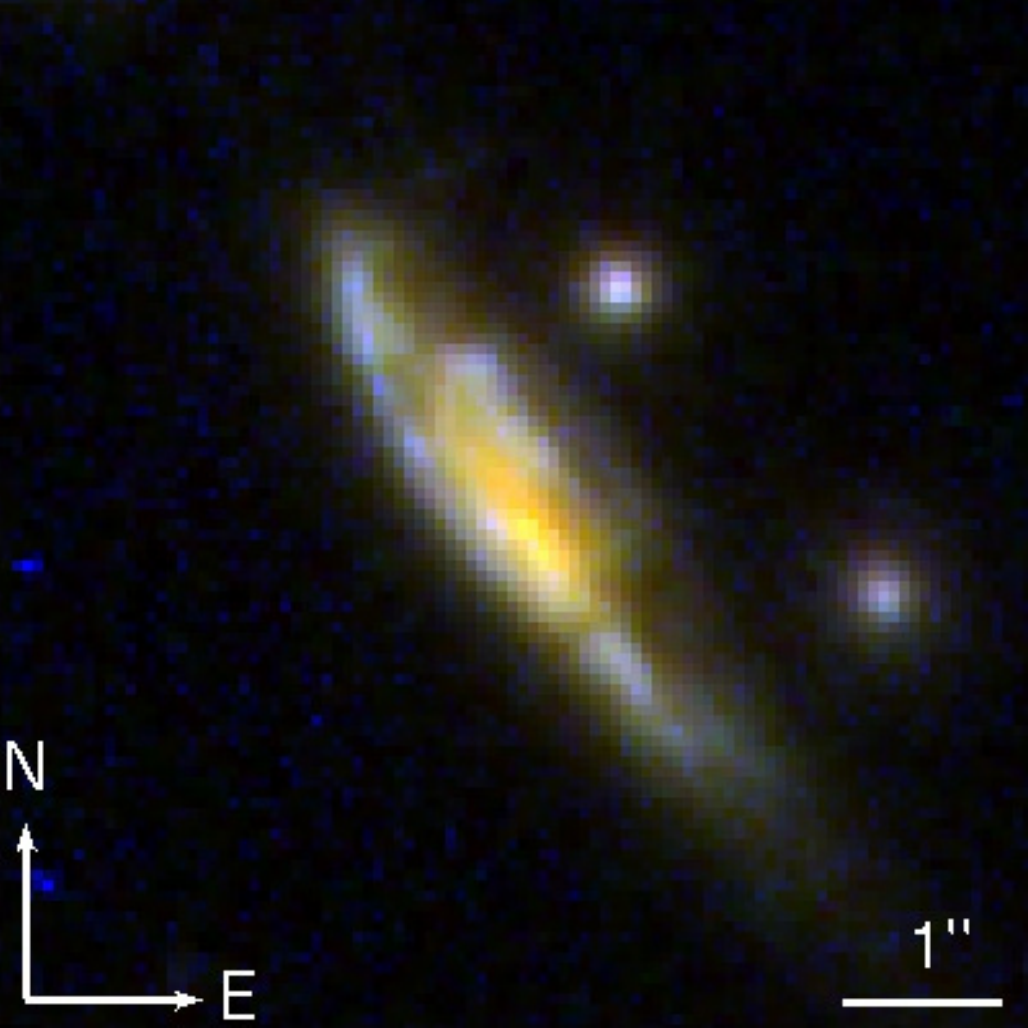}}
\caption{These  false color images use the F775W, F125W and F160W frames,
  corresponding to approximate BRI restframe colors. The left image is the HST observation of the
  counterimage, the right one is the unlensed source at a pixel size observed by HST for
  the unlensed source. The unlensed source is convolved with a
  Gaussian function in each filter representing the approximate
  PSF. In the source plane $1\arcsec$ corresponds to
    $8.13\rm kpc$. We gain an increase in spatial resolution by the
    gravitational telescope of $\sim\sqrt{5.8}$.}
\label{discussion:CANDLES_like_BRI}
\end{figure}

Since not all of the arc source is lensed into the
  giant arc -- basically, all parts above image 1c.1 on the
    counterimage are outside of the caustic and therefore only imaged one
    time in the counterimage and not in the arc -- only the observed
  counterimage can be used to obtain the  source properties.
The observed counterimage and the best-fit source model can be seen in
Fig. \ref{modelling:extended:reconstruction:colors}, 
both at HST resolution and better than HST resolution. Comparing the
observed counterimage and the source at HST resolution, the increase
in the level of detail due to lensing in this case can be seen.
The observed counterimage (Fig. \ref{modelling:extended:crit line
  count}) and the high resolution delensed
counterimage (Fig. \ref{modelling:extended:source
50 pix}) reveal
the magnification of the source due to lensing. The magnification is
approximately equal to $\sim5.8$, this corresponds to a flux
brightening by about 2 magnitudes.
\\
A three color representation of the counterimage in the
F775W, F125W and F160W filters and an approximately delensed 
version of it is shown in Fig. \ref{discussion:CANDLES_like_BRI}.  The filters are
chosen to be equal to the restframe B,
R and I band filters. The color image suggests that the source is a
fairly inclined, spiral star-forming galaxy with a core hosting more
evolved stars. Comparing with CANDLES results (Fig.2 of \citealt{wuyts2012})
we conclude that the lensed galaxy is a fairly normal redshift one
galaxy. Results of the 3D-HST project indicate that about half of the
$1<z<1.5$ galaxies have $H_{\alpha}$ emission lines width with  rest-frame equivalent
widths for the detected galaxies within a $10\rm \AA$ to $130\rm \AA$ for the detected galaxies
\citep{vanDokkum2011}  and that star formation occurs inside out with
$H_{\alpha}$-emission lines in the outskirts of galaxies and continuum
emission from their centers, \cite{nelson2012}. Thus it is likely that our
source has emission lines, too. This makes the galaxy an ideal target for measuring the 2D
 kinematics with the ground based NIR IFUs of KMOS at the VLT.

In Table \ref{discussion:apparent magnitudes of ci and
    sr}, the magnitudes of the counterimage and the source are
  stated. The increase in brightness due to the lensing effect makes
  this galaxy at $z=1.036$ much easier to observe than the unlensed
  source would be.

\begin{table*}
\centering
\caption{apparent magnitudes of the counterimage (CI) and modeled source
  (SR)in AB}
\begin{tabular}{p{0.8cm}|p{0.8cm}p{0.8cm}p{0.8cm}p{0.8cm}p{0.8cm}p{0.8cm}p{0.8cm}p{0.8cm}p{0.8cm}p{0.8cm}p{0.8cm}p{0.8cm}}
\hline
Filter & F435W & F475W & F606W & F625W & F775W & F814W & F850LP &
F105W & F110W & F125W & F140W & F160W \\
\hline
CI & 22.20 & 22.14 & 21.73 & 21.52 & 20.92 & 20.76 & 20.39 & 20.25 &
20.06 & 19.93 & 19.81 & 19.72 \\
SR & 24.11 & 24.05 & 23.64 & 23.43 & 22.83 & 22.67 & 22.30 & 22.16 &
21.97 & 21.84 & 21.72 & 21.63\\
\hline
\end{tabular}
\label{discussion:apparent magnitudes of ci and sr}
\end{table*}

\section{Summary and Conclusions}

In this work, we measure the sizes of galaxies in the center of the
galaxy cluster MACSJ1206.2-0847 using strong lensing.
Measurements of the dark matter halo sizes of distant galaxies are
  rare, since dynamical methods are not yet sensitive enough to achieve this. Hence,
  we rely on the gravitational lensing signal to study truncation of
  elliptical galaxies in a galaxy cluster.
We first build a model
for the cluster mass distribution based on the 12 multiple image
systems with 52 multiple images stated in \citet{zitrin2011}. We model the cluster
  galaxies employing scaling laws based on the NIR fluxes. We then
  derive the average truncation of the galaxy halos by optimization of
  the normalization of
  these scaling laws. Based on this, we
reconstruct the full surface brightness distribution of the giant arc
and its counterimage in this cluster by modeling the
  truncation of the cluster galaxies surrounding the arc separately,
  giving agreeing results for both approaches. In detail, our results are:

\begin{itemize}

\item We get a mean distance of the model predicted
multiple image positions from its input positions of $\sim
0.85\arcsec$.

\item  We measure a mass of
  $\rm M_{tot}\sim7\times 10^{13}M_{\odot}$ within a (cylindrical) radius of $100
  \rm kpc$, which is in good agreement with other studies of this cluster.

\item We model the individual galaxies
  assuming scaling relations with the F160W band flux of each
  galaxy, using the normalizations of these scaling laws as free
  parameters. We refer these normalizations to one reference galaxy and
  calculate values of $r_{\rm t,GR}=41_{-18}^{+34}\rm kpc$ and $\sigma_{\rm
  GR}=236_{-32}^{+29}\rm kms^{-1}$ for it. We constrain the mass distribution
  of cluster galaxies best at $\sim5$ effective radii. Assuming
  passive luminosity evolution for the absolute B-band
  luminosity of the cluster galaxies, we show that our lensing
  derived velocity dispersions agree well with values
  given in \cite{kormendy2013} for
  local elliptical galaxies. 

\item We reconstruct the full surface brightness of the giant arc and its
  counterimage by individually modeling the 4 cluster galaxies closest
  to it. For these 4 galaxies, we calculate values for the individual
  velocity dispersions that agree
  with those derived from the scaling relations. The derived sizes of
  the 4
  galaxies are similar to the sizes derived from the point--like
  lensing model. 
We derive the following truncation law for cluster members when
  reconstructing the full surface brightness distribution of the arc: $$
r_{\rm t}=(35\pm8) \rm kpc\left(\frac{\sigma}{186\rm kms^{-1}}\right)^{\frac{4}{3}}.
$$ This truncation law agrees with predictions from simulations and with
other measurements carried out in dense environments. Testing
different exponents of the truncation law gives agreeing results for
the sizes of the individual galaxies within the error ranges, meaning that
we cannot constrain the exponent of the scaling law.

\item The above stated truncation law means that large fractions of
  the dark matter halos of the cluster galaxies in this cluster have
  been stripped from their host galaxies when compared to field galaxies of the
  same velocity dispersion. Again, this agrees with expectations from
  simulations.

\end{itemize}

In summary, the investigated galaxies in MACSJ1206.2-0847 have 
shrunk significantly, which is  consistently derived from both point-like modeling of all
multiple image systems and from modeling the full surface brightness of
the arc and its counterimage. The results for the
  sizes of the galaxies in the center of this cluster at $\rm z=0.44$
  agree with results derived for other clusters at lower
  redshifts, e.g. Abell 1689 (z=0.183) or the Coma cluster, indicating that most of
  the truncation of galaxies close to the cluster center has already
  been completed for MACSJ1206.2-0847 at $\rm z=0.44$.\\
The analysis presented here can be extended to other clusters in the
CLASH survey, e.g. MACSJ1149.6+2223 and Abell 383, leading to a more
complete picture of galaxy sizes in dense environments and -- closely
connected -- their relation with the cluster--scale dark matter halo.


\acknowledgements
This work is supported by the Transregional Collaborative Research Centre TRR
33 - The Dark Universe and the DFG cluster of excellence ``Origin
and Structure of the Universe''.
We thank Jens Thomas, Ralf Bender and Roberto P. Saglia on fruitful
discussions of the properties of early--type galaxies.
We thank the anonymous referee for his/her comments
  and suggestions to improve the text.
Based on observations made with the NASA/ESA Hubble Space Telescope,
obtained from the data archive at the Space Telescope Science
Institute. STScI is operated by the Association of Universities for
Research in Astronomy, Inc. under NASA contract NAS 5-26555.The CLASH
Multi-Cycle Treasury Program (GO-12065) is based on observations made
with the NASA/ESA Hubble Space Telescope. The Space Telescope Science
Institute is operated by the Association of Universities for Research
in Astronomy, Inc. under NASA contract NAS 5-26555. Part of this work
is based on data collected at
the Very Large Telescope at the ESO Paranal Observatory, under Programme ID 186.A-0798.
K.U. acknowledges partial support from the National Science Council of
Taiwan grant NSC100-2112-M-001-008-MY3 and from the Academia Sinica
Career Development Award. AZ is supported by contract
research ``Internationale Spitzenforschung II-1'' of the Baden
W\" urttemberg Stiftung. The Dark
Cosmology Centre is funded by the DNRF
\clearpage
    \bibliography{MACSJ1206.2_truncation_draft_v5a.bbl}

\begin{thebibliography}{96}
\expandafter\ifx\csname natexlab\endcsname\relax\def\natexlab#1{#1}\fi

\bibitem[{{Andreon}(1996)}]{andreon1996}
{Andreon}, S. 1996, A\&A, 314, 763

\bibitem[{{Auger} {et~al.}(2010){Auger}, {Treu}, {Bolton}, {Gavazzi},
  {Koopmans}, {Marshall}, {Moustakas}, \& {Burles}}]{slacs10}
{Auger}, M.~W., {Treu}, T., {Bolton}, A.~S., {Gavazzi}, R., {Koopmans},
  L.~V.~E., {Marshall}, P.~J., {Moustakas}, L.~A., \& {Burles}, S. 2010, ApJ,
  724, 511

\bibitem[{{Barbera} {et~al.}(2011){Barbera}, {Lopes}, \& {de
  Carvalho}}]{barbera2011}
{Barbera}, F.~L., {Lopes}, P.~A.~A., \& {de Carvalho}, R.~R. 2011, {The
  Fundamental Plane of Early-Type Galaxies: Environmental Dependence from g
  Through K}, ed. I.~{Ferreras} \& A.~{Pasquali}, 79

\bibitem[{{Bender} {et~al.}(1992){Bender}, {Burstein}, \& {Faber}}]{bender1992}
{Bender}, R., {Burstein}, D., \& {Faber}, S.~M. 1992, ApJ, 399, 462

\bibitem[{{Bender} {et~al.}(1998){Bender}, {Saglia}, {Ziegler}, {Belloni},
  {Greggio}, {Hopp}, \& {Bruzual}}]{bender1998}
{Bender}, R., {Saglia}, R.~P., {Ziegler}, B., {Belloni}, P., {Greggio}, L.,
  {Hopp}, U., \& {Bruzual}, G. 1998, ApJ, 493, 529

\bibitem[{{Ben{\'{\i}}tez}(2000)}]{bpz2000}
{Ben{\'{\i}}tez}, N. 2000, ApJ, 536, 571

\bibitem[{{Ben{\'{\i}}tez} {et~al.}(2004){Ben{\'{\i}}tez}, {Ford}, {Bouwens},
  {Menanteau}, {Blakeslee}, {Gronwall}, {Illingworth}, {Meurer}, {Broadhurst},
  {Clampin}, {Franx}, {Hartig}, {Magee}, {Sirianni}, {Ardila}, {Bartko},
  {Brown}, {Burrows}, {Cheng}, {Cross}, {Feldman}, {Golimowski}, {Infante},
  {Kimble}, {Krist}, {Lesser}, {Levay}, {Martel}, {Miley}, {Postman}, {Rosati},
  {Sparks}, {Tran}, {Tsvetanov}, {White}, \& {Zheng}}]{bpz2004}
{Ben{\'{\i}}tez}, N., {et~al.} 2004, ApJS, 150, 1

\bibitem[{{Bernardi} {et~al.}(2003){Bernardi}, {Sheth}, {Annis}, {Burles},
  {Eisenstein}, {Finkbeiner}, {Hogg}, {Lupton}, {Schlegel}, {SubbaRao},
  {Bahcall}, {Blakeslee}, {Brinkmann}, {Castander}, {Connolly}, {Csabai},
  {Doi}, {Fukugita}, {Frieman}, {Heckman}, {Hennessy}, {Ivezi{\'c}}, {Knapp},
  {Lamb}, {McKay}, {Munn}, {Nichol}, {Okamura}, {Schneider}, {Thakar}, \&
  {York}}]{bernardi2003a}
{Bernardi}, M., {et~al.} 2003, AJ, 125, 1849

\bibitem[{{Bertin} \& {Arnouts}(1996)}]{sextractor}
{Bertin}, E., \& {Arnouts}, S. 1996, A\&AS, 117, 393

\bibitem[{{Bolton} {et~al.}(2008){Bolton}, {Treu}, {Koopmans}, {Gavazzi},
  {Moustakas}, {Burles}, {Schlegel}, \& {Wayth}}]{slacs7}
{Bolton}, A.~S., {Treu}, T., {Koopmans}, L.~V.~E., {Gavazzi}, R., {Moustakas},
  L.~A., {Burles}, S., {Schlegel}, D.~J., \& {Wayth}, R. 2008, ApJ, 684, 248

\bibitem[{{Brainerd} {et~al.}(1996){Brainerd}, {Blandford}, \&
  {Smail}}]{bbs_profile_ref}
{Brainerd}, T.~G., {Blandford}, R.~D., \& {Smail}, I. 1996, ApJ, 466, 623

\bibitem[{{Brimioulle} {et~al.}(2013){Brimioulle}, {Seitz}, {Lerchster},
  {Bender}, \& {Snigula}}]{fabrice12}
{Brimioulle}, F., {Seitz}, S., {Lerchster}, M., {Bender}, R., \& {Snigula}, J.
  2013, MNRAS, 432, 1046

\bibitem[{{Broadhurst} {et~al.}(2005){Broadhurst}, {Ben{\'{\i}}tez}, {Coe},
  {Sharon}, {Zekser}, {White}, {Ford}, {Bouwens}, {Blakeslee}, {Clampin},
  {Cross}, {Franx}, {Frye}, {Hartig}, {Illingworth}, {Infante}, {Menanteau},
  {Meurer}, {Postman}, {Ardila}, {Bartko}, {Brown}, {Burrows}, {Cheng},
  {Feldman}, {Golimowski}, {Goto}, {Gronwall}, {Herranz}, {Holden}, {Homeier},
  {Krist}, {Lesser}, {Martel}, {Miley}, {Rosati}, {Sirianni}, {Sparks},
  {Steindling}, {Tran}, {Tsvetanov}, \& {Zheng}}]{broadhurst2005}
{Broadhurst}, T., {et~al.} 2005, ApJ, 621, 53

\bibitem[{{Budzynski} {et~al.}(2012){Budzynski}, {Koposov}, {McCarthy},
  {McGee}, \& {Belokurov}}]{budzynski2012}
{Budzynski}, J.~M., {Koposov}, S.~E., {McCarthy}, I.~G., {McGee}, S.~L., \&
  {Belokurov}, V. 2012, MNRAS, 423, 104

\bibitem[{{Coe} {et~al.}(2010){Coe}, {Ben{\'{\i}}tez}, {Broadhurst}, \&
  {Moustakas}}]{coe2010}
{Coe}, D., {Ben{\'{\i}}tez}, N., {Broadhurst}, T., \& {Moustakas}, L.~A. 2010,
  ApJ, 723, 1678

\bibitem[{{Coe} {et~al.}(2006){Coe}, {Ben{\'{\i}}tez}, {S{\'a}nchez}, {Jee},
  {Bouwens}, \& {Ford}}]{coe2006}
{Coe}, D., {Ben{\'{\i}}tez}, N., {S{\'a}nchez}, S.~F., {Jee}, M., {Bouwens},
  R., \& {Ford}, H. 2006, AJ, 132, 926

\bibitem[{{Colley} {et~al.}(1996){Colley}, {Tyson}, \& {Turner}}]{colley1996}
{Colley}, W.~N., {Tyson}, J.~A., \& {Turner}, E.~L. 1996, ApJ, 461, L83

\bibitem[{{de Vaucouleurs}(1948)}]{deVauc48}
{de Vaucouleurs}, G. 1948, Annales d'Astrophysique, 11, 247

\bibitem[{{Diego} {et~al.}(2005){Diego}, {Sandvik}, {Protopapas}, {Tegmark},
  {Ben{\'{\i}}tez}, \& {Broadhurst}}]{diego2005}
{Diego}, J.~M., {Sandvik}, H.~B., {Protopapas}, P., {Tegmark}, M.,
  {Ben{\'{\i}}tez}, N., \& {Broadhurst}, T. 2005, MNRAS, 362, 1247

\bibitem[{{Diemand} {et~al.}(2007){Diemand}, {Kuhlen}, \&
  {Madau}}]{diemand2007}
{Diemand}, J., {Kuhlen}, M., \& {Madau}, P. 2007, ApJ, 667, 859

\bibitem[{{Donnarumma} {et~al.}(2011){Donnarumma}, {Ettori}, {Meneghetti},
  {Gavazzi}, {Fort}, {Moscardini}, {Romano}, {Fu}, {Giordano}, {Radovich},
  {Maoli}, {Scaramella}, \& {Richard}}]{donnarumma11}
{Donnarumma}, A., {et~al.} 2011, A\&A, 528, A73

\bibitem[{{Dressler}(1980)}]{dressler80}
{Dressler}, A. 1980, ApJ, 236, 351

\bibitem[{{Dressler} {et~al.}(1997){Dressler}, {Oemler}, {Couch}, {Smail},
  {Ellis}, {Barger}, {Butcher}, {Poggianti}, \& {Sharples}}]{dressler1997}
{Dressler}, A., {et~al.} 1997, ApJ, 490, 577

\bibitem[{{Dunkley} {et~al.}(2005){Dunkley}, {Bucher}, {Ferreira}, {Moodley},
  \& {Skordis}}]{dunkley2005}
{Dunkley}, J., {Bucher}, M., {Ferreira}, P.~G., {Moodley}, K., \& {Skordis}, C.
  2005, MNRAS, 356, 925

\bibitem[{{Ebeling} {et~al.}(2009){Ebeling}, {Ma}, {Kneib}, {Jullo},
  {Courtney}, {Barrett}, {Edge}, \& {Le Borgne}}]{ebeling2009}
{Ebeling}, H., {Ma}, C.~J., {Kneib}, J.-P., {Jullo}, E., {Courtney}, N.~J.~D.,
  {Barrett}, E., {Edge}, A.~C., \& {Le Borgne}, J.-F. 2009, MNRAS, 395, 1213

\bibitem[{{Eichner} {et~al.}(2012){Eichner}, {Seitz}, \& {Bauer}}]{sdss1430}
{Eichner}, T., {Seitz}, S., \& {Bauer}, A. 2012, MNRAS, 427, 1918

\bibitem[{{El{\'{\i}}asd{\'o}ttir} {et~al.}(2007){El{\'{\i}}asd{\'o}ttir},
  {Limousin}, {Richard}, {Hjorth}, {Kneib}, {Natarajan}, {Pedersen}, {Jullo},
  \& {Paraficz}}]{eliasdottir2007}
{El{\'{\i}}asd{\'o}ttir}, {\'A}., {et~al.} 2007, ArXiv e-prints

\bibitem[{{Faber} \& {Jackson}(1976)}]{faber76}
{Faber}, S.~M., \& {Jackson}, R.~E. 1976, ApJ, 204, 668

\bibitem[{{Focardi} \& {Malavasi}(2012)}]{focardi2012}
{Focardi}, P., \& {Malavasi}, N. 2012, ApJ, 756, 117

\bibitem[{{Fritz} {et~al.}(2009){Fritz}, {B{\"o}hm}, \& {Ziegler}}]{fritz2009}
{Fritz}, A., {B{\"o}hm}, A., \& {Ziegler}, B.~L. 2009, MNRAS, 393, 1467

\bibitem[{{Gao} {et~al.}(2004{\natexlab{a}}){Gao}, {De Lucia}, {White}, \&
  {Jenkins}}]{gao2004b}
{Gao}, L., {De Lucia}, G., {White}, S.~D.~M., \& {Jenkins}, A.
  2004{\natexlab{a}}, \mnras, 352, L1

\bibitem[{{Gao} {et~al.}(2004{\natexlab{b}}){Gao}, {White}, {Jenkins},
  {Stoehr}, \& {Springel}}]{gao2004}
{Gao}, L., {White}, S.~D.~M., {Jenkins}, A., {Stoehr}, F., \& {Springel}, V.
  2004{\natexlab{b}}, MNRAS, 355, 819

\bibitem[{{Gavazzi} {et~al.}(2007){Gavazzi}, {Treu}, {Rhodes}, {Koopmans},
  {Bolton}, {Burles}, {Massey}, \& {Moustakas}}]{slacs4}
{Gavazzi}, R., {Treu}, T., {Rhodes}, J.~D., {Koopmans}, L.~V.~E., {Bolton},
  A.~S., {Burles}, S., {Massey}, R.~J., \& {Moustakas}, L.~A. 2007, ApJ, 667,
  176

\bibitem[{{Geiger} \& {Schneider}(1999)}]{geiger1999}
{Geiger}, B., \& {Schneider}, P. 1999, MNRAS, 302, 118

\bibitem[{{Gerhard} {et~al.}(2001){Gerhard}, {Kronawitter}, {Saglia}, \&
  {Bender}}]{ortwin_giant_ellitpcals}
{Gerhard}, O., {Kronawitter}, A., {Saglia}, R.~P., \& {Bender}, R. 2001, AJ,
  121, 1936

\bibitem[{{Ghigna} {et~al.}(1998){Ghigna}, {Moore}, {Governato}, {Lake},
  {Quinn}, \& {Stadel}}]{ghigna98}
{Ghigna}, S., {Moore}, B., {Governato}, F., {Lake}, G., {Quinn}, T., \&
  {Stadel}, J. 1998, MNRAS, 300, 146

\bibitem[{{Golse} \& {Kneib}(2002)}]{golse2002}
{Golse}, G., \& {Kneib}, J.-P. 2002, A\&A, 390, 821

\bibitem[{{Grillo} {et~al.}(2010){Grillo}, {Eichner}, {Seitz}, {Bender},
  {Lombardi}, {Gobat}, \& {Bauer}}]{sdss1538}
{Grillo}, C., {Eichner}, T., {Seitz}, S., {Bender}, R., {Lombardi}, M.,
  {Gobat}, R., \& {Bauer}, A. 2010, ApJ, 710, 372

\bibitem[{{Grillo} {et~al.}(2009){Grillo}, {Gobat}, {Lombardi}, \&
  {Rosati}}]{gri09}
{Grillo}, C., {Gobat}, R., {Lombardi}, M., \& {Rosati}, P. 2009, A\&A, 501, 461

\bibitem[{{Halkola} {et~al.}(2008){Halkola}, {Hildebrandt}, {Schrabback},
  {Lombardi}, {Brada{\v c}}, {Erben}, {Schneider}, \& {Wuttke}}]{halkola2008}
{Halkola}, A., {Hildebrandt}, H., {Schrabback}, T., {Lombardi}, M., {Brada{\v
  c}}, M., {Erben}, T., {Schneider}, P., \& {Wuttke}, D. 2008, A\&A, 481, 65

\bibitem[{{Halkola} {et~al.}(2006){Halkola}, {Seitz}, \&
  {Pannella}}]{halkola06}
{Halkola}, A., {Seitz}, S., \& {Pannella}, M. 2006, MNRAS, 372, 1425

\bibitem[{{Halkola} {et~al.}(2007){Halkola}, {Seitz}, \&
  {Pannella}}]{halkola07}
---. 2007, ApJ, 656, 739

\bibitem[{{Hoekstra} {et~al.}(2003){Hoekstra}, {Franx}, {Kuijken}, {Carlberg},
  \& {Yee}}]{hoekstra2003b}
{Hoekstra}, H., {Franx}, M., {Kuijken}, K., {Carlberg}, R.~G., \& {Yee},
  H.~K.~C. 2003, \mnras, 340, 609

\bibitem[{{Hoekstra} {et~al.}(2004){Hoekstra}, {Yee}, \&
  {Gladders}}]{hoekstra2004}
{Hoekstra}, H., {Yee}, H.~K.~C., \& {Gladders}, M.~D. 2004, ApJ, 606, 67

\bibitem[{{Host}(2012)}]{host12}
{Host}, O. 2012, MNRAS, 420, L18

\bibitem[{{Jullo} {et~al.}(2010){Jullo}, {Natarajan}, {Kneib}, {D'Aloisio},
  {Limousin}, {Richard}, \& {Schimd}}]{jullo2010}
{Jullo}, E., {Natarajan}, P., {Kneib}, J.-P., {D'Aloisio}, A., {Limousin}, M.,
  {Richard}, J., \& {Schimd}, C. 2010, Science, 329, 924

\bibitem[{{Knebe} {et~al.}(2008){Knebe}, {Yahagi}, {Kase}, {Lewis}, \&
  {Gibson}}]{knebe2008}
{Knebe}, A., {Yahagi}, H., {Kase}, H., {Lewis}, G., \& {Gibson}, B.~K. 2008,
  MNRAS, 388, L34

\bibitem[{{Komatsu} {et~al.}(2011){Komatsu}, {Smith}, {Dunkley}, {Bennett},
  {Gold}, {Hinshaw}, {Jarosik}, {Larson}, {Nolta}, {Page}, {Spergel},
  {Halpern}, {Hill}, {Kogut}, {Limon}, {Meyer}, {Odegard}, {Tucker}, {Weiland},
  {Wollack}, \& {Wright}}]{wmap7}
{Komatsu}, E., {et~al.} 2011, ApJS, 192, 18

\bibitem[{{Koopmans} {et~al.}(2006){Koopmans}, {Treu}, {Bolton}, {Burles}, \&
  {Moustakas}}]{slacs3}
{Koopmans}, L.~V.~E., {Treu}, T., {Bolton}, A.~S., {Burles}, S., \&
  {Moustakas}, L.~A. 2006, ApJ, 649, 599

\bibitem[{{Kormendy} \& {Bender}(2013)}]{kormendy2013}
{Kormendy}, J., \& {Bender}, R. 2013, ApJ, 769, L5

\bibitem[{{Limousin} {et~al.}(2007{\natexlab{a}}){Limousin}, {Kneib},
  {Bardeau}, {Natarajan}, {Czoske}, {Smail}, {Ebeling}, \&
  {Smith}}]{limousin07}
{Limousin}, M., {Kneib}, J.~P., {Bardeau}, S., {Natarajan}, P., {Czoske}, O.,
  {Smail}, I., {Ebeling}, H., \& {Smith}, G.~P. 2007{\natexlab{a}}, A\&A, 461,
  881

\bibitem[{{Limousin} {et~al.}(2009){Limousin}, {Sommer-Larsen}, {Natarajan}, \&
  {Milvang-Jensen}}]{limousin2009}
{Limousin}, M., {Sommer-Larsen}, J., {Natarajan}, P., \& {Milvang-Jensen}, B.
  2009, ApJ, 696, 1771

\bibitem[{{Limousin} {et~al.}(2007{\natexlab{b}}){Limousin}, {Richard},
  {Jullo}, {Kneib}, {Fort}, {Soucail}, {El{\'{\i}}asd{\'o}ttir}, {Natarajan},
  {Ellis}, {Smail}, {Czoske}, {Smith}, {Hudelot}, {Bardeau}, {Ebeling},
  {Egami}, \& {Knudsen}}]{limousin2007}
{Limousin}, M., {et~al.} 2007{\natexlab{b}}, ApJ, 668, 643

\bibitem[{{Limousin} {et~al.}(2008){Limousin}, {Richard}, {Kneib}, {Brink},
  {Pell{\'o}}, {Jullo}, {Tu}, {Sommer-Larsen}, {Egami}, {Micha{\l}owski},
  {Cabanac}, \& {Stark}}]{lim07b}
---. 2008, A\&A, 489, 23

\bibitem[{{Mandelbaum} {et~al.}(2006){Mandelbaum}, {Seljak}, {Cool}, {Blanton},
  {Hirata}, \& {Brinkmann}}]{mandelbaum06}
{Mandelbaum}, R., {Seljak}, U., {Cool}, R.~J., {Blanton}, M., {Hirata}, C.~M.,
  \& {Brinkmann}, J. 2006, MNRAS, 372, 758

\bibitem[{{Matkovi{\'c}} \& {Guzm{\'a}n}(2005)}]{matkovic2005}
{Matkovi{\'c}}, A., \& {Guzm{\'a}n}, R. 2005, MNRAS, 362, 289

\bibitem[{{Merritt}(1983)}]{merritt83}
{Merritt}, D. 1983, ApJ, 264, 24

\bibitem[{{Merritt}(1984)}]{merritt1984}
---. 1984, ApJ, 276, 26

\bibitem[{{Narayan}(1998)}]{narayan1998}
{Narayan}, R. 1998, New Astronomy Reviews, 42, 73

\bibitem[{{Natarajan} {et~al.}(2002{\natexlab{a}}){Natarajan}, {Kneib}, \&
  {Smail}}]{natarajan2002a}
{Natarajan}, P., {Kneib}, J.-P., \& {Smail}, I. 2002{\natexlab{a}}, ApJ, 580,
  L11

\bibitem[{{Natarajan} {et~al.}(2002{\natexlab{b}}){Natarajan}, {Loeb}, {Kneib},
  \& {Smail}}]{natarajan2002b}
{Natarajan}, P., {Loeb}, A., {Kneib}, J.-P., \& {Smail}, I. 2002{\natexlab{b}},
  ApJ, 580, L17

\bibitem[{{Navarro} {et~al.}(1997){Navarro}, {Frenk}, \& {White}}]{nfw}
{Navarro}, J.~F., {Frenk}, C.~S., \& {White}, S.~D.~M. 1997, ApJ, 490, 493

\bibitem[{{Nelson} {et~al.}(2012){Nelson}, {van Dokkum}, {Brammer},
  {F{\"o}rster Schreiber}, {Franx}, {Fumagalli}, {Patel}, {Rix}, {Skelton},
  {Bezanson}, {Da Cunha}, {Kriek}, {Labbe}, {Lundgren}, {Quadri}, \&
  {Schmidt}}]{nelson2012}
{Nelson}, E.~J., {et~al.} 2012, ApJ, 747, L28

\bibitem[{{Nigoche-Netro} {et~al.}(2011){Nigoche-Netro}, {Ruelas}, \&
  {S{\'a}nchez}}]{nigoche2011}
{Nigoche-Netro}, A., {Ruelas}, A., \& {S{\'a}nchez}, L.~J. 2011, in Revista
  Mexicana de Astronomia y Astrofisica Conference Series, Vol.~40, Revista
  Mexicana de Astronomia y Astrofisica Conference Series, 126--126

\bibitem[{{Peng} {et~al.}(2010){Peng}, {Ho}, {Impey}, \& {Rix}}]{galfit3}
{Peng}, C.~Y., {Ho}, L.~C., {Impey}, C.~D., \& {Rix}, H.-W. 2010, AJ, 139, 2097

\bibitem[{{Postman} {et~al.}(2012){Postman}, {Coe}, {Ben{\'{\i}}tez},
  {Bradley}, {Broadhurst}, {Donahue}, {Ford}, {Graur}, {Graves}, {Jouvel},
  {Koekemoer}, {Lemze}, {Medezinski}, {Molino}, {Moustakas}, {Ogaz}, {Riess},
  {Rodney}, {Rosati}, {Umetsu}, {Zheng}, {Zitrin}, {Bartelmann}, {Bouwens},
  {Czakon}, {Golwala}, {Host}, {Infante}, {Jha}, {Jimenez-Teja}, {Kelson},
  {Lahav}, {Lazkoz}, {Maoz}, {McCully}, {Melchior}, {Meneghetti}, {Merten},
  {Moustakas}, {Nonino}, {Patel}, {Reg{\"o}s}, {Sayers}, {Seitz}, \& {Van der
  Wel}}]{postman_clash2011}
{Postman}, M., {et~al.} 2012, ApJS, 199, 25

\bibitem[{{Pu} {et~al.}(2010){Pu}, {Saglia}, {Fabricius}, {Thomas}, {Bender},
  \& {Han}}]{pu2010}
{Pu}, S.~B., {Saglia}, R.~P., {Fabricius}, M.~H., {Thomas}, J., {Bender}, R.,
  \& {Han}, Z. 2010, A\&A, 516, A4

\bibitem[{{Richard} {et~al.}(2010{\natexlab{a}}){Richard}, {Kneib}, {Limousin},
  {Edge}, \& {Jullo}}]{richard2010}
{Richard}, J., {Kneib}, J.-P., {Limousin}, M., {Edge}, A., \& {Jullo}, E.
  2010{\natexlab{a}}, MNRAS, 402, L44

\bibitem[{{Richard} {et~al.}(2010{\natexlab{b}}){Richard}, {Smith}, {Kneib},
  {Ellis}, {Sanderson}, {Pei}, {Targett}, {Sand}, {Swinbank}, {Dannerbauer},
  {Mazzotta}, {Limousin}, {Egami}, {Jullo}, {Hamilton-Morris}, \&
  {Moran}}]{richard2010Locuss}
{Richard}, J., {et~al.} 2010{\natexlab{b}}, MNRAS, 404, 325

\bibitem[{{Romanishin}(1986)}]{romanishin1986}
{Romanishin}, W. 1986, AJ, 91, 76

\bibitem[{{Rusin} {et~al.}(2003){Rusin}, {Kochanek}, {Falco}, {Keeton},
  {McLeod}, {Impey}, {Leh{\'a}r}, {Mu{\~n}oz}, {Peng}, \& {Rix}}]{rusin03}
{Rusin}, D., {et~al.} 2003, ApJ, 587, 143

\bibitem[{{Saglia} {et~al.}(2010){Saglia}, {S{\'a}nchez-Bl{\'a}zquez},
  {Bender}, {Simard}, {Desai}, {Arag{\'o}n-Salamanca}, {Milvang-Jensen},
  {Halliday}, {Jablonka}, {Noll}, {Poggianti}, {Clowe}, {De Lucia},
  {Pell{\'o}}, {Rudnick}, {Valentinuzzi}, {White}, \& {Zaritsky}}]{saglia2010}
{Saglia}, R.~P., {et~al.} 2010, A\&A, 524, A6

\bibitem[{{Sand} {et~al.}(2004){Sand}, {Treu}, {Smith}, \& {Ellis}}]{sand2004}
{Sand}, D.~J., {Treu}, T., {Smith}, G.~P., \& {Ellis}, R.~S. 2004, ApJ, 604, 88

\bibitem[{{Schneider} \& {Rix}(1997)}]{schneider1997}
{Schneider}, P., \& {Rix}, H.-W. 1997, ApJ, 474, 25

\bibitem[{{Seitz} {et~al.}(1998){Seitz}, {Saglia}, {Bender}, {Hopp}, {Belloni},
  \& {Ziegler}}]{sei98}
{Seitz}, S., {Saglia}, R.~P., {Bender}, R., {Hopp}, U., {Belloni}, P., \&
  {Ziegler}, B. 1998, MNRAS, 298, 945

\bibitem[{{S{\'e}rsic}(1963)}]{sersic63}
{S{\'e}rsic}, J.~L. 1963, Boletin de la Asociacion Argentina de Astronomia La
  Plata Argentina, 6, 41

\bibitem[{{Suyu} \& {Halkola}(2010)}]{suyuhalkola2010}
{Suyu}, S.~H., \& {Halkola}, A. 2010, A\&A, 524, A94

\bibitem[{{Suyu} {et~al.}(2006){Suyu}, {Marshall}, {Hobson}, \&
  {Blandford}}]{suyu2006}
{Suyu}, S.~H., {Marshall}, P.~J., {Hobson}, M.~P., \& {Blandford}, R.~D. 2006,
  MNRAS, 371, 983

\bibitem[{{Suyu} {et~al.}(2012){Suyu}, {Hensel}, {McKean}, {Fassnacht}, {Treu},
  {Halkola}, {Norbury}, {Jackson}, {Schneider}, {Thompson}, {Auger},
  {Koopmans}, \& {Matthews}}]{suyu2012}
{Suyu}, S.~H., {et~al.} 2012, ApJ, 750, 10

\bibitem[{{Thomas} {et~al.}(2005){Thomas}, {Maraston}, {Bender}, \& {Mendes de
  Oliveira}}]{thomas05}
{Thomas}, D., {Maraston}, C., {Bender}, R., \& {Mendes de Oliveira}, C. 2005,
  ApJ, 621, 673

\bibitem[{{Thomas} {et~al.}(2009){Thomas}, {Saglia}, {Bender}, {Thomas},
  {Gebhardt}, {Magorrian}, {Corsini}, \& {Wegner}}]{tho09}
{Thomas}, J., {Saglia}, R.~P., {Bender}, R., {Thomas}, D., {Gebhardt}, K.,
  {Magorrian}, J., {Corsini}, E.~M., \& {Wegner}, G. 2009, ApJ, 691, 770

\bibitem[{{Umetsu} {et~al.}(2012){Umetsu}, {Medezinski}, {Nonino}, {Merten},
  {Zitrin}, {Molino}, {Grillo}, {Carrasco}, {Donahue}, {Mahdavi}, {Coe},
  {Postman}, {Koekemoer}, {Czakon}, {Sayers}, {Mroczkowski}, {Golwala}, {Koch},
  {Lin}, {Molnar}, {Rosati}, {Balestra}, {Mercurio}, {Scodeggio}, {Biviano},
  {Anguita}, {Infante}, {Seidel}, {Sendra}, {Jouvel}, {Host}, {Lemze},
  {Broadhurst}, {Meneghetti}, {Moustakas}, {Bartelmann}, {Ben{\'{\i}}tez},
  {Bouwens}, {Bradley}, {Ford}, {Jim{\'e}nez-Teja}, {Kelson}, {Lahav},
  {Melchior}, {Moustakas}, {Ogaz}, {Seitz}, \& {Zheng}}]{ume12}
{Umetsu}, K., {et~al.} 2012, ApJ, 755, 56

\bibitem[{{van Dokkum} \& {van der Marel}(2007)}]{vanDokkum2007}
{van Dokkum}, P.~G., \& {van der Marel}, R.~P. 2007, ApJ, 655, 30

\bibitem[{{van Dokkum} {et~al.}(2011){van Dokkum}, {Brammer}, {Fumagalli},
  {Nelson}, {Franx}, {Rix}, {Kriek}, {Skelton}, {Patel}, {Schmidt}, {Bezanson},
  {Bian}, {da Cunha}, {Erb}, {Fan}, {F{\"o}rster Schreiber}, {Illingworth},
  {Labb{\'e}}, {Lundgren}, {Magee}, {Marchesini}, {McCarthy}, {Muzzin},
  {Quadri}, {Steidel}, {Tal}, {Wake}, {Whitaker}, \&
  {Williams}}]{vanDokkum2011}
{van Dokkum}, P.~G., {et~al.} 2011, ApJ, 743, L15

\bibitem[{{Ventimiglia} {et~al.}(2011){Ventimiglia}, {Arnaboldi}, \&
  {Gerhard}}]{ventimiglia2011}
{Ventimiglia}, G., {Arnaboldi}, M., \& {Gerhard}, O. 2011, A\&A, 528, A24

\bibitem[{{Warnick} {et~al.}(2008){Warnick}, {Knebe}, \& {Power}}]{warnick2008}
{Warnick}, K., {Knebe}, A., \& {Power}, C. 2008, MNRAS, 385, 1859

\bibitem[{{Warren} \& {Dye}(2003)}]{war03}
{Warren}, S.~J., \& {Dye}, S. 2003, ApJ, 590, 673

\bibitem[{{Wegner} {et~al.}(2012){Wegner}, {Corsini}, {Thomas}, {Saglia},
  {Bender}, \& {Pu}}]{wegner2012}
{Wegner}, G.~A., {Corsini}, E.~M., {Thomas}, J., {Saglia}, R.~P., {Bender}, R.,
  \& {Pu}, S.~B. 2012, AJ, 144, 78

\bibitem[{{Wilman} \& {Erwin}(2012)}]{wilman2012}
{Wilman}, D.~J., \& {Erwin}, P. 2012, ApJ, 746, 160

\bibitem[{{Wright} \& {Brainerd}(2000)}]{wb00}
{Wright}, C.~O., \& {Brainerd}, T.~G. 2000, ApJ, 534, 34

\bibitem[{{Wuyts} {et~al.}(2012){Wuyts}, {F{\"o}rster Schreiber}, {Genzel},
  {Guo}, {Barro}, {Bell}, {Dekel}, {Faber}, {Ferguson}, {Giavalisco}, {Grogin},
  {Hathi}, {Huang}, {Kocevski}, {Koekemoer}, {Koo}, {Lotz}, {Lutz}, {McGrath},
  {Newman}, {Rosario}, {Saintonge}, {Tacconi}, {Weiner}, \& {van der
  Wel}}]{wuyts2012}
{Wuyts}, S., {et~al.} 2012, ApJ, 753, 114

\bibitem[{{Ziegler} \& {Bender}(1997)}]{ziegler1997}
{Ziegler}, B.~L., \& {Bender}, R. 1997, MNRAS, 291, 527

\bibitem[{{Zitrin} {et~al.}(2012{\natexlab{a}}){Zitrin}, {Bartelmann},
  {Umetsu}, {Oguri}, \& {Broadhurst}}]{zitrin2012}
{Zitrin}, A., {Bartelmann}, M., {Umetsu}, K., {Oguri}, M., \& {Broadhurst}, T.
  2012{\natexlab{a}}, ArXiv e-prints

\bibitem[{{Zitrin} {et~al.}(2011){Zitrin}, {Broadhurst}, {Barkana}, {Rephaeli},
  \& {Ben{\'{\i}}tez}}]{zitrin2011A}
{Zitrin}, A., {Broadhurst}, T., {Barkana}, R., {Rephaeli}, Y., \&
  {Ben{\'{\i}}tez}, N. 2011, MNRAS, 410, 1939

\bibitem[{{Zitrin} {et~al.}(2012{\natexlab{b}}){Zitrin}, {Rosati}, {Nonino},
  {Grillo}, {Postman}, {Coe}, {Seitz}, {Eichner}, {Broadhurst}, {Jouvel},
  {Balestra}, {Mercurio}, {Scodeggio}, {Ben{\'{\i}}tez}, {Bradley}, {Ford},
  {Host}, {Jimenez-Teja}, {Koekemoer}, {Zheng}, {Bartelmann}, {Bouwens},
  {Czoske}, {Donahue}, {Graur}, {Graves}, {Infante}, {Jha}, {Kelson}, {Lahav},
  {Lazkoz}, {Lemze}, {Lombardi}, {Maoz}, {McCully}, {Medezinski}, {Melchior},
  {Meneghetti}, {Merten}, {Molino}, {Moustakas}, {Ogaz}, {Patel}, {Regoes},
  {Riess}, {Rodney}, {Umetsu}, \& {Van der Wel}}]{zitrin2011}
{Zitrin}, A., {et~al.} 2012{\natexlab{b}}, ApJ, 749, 97

\bibitem[{{Zitrin} {et~al.}(2013){Zitrin}, {Meneghetti}, {Umetsu},
  {Broadhurst}, {Bartelmann}, {Bouwens}, {Bradley}, {Carrasco}, {Coe}, {Ford},
  {Kelson}, {Koekemoer}, {Medezinski}, {Moustakas}, {Moustakas}, {Nonino},
  {Postman}, {Rosati}, {Seidel}, {Seitz}, {Sendra}, {Shu}, {Vega}, \&
  {Zheng}}]{zitrin2013}
---. 2013, ApJ, 762, L30

\end{thebibliography}

\appendix


\section{Galaxy lenses list}
\label{sec:galaxy lenses:list}

In this appendix, we present the list of derived galaxy lenses used
for the strong lensing model in Table \ref{modelling:point like:galaxy lenses list}. We show the position relative to the
BCG, the ellipticity and orientation and the best fit Einstein
and truncation radius from the best-fit model presented in
Sec. \ref{sec:point-like modeling:model output}. The positions are
again given relative to the BCG.

\begin{longtable}{cccccc}
\caption{Derived galaxy lenses}\\
\hline
x\footnotemark[1] & y\footnotemark[1] & q & $\Theta_{\rm q}$ & $\sigma$ & $r_{\rm t}$ \\
$(\arcsec)$ & $(\arcsec)$ & & $(^\circ)$ & $(\rm kms^{-1})$ & $(\rm kpc)$\\
\endfirsthead
x\footnotemark[1] & y\footnotemark[1] & q & $\Theta_{\rm q}$ & $\sigma$ & $r_{\rm t}$ \\
\hline
\endhead
\multicolumn{6}{c}{{Continued on next page}} \\
\endfoot
\endlastfoot
\hline
0.000 & 0.000 & 0.59 & 14.6 & 296.3 & 57.8 \\ 
-51.961 & -18.520 & 0.85 & 28.2 & 235.7 & 42.6 \\ 
52.028 & -23.567 & 0.87 & -47.1 & 196.6 & 33.4 \\ 
70.454 & -24.260 & 0.42 & -78.3 & 191.3 & 32.2 \\ 
79.369 & -16.726 & 0.49 & 62.4 & 188.9 & 31.7 \\ 
-75.133 & -19.806 & 0.56 & 36.5 & 185.7 & 31.0 \\ 
17.846 & 4.499 & 0.68 & -47.3 & 182.3 & 30.2 \\ 
16.918 & -18.981 & 0.37 & 36.7 & 182.3 & 30.2 \\ 
-19.523 & -12.957 & 0.92 & 89.3 & 181.9 & 30.2 \\ 
-58.427 & -2.920 & 0.66 & 8.6 & 180.5 & 29.8 \\ 
36.181 & 44.994 & 0.83 & 57.2 & 179.8 & 29.7 \\ 
-27.249 & 52.149 & 0.53 & 58.0 & 174.4 & 28.5 \\ 
-25.792 & 57.782 & 0.75 & 56.9 & 167.0 & 26.9 \\ 
60.244 & -20.431 & 0.77 & 28.4 & 165.9 & 26.7 \\ 
65.420 & 45.366 & 0.24 & -70.9 & 159.9 & 25.4 \\ 
-8.492 & 75.149 & 0.76 & 31.2 & 154.6 & 24.3 \\ 
56.269 & 39.858 & 0.67 & -88.2 & 147.2 & 22.7 \\ 
-78.600 & -29.236 & 0.81 & 62.5 & 146.7 & 22.6 \\ 
-5.413 & 26.620 & 0.86 & 59.3 & 144.0 & 22.1 \\ 
9.396 & 8.386 & 0.69 & -10.1 & 135.6 & 20.4 \\ 
19.473 & -3.083 & 0.80 & 25.5 & 133.9 & 20.0 \\ 
93.544 & -8.950 & 0.61 & -29.5 & 133.3 & 19.9 \\ 
-8.863 & -49.294 & 0.82 & 39.6 & 133.0 & 19.9 \\ 
17.747 & 51.281 & 0.31 & -6.5 & 130.1 & 19.3 \\ 
-23.885 & 17.186 & 0.84 & -8.0 & 128.2 & 18.9 \\ 
-6.973 & -48.796 & 0.55 & -13.6 & 128.1 & 18.9 \\ 
-1.892 & 7.503 & 0.67 & 34.0 & 124.3 & 18.2 \\ 
10.517 & -52.862 & 0.91 & 44.5 & 121.7 & 17.6 \\ 
22.649 & -53.029 & 0.94 & -7.4 & 120.6 & 17.4 \\ 
9.135 & -16.001 & 0.56 & 67.1 & 117.3 & 16.8 \\ 
-8.101 & -1.448 & 0.77 & -41.9 & 116.4 & 16.6 \\ 
6.194 & 14.554 & 0.75 & -37.1 & 116.4 & 16.6 \\ 
21.592 & 5.996 & 0.79 & 18.5 & 115.7 & 16.5 \\ 
59.033 & 14.592 & 0.76 & 54.0 & 114.5 & 16.3 \\ 
-29.284 & 9.803 & 0.63 & 61.5 & 111.5 & 15.7 \\ 
-9.271 & -10.396 & 0.81 & -52.7 & 111.5 & 15.7 \\ 
67.743 & 5.471 & 0.48 & -81.6 & 111.2 & 15.6 \\ 
46.041 & -9.524 & 0.75 & 83.3 & 110.8 & 15.6 \\ 
-13.786 & 79.453 & 0.95 & 13.8 & 109.3 & 15.3 \\ 
-52.979 & -8.875 & 0.83 & 60.4 & 106.9 & 14.8 \\ 
-66.483 & -38.795 & 0.87 & 38.7 & 105.6 & 14.6 \\ 
52.678 & -9.056 & 0.82 & -35.2 & 105.4 & 14.6 \\ 
-9.375 & -2.275 & 0.82 & -17.6 & 105.2 & 14.5 \\ 
-10.290 & -55.387 & 0.57 & -2.6 & 104.3 & 14.4 \\ 
-72.341 & -21.512 & 0.71 & 13.0 & 104.0 & 14.3 \\ 
-27.184 & 9.263 & 0.46 & 39.5 & 102.2 & 14.0 \\ 
-1.316 & 22.373 & 0.92 & 4.4 & 102.2 & 14.0 \\ 
33.113 & -24.055 & 0.78 & 27.2 & 100.8 & 13.7 \\ 
-56.983 & -13.861 & 0.79 & 19.6 & 100.5 & 13.7 \\ 
20.862 & -6.007 & 0.71 & -74.9 & 98.9 & 13.4 \\ 
77.397 & -10.026 & 0.74 & -76.1 & 97.9 & 13.2 \\ 
-64.503 & 18.440 & 0.74 & -66.3 & 97.6 & 13.1 \\ 
-20.902 & 16.033 & 0.53 & -34.5 & 96.0 & 12.9 \\ 
-21.888 & -34.820 & 0.86 & 16.5 & 93.7 & 12.4 \\ 
-29.310 & -23.851 & 0.82 & -68.7 & 92.7 & 12.3 \\ 
3.169 & 25.294 & 0.68 & -1.7 & 92.3 & 12.2 \\ 
-19.579 & 16.272 & 0.44 & 77.8 & 91.7 & 12.1 \\ 
34.186 & -5.002 & 0.54 & -82.8 & 91.2 & 12.0 \\ 
84.412 & -3.159 & 0.90 & 67.9 & 89.5 & 11.7 \\ 
-10.054 & -32.007 & 0.92 & 76.0 & 85.3 & 11.0 \\ 
-51.558 & -28.504 & 0.63 & -35.0 & 85.1 & 10.9 \\ 
-11.110 & -10.114 & 0.56 & 28.8 & 85.0 & 10.9 \\ 
-56.052 & -46.963 & 0.82 & -29.5 & 84.0 & 10.8 \\ 
-58.147 & 16.382 & 0.91 & 89.8 & 83.3 & 10.6 \\ 
19.309 & 18.352 & 0.54 & -55.4 & 82.6 & 10.5 \\ 
-62.885 & 19.793 & 0.59 & 59.2 & 81.9 & 10.4 \\ 
44.739 & 40.229 & 0.71 & -82.0 & 78.1 & 9.8 \\ 
-6.185 & 23.563 & 0.69 & -4.0 & 77.7 & 9.7 \\ 
-47.318 & 20.110 & 0.42 & 5.7 & 77.3 & 9.6 \\ 
-42.827 & -33.806 & 0.61 & 21.9 & 77.2 & 9.6 \\ 
18.023 & -83.677 & 0.58 & 65.9 & 76.0 & 9.4 \\ 
44.578 & -39.135 & 0.80 & -89.8 & 75.7 & 9.4 \\ 
45.326 & -39.482 & 0.76 & -40.3 & 75.7 & 9.4 \\ 
-54.643 & 12.139 & 0.70 & -40.9 & 75.7 & 9.4 \\ 
-68.528 & 12.943 & 0.95 & 38.1 & 75.3 & 9.3 \\ 
-8.039 & -23.346 & 0.70 & 11.6 & 74.7 & 9.2 \\ 
-63.008 & 9.546 & 0.18 & -37.4 & 74.0 & 9.1 \\ 
-20.924 & 34.030 & 0.37 & 41.5 & 71.6 & 8.7 \\ 
-54.231 & -35.225 & 0.82 & -35.1 & 70.5 & 8.5 \\ 
-25.998 & 7.123 & 0.54 & 33.7 & 68.1 & 8.1 \\ 
-34.878 & -43.036 & 0.49 & 71.2 & 67.7 & 8.1 \\ 
55.964 & -49.356 & 0.58 & -83.0 & 64.8 & 7.6 \\ 
28.577 & 28.456 & 0.82 & 0.9 & 64.7 & 7.6 \\ 
26.103 & -30.477 & 0.97 & 5.7 & 64.4 & 7.5 \\ 
41.714 & 41.585 & 0.67 & 40.0 & 62.5 & 7.3 \\ 
43.372 & -11.300 & 0.70 & -73.7 & 58.1 & 6.6 \\ 
20.281 & 19.628 & 0.77 & -9.3 & 57.2 & 6.5 \\ 
0.946 & 32.097 & 0.77 & 11.2 & 53.1 & 5.8 \\ 
-58.728 & 33.022 & 0.61 & 80.6 & 51.4 & 5.6 \\ 
44.865 & 52.272 & 0.60 & -11.5 & 45.3 & 4.7 \\ 
55.581 & -41.990 & 0.60 & -87.2 & 42.8 & 4.4 \\ 

\hline \\
\footnotetext[1]{relative to the center of the BCG at 12:06:12.134 RA
  (J2000) -08:48:03.35 DEC (J2000)}\\
\label{modelling:point like:galaxy lenses list}
\end{longtable}

\label{lastpage}
\end{document}